\documentclass[preprint2]{aastex}
\usepackage{amsmath}
\usepackage{color}
\usepackage{comment}
\usepackage{url}

\begin{document}
\shorttitle{Candidate Water Vapor Lines to Locate the $\mathrm{H_2O}$ Snowline through High-Dispersion Spectroscopic Observations II. The Case of a Herbig Ae Star}
\shortauthors{Notsu et al.}

\received{2016/10/27} 
\accepted{2017/1/13}

\title{Candidate Water Vapor Lines to Locate the $\mathrm{H_2O}$ Snowline through High-Dispersion Spectroscopic Observations II. The Case of a Herbig Ae Star}
\author{Shota Notsu\altaffilmark{1,8},  Hideko Nomura\altaffilmark{2}, Daiki Ishimoto\altaffilmark{1,2}, Catherine Walsh\altaffilmark{3,4}, Mitsuhiko Honda\altaffilmark{5}, Tomoya Hirota\altaffilmark{6}, and T. J. Millar\altaffilmark{7}}
\affil{\altaffilmark{1}Department of Astronomy, Graduate School of Science, Kyoto University, Kitashirakawa-Oiwake-cho, Sakyo-ku, Kyoto 606-8502, Japan}
\affil{\altaffilmark{2}Department of Earth and Planetary Science, Tokyo Institute of Technology, 2-12-1 Ookayama, Meguro-ku, Tokyo 152-8551, Japan}
\affil{\altaffilmark{3}Leiden Observatory, Leiden University, P.O. Box 9513, 2300 RA Leiden, The Netherlands}
\affil{\altaffilmark{4}School of Physics and Astronomy, University of Leeds, Leeds, LS2 9JT, UK}
\affil{\altaffilmark{5}Department of Physics, School of Medicine, Kurume University, 67 Asahi-machi, Kurume, Fukuoka 830-0011, Japan}
\affil{\altaffilmark{6}National Astronomical Observatory of Japan, 2-21-1 Osawa, Mitaka, Tokyo 181-8588, Japan}
\affil{\altaffilmark{7}Astrophysics Research Centre, School of Mathematics and Physics, Queen's University Belfast, University Road, Belfast, BT7 1NN, UK}
\affil{\altaffilmark{8}Research Fellow of Japan Society for the Promotion of Science (DC1)}
\email{snotsu@kusastro.kyoto-u.ac.jp}
\affil{}
\affil{This paper was received by The Astrophysical Journal (ApJ) on October 27th, 2016, \\ and was accepted on January 13th, 2017.}

\begin{abstract}
\noindent
Observationally measuring the location of the $\mathrm{H_2O}$ snowline is crucial for understanding the planetesimal and planet formation processes, and the origin of water on Earth.
In disks around Herbig Ae stars ($T_{\mathrm{*}}\sim$10,000K$, M_{\mathrm{*}}\gtrsim$2.5$M_{\bigodot}$), the position of the $\mathrm{H_2O}$ snowline is further from the central star compared with that around cooler, and less massive T Tauri stars. Thus, the $\mathrm{H_2O}$ emission line fluxes from the region within the $\mathrm{H_2O}$ snowline are expected to be stronger.
In this paper, we calculate the chemical composition of a Herbig Ae disk using chemical kinetics.
Next, we calculate the $\mathrm{H_2O}$ emission line profiles, 
and investigate the properties of candidate water lines 
across a wide range of wavelengths (from mid-infrared to sub-millimeter) that can locate the position of the $\mathrm{H_2O}$ snowline.
Those line identified have small Einstein $A$ coefficients ($\sim 10^{-6} -10^{-3}$ s$^{-1}$) and relatively high upper state energies ($\sim$ 1000K).
The total fluxes tend to increase with decreasing wavelengths.
We investigate the possibility of future observations (e.g., ALMA, SPICA/SMI-HRS) to locate the position of the $\mathrm{H_2O}$ snowline.
Since the fluxes of those identified lines from Herbig Ae disks are stronger than those from T Tauri disks, the possibility of a successful detection is expected to increase for a Herbig Ae disk. 
\end{abstract}

\keywords{astrochemistry--- protoplanetary disks--- ISM: molecules--- sub-millimeter \& infrared: planetary systems---  stars: formation}

\section{Introduction}
\noindent 
Observationally locating the position of the $\mathrm{H_2O}$ snowline \citep{Hayashi1981,Hayashi1985} in a protoplanetary disk is important. 
It will provide information on the physical and chemical conditions in disks, such as the temperature structure, the dust-grain size distribution, and the water vapor distribution in the disk midplane (e.g., \citealt{Oka2011, Piso2015}), and will give constraints on the current formation theories of planetesimals and planets (e.g., \citealt{Oberg2011, Okuzumi2012, Ros2013}). 
It will help clarify the origin of water on rocky planets including the Earth (e.g., \citealt{Morbidelli2000, Morbidelli2012, Morbidelli2016, Sato2016}).
\citet{Banzatti2015} and \citet{Cieza2016} recently showed that the presence of the $\mathrm{H_2O}$ snowline leads to a sharp discontinuity in the radial profile of the dust emission spectral index, due to the replenishment of small grains through fragmentation because of the change in fragmentation velocities
across the $\mathrm{H_2O}$ snowline.
Through recent space and ground infrared spectroscopic observations for protoplanetary disks, some infrared $\mathrm{H_2O}$ lines, which mainly trace the disk surface, have been detected (for more details, see e.g., \citealt{Pontoppidan2010b, vanDishoeck2014, Blevins2016, Banzatti2016, Notsu2016}).
\\ \\
The velocity profiles of emission lines from protoplanetary disks are usually affected by Doppler shift due to Keplerian rotation and thermal broadening. 
Therefore, the velocity profiles are sensitive to the radial distribution of the line-emitting regions.
In our previous paper (paper I, \citealt{Notsu2016}),
we calculated the chemical composition and the $\mathrm{H_2O}$ line profiles in a T Tauri disk\footnote[1]{In the remainder of this paper, we define the protoplanetary disks around T Tauri/Herbig Ae stars as ``T Tauri/Herbig Ae disks".}, and identified candidate $\mathrm{H_2O}$ lines especially at sub-millimeter wavelengths, to locate the position of the $\mathrm{H_2O}$ snowline through future high-dispersion spectroscopic observations.
Our calculations showed that the fluxes of $\mathrm{H_2O}$ lines with small Einstein $A$ coefficients ($A_{ul}$$\sim 10^{-6} -10^{-3}$ s$^{-1}$) and relatively high upper state energies (E$_{\mathrm{up}}$$\sim$ 1000K) are dominated by the disk region inside the $\mathrm{H_2O}$ snowline. Therefore, their profiles could be used to locate the position of the $\mathrm{H_2O}$ snowline.
This is because the water gas column density of the region inside the $\mathrm{H_2O}$ snowline is high enough that all lines are optically thick as long as $A_{ul} > 10^{-6}$ s$^{-1}$.
On the other hand, the region outside the $\mathrm{H_2O}$ snowline has lower water gas column densities and lines with larger Einstein $A$ coefficients have a more significant contribution to their fluxes since the lines are optically thin.
The wavelengths of those candidate lines we identified to locate the position of the $\mathrm{H_2O}$ snowline overlap with the capabilities of ALMA.
In addition, we calculated the profiles of lines which have been detected by previous spectroscopic observations using $Herschel$ (e.g., the ortho-$\mathrm{H_2O}$ 63.32$\mu$m and 538.29$\mu$m lines).
These lines are less suited to locate the position of the $\mathrm{H_2O}$ snowline, because they are not dominated in flux by the region inside the  snowline.
\\
\\
In this work (paper II), we extend our disk chemical model and the $\mathrm{H_2O}$ line profile calculations to the case of a Herbig Ae disk.
We discuss the differences in disk chemical structures and line properties between the cases of a T Tauri disk (paper I) and a Herbig Ae disk (this paper).
We investigate the line properties in detail for candidate water lines to locate the position of the $\mathrm{H_2O}$ snowline over a wide wavelength range from mid-infrared to sub-millimeter, and discuss the possibility of detecting such lines with future observations.
The methods are outlined in Section 2. The results and discussions are described in Sections 3 and 4, respectively, and the conclusions are listed in Section 5.
\\
\section{Methods}
\begin{figure*}[htbp]
\begin{center}
\includegraphics[scale=0.6]{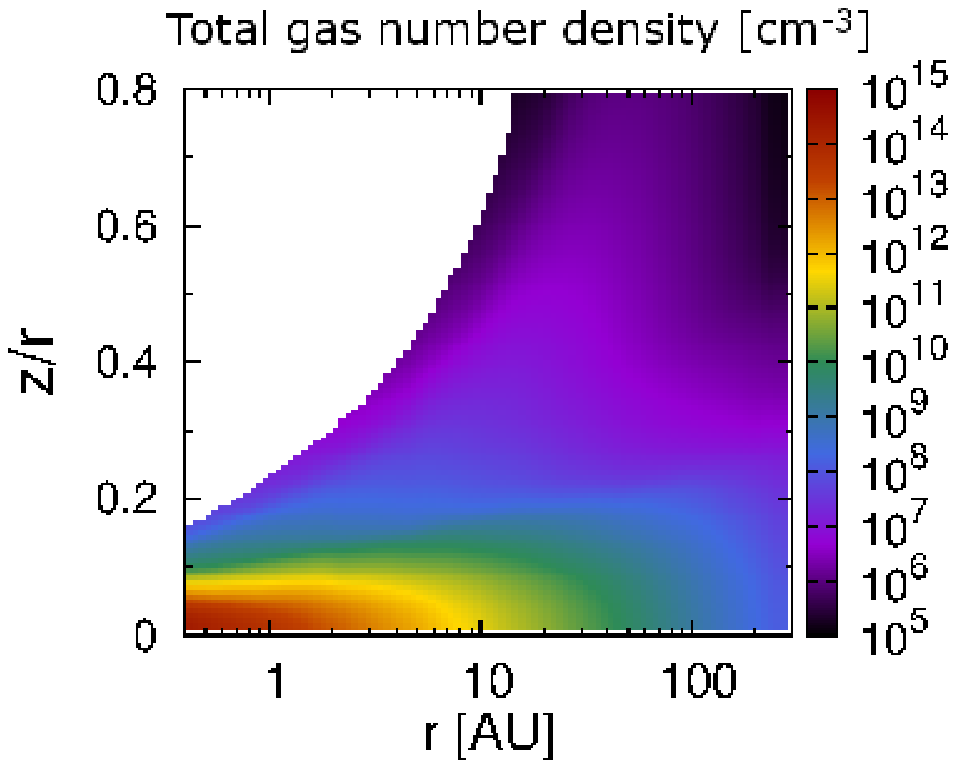}
\includegraphics[scale=0.6]{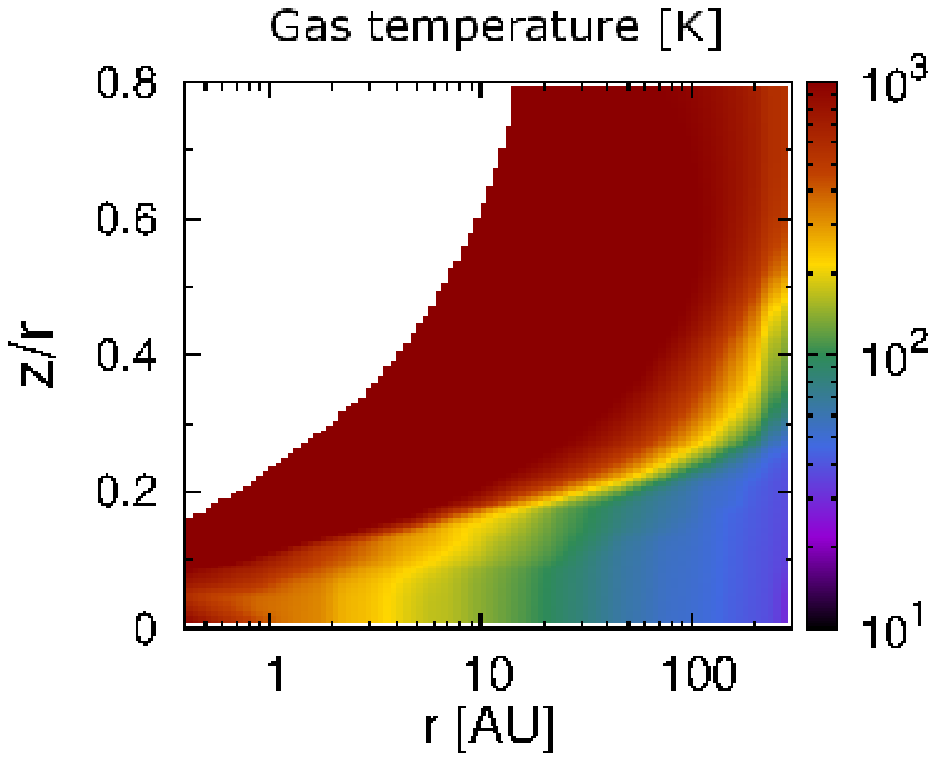}
\includegraphics[scale=0.6]{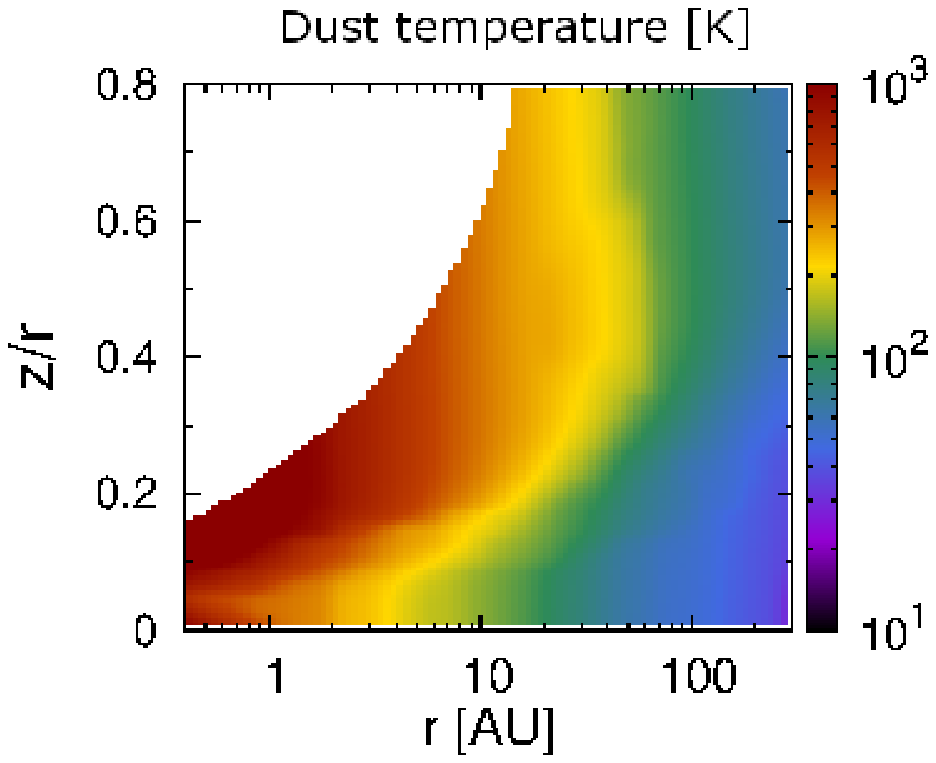}
\includegraphics[scale=0.6]{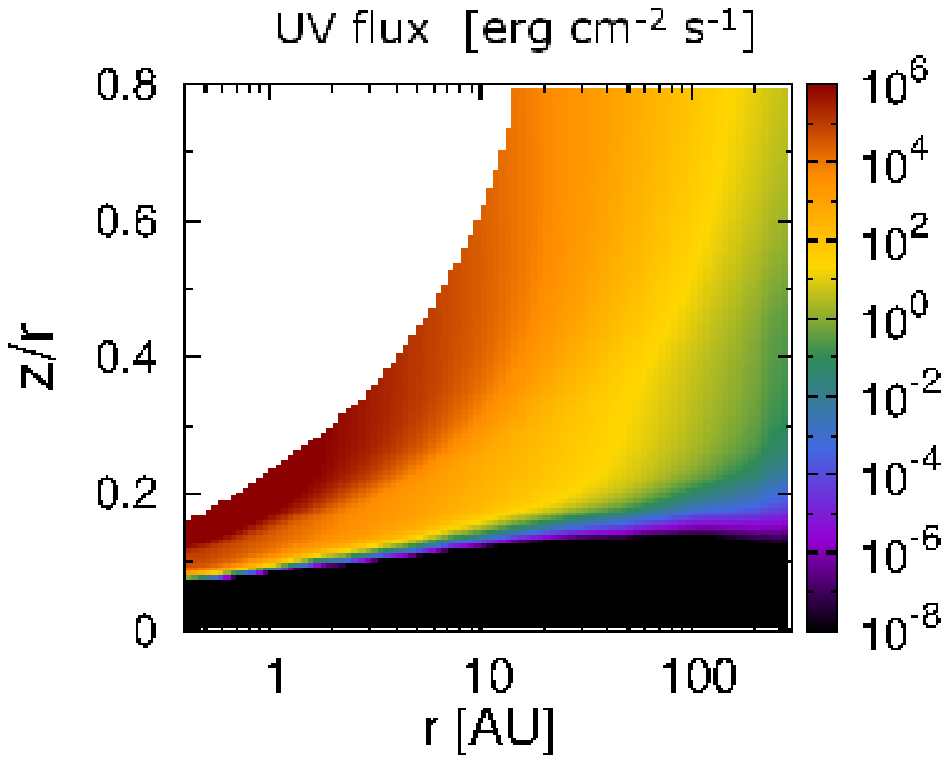}
\end{center}
\caption{The total gas number density in $\mathrm{cm}^{-3}$ (top left), the gas temperature in Kelvin (top right), the dust temperature in Kelvin (bottom left), and the UV flux in erg $\mathrm{cm}^{-2}$ s$^{-1}$ (bottom right) of a Herbig Ae disk as a function of the disk radius in au and height (scaled by the radius, $z/r$) up to maximum radius of $r=$300 au.}\label{Figure1_HerbigAe}
\end{figure*}

\noindent 
The physical structures of the protoplanetary disk models used here are calculated using the methods in 
\citet{NomuraMillar2005} including X-ray heating \citep{Nomura2007}. 
A more detailed description of the background theory and computation of this physical model 
is described in the original papers \citep{NomuraMillar2005, Nomura2007} and paper I \citep{Notsu2016}.
\citet{Walsh2010, Walsh2012, Walsh2014a, Walsh2015}, \citet{Heinzeller2011}, \citet{Furuya2013}, \citet{Notsu2015}, and \citet{Notsu2016}
used the same physical models for a T Tauri disk and a Herbig Ae disk to study various chemical and physical effects, 
and they also describe the calculation of the physical structures in detail.
\\
\\
In paper I \citep{Notsu2016}, we adopted the physical model of a steady, axisymmetric Keplerian disk surrounding a T Tauri star with mass $M_{\mathrm{*}}$=0.5$M_{\bigodot}$, radius $R_{\mathrm{*}}$=2.0$R_{\bigodot}$, and effective temperature $T_{\mathrm{*}}$=4000K \citep{KenyonHartmann1995}.
In this paper, we adopt the physical model of a disk surrounding a 
Herbig Ae star with $M_{\mathrm{*}}$=2.5$M_{\bigodot}$, $R_{\mathrm{*}}$=2.0$R_{\bigodot}$, and $T_{\mathrm{*}}$=10,000K (see also \citealt{Walsh2015}).
In our disk physical models, we adopt a viscous parameter $\alpha$=$10^{-2}$, a mass accretion rate $\dot{M}$=$10^{-8}M_{\bigodot}$ yr$^{-1}$, and gas-to-dust mass ratio $g/d=100$.
The values of total disk mass are $M_{\mathrm{disk}}\sim 2.4\times 10^{-2}$$M_{\bigodot}$ for the T Tauri disk \citep{Heinzeller2011}, and $M_{\mathrm{disk}}\sim 2.5\times 10^{-2}$$M_{\bigodot}$ for the Herbig Ae disk.
We adopt the same compact and spherical dust-grain model of \citet{NomuraMillar2005}.
They assume that dust and gas are well mixed, and that the dust grains consist of silicate grains, carbonaceous grains, and water ices.
They adopt the dust-grain size distribution which is consistent with the extinction curve 
observed in dense clouds \citep{Mathis1977, WeingartnerDraine2001}.
The stellar UV radiation field in our Herbig Ae disk model has no excess emission components (e.g., optically thin hydrogenic bremsstrahlung radiation and Lyman-$\alpha$ line emission), although that in our T Tauri disk model has such excess emission components (for more detail, see \citealt{NomuraMillar2005}, \citealt{Walsh2015} and  \citealt{Notsu2016}).
In Figure \ref{Figure1_HerbigAe}, we display the gas number density in $\mathrm{cm}^{-3}$ (top left), the gas temperature in K (top right, $T_{g}$), the dust-grain temperature in K (bottom left, $T_{d}$), 
and the wavelength-integrated UV flux in erg $\mathrm{cm}^{-2}$ s$^{-1}$ (bottom right) of a Herbig Ae disk as a function of disk radius in au and height (scaled by the radius, $z/r$).
\\ \\
Here we focus on the differences between the physical structures of the T Tauri disk (see Figure 1 of paper I, \citealt{Notsu2016}) and the Herbig Ae disk.
The density in the atmosphere of the Herbig Ae disk (e.g., $n_{\mathrm{H}}= 6\times 10^{10}$ cm$^{-3}$ at $r = 5$ au and $z/r = 0.1$) is lower than that of the T Tauri disk (e.g., $n_{\mathrm{H}}= 2\times 10^{11}$ cm$^{-3}$ at $r = 5$ au and $z/r = 0.1$), because the scale height\footnote[2]{$H=c_{s}/\Omega \propto M_{\mathrm{*}}^{-0.5}$$T_{g}^{0.5}$, where $c_{s}$ and $\Omega$ are the sound speed and Keplerian angular velocity, respectively.} $H$
of the Herbig Ae disk (e.g., $H/r \sim 1.2$ at $r = 5$ au) is smaller than that for the disks around the T Tauri disk (e.g., $H/r \sim 1.7$ at $r = 5$ au).
The gas density and temperature distributions of the disks are obtained self-consistently by iteratively solving the equations for hydrostatic equilibrium in the vertical direction and local thermal balance between heating and cooling of gas \citep{NomuraMillar2005}.
The gas and dust temperatures throughout most of the Herbig Ae disk, and the strength of the UV flux in the disk
surface of the Herbig Ae disk are higher compared with those of the T Tauri disk, although the stellar UV radiation field in our Herbig Ae disk model has no excess emission components, apart from that in our T Tauri disk model. This is because the photospheric blackbody radiative flux from the central Herbig Ae star is larger than that from the central T Tauri star.
The strength of the X-ray flux in the disk surface of the Herbig Ae disk is lower compared with that of the T Tauri disk, since we adopted a smaller value of X-ray luminosity in the Herbig Ae disk ($L_{\mathrm{X}}\sim 3\times 10^{29}$ erg s$^{-1}$) compared with that in the T Tauri disk ($L_{\mathrm{X}}\sim 10^{30}$ erg s$^{-1}$).
\\
\noindent
To investigate the chemical structure of the Herbig Ae disk, we use a large chemical network which includes gas-phase reactions and gas-grain interactions (freeze-out of gas molecules on dust grains, and thermal and non-thermal desorption from dust grains). 
The initial elemental fractional abundances (relative to total hydrogen nuclei density) we
use are the set of atomic oxygen-rich low-metallicity abundances from \citet{Graedel1982}, 
listed in Table 8 of \citet{Woodall2007}, which is the same set as used in paper I \citep{Notsu2016}.
We adopt the same chemical network as described in paper I \citep{Notsu2016}.
\citet{Henning2013}, \citet{Dutrey2014}, and \citet{Haworth2016} reviewed the recent development of calculations for chemical structure in protoplanetary disks.
\\ \\
\noindent Using the $\mathrm{H_2O}$ gas abundance distribution obtained from our chemical calculation described in the previous paragraph, we
calculate the $\mathrm{H_2O}$ emission line profiles ranging from near-infrared to sub-millimeter wavelengths from the Herbig Ae disk assuming Keplerian rotation, and identify the lines which are the best candidates for probing emission from the inner thermally desorbed water reservoir, i.e., within the $\mathrm{H_2O}$ snowline.
We also study how the line fluxes and profile shapes depend on the position of the $\mathrm{H_2O}$ snowline.
In paper I \citep{Notsu2016}, we adopted the same calculation method to determine the $\mathrm{H_2O}$ emission line profiles from a T Tauri disk (based on \citealt{Rybicki1986}, \citealt{Hogerheijde2000}, \citealt{NomuraMillar2005}, and \citealt{Schoier2005}), with the detailed model explained in Section 2.3 of paper I.
The code which we have built for calculating emission line profiles 
is a modification of the original 1D code called RATRAN\footnote[3]{\url{http://home.strw.leidenuniv.nl/~michiel/ratran/}} \citep{Hogerheijde2000}.
We adopt the data of line parameters in the Leiden Atomic and Molecular Database LAMDA\footnote[4]{\url{http://home.strw.leidenuniv.nl/~moldata/}} \citep{Schoier2005}.
Here we note that in our method, we adopt the assumption of local thermal equilibrium (LTE) to obtain the level populations of the water molecule ($n_{u}$ and $n_{l}$).
In Section 4.2, we discuss the validity of this assumption.
In addition, we set the ortho to para ratio (OPR) of water to its high-temperature value of 3 throughout the disk.
\\
\section{Results}
\subsection{The distributions of $\mathrm{H_2O}$ gas and ice}
\begin{figure}[htbp]
\begin{center}
\includegraphics[scale=0.65]{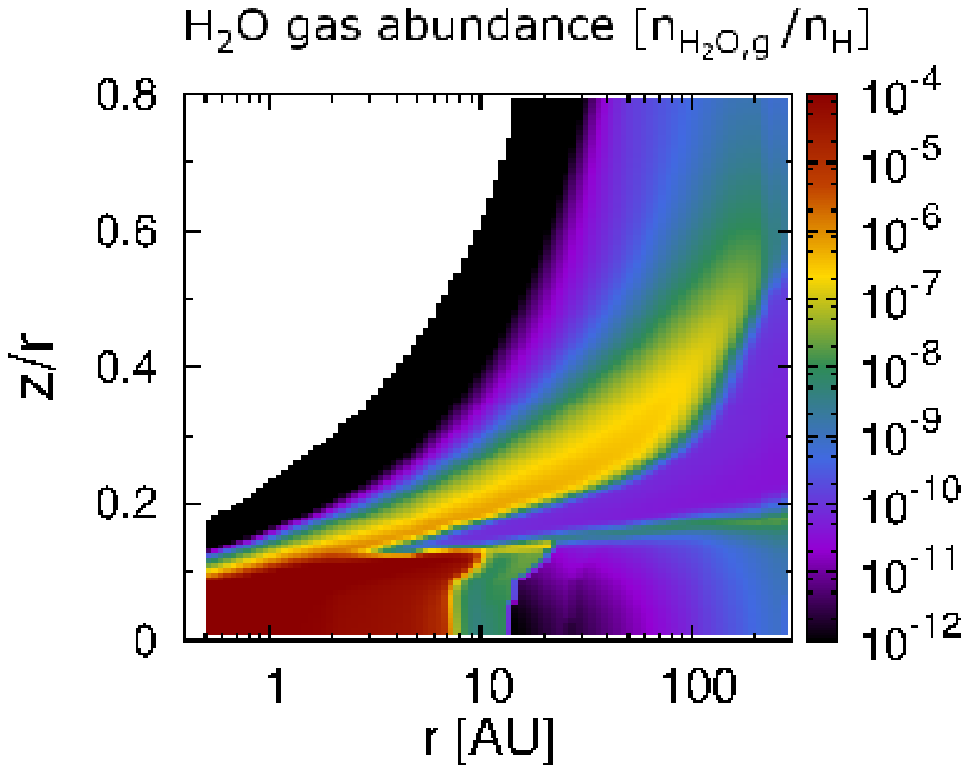}
\includegraphics[scale=0.65]{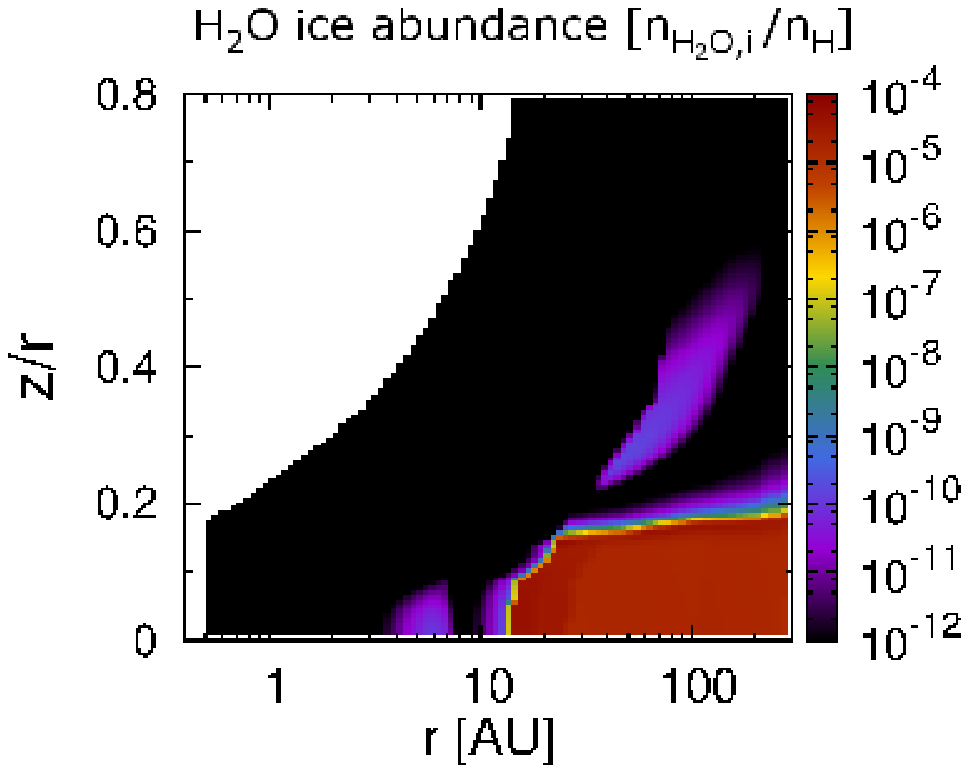}
\end{center}
\caption{\noindent The fractional abundance (relative to total hydrogen nuclei density) distributions of $\mathrm{H_2O}$ gas (top) and $\mathrm{H_2O}$ ice (bottom) of a Herbig Ae disk as a function of disk radius and height (scaled by the radius, $z/r$) up to maximum radius of $r=$300au.}\label{Figure2_HerbigAe}
\end{figure}

\noindent Figure \ref{Figure2_HerbigAe} shows the fractional abundances (relative to total gas hydrogen nuclei density, $n_{\mathrm{H}}$) of $\mathrm{H_2O}$ gas and ice in a Herbig Ae disk as a function of disk radius $r$ and height scaled by the radius ($z/r$).
Here we focus on the differences in $\mathrm{H_2O}$ distributions between the cases of a Herbig Ae disk and a T Tauri disk (see Figure 2 of paper I, \citealt{Notsu2016}).  
 \\ \\
The $\mathrm{H_2O}$ snowline of the Herbig Ae disk exists at a radius of $r\sim$ 14 au in the midplane ($T_{g} \sim T_{d} \sim 120$K), which is significantly larger than that for the T Tauri disk model ($r\sim$ 1.6 au, see Figure 2 of paper I, \citealt{Notsu2016}). This is because the gas and dust 
temperatures, which are coupled in the midplane of the Herbig Ae disk, are higher than that of our T Tauri disk.
Inside the $\mathrm{H_2O}$ snowline, the temperature exceeds the sublimation temperature under the pressure conditions in the midplane of the Herbig Ae disk ($T_{g} \sim T_{d} \sim 120$K), and most of the $\mathrm{H_2O}$ is released into the gas-phase by thermal desorption. 
Here we note that the sublimation temperature under the pressure conditions in the midplane of the Herbig Ae disk ($T_{g} \sim T_{d} \sim 120$K) is lower than that in the midplane of the T Tauri disk ($T_{g} \sim T_{d} \sim 150-160$K, see paper I \citealt{Notsu2016}). The region in the midplane of the Herbig Ae disk where the temperature is around 100$-$200K is at a larger radius compared with that in the midplane of T Tauri disk, and the gas number density of such a region in the midplane of the Herbig Ae disk is lower ($n_{\mathrm{H}}\sim 10^{11}-10^{12}$ cm$^{-3}$) versus ($n_{\mathrm{H}}\sim 10^{12}-10^{13}$ cm$^{-3}$). According to Eq. (3)-(5) in Section 2.2.2 of paper I \citep{Notsu2016}, the sublimation temperature is higher if the gas number density is also higher. 
\\ \\
The temperature of the region just inside the $\mathrm{H_2O}$ snowline in the Herbig Ae disk (between $7-8$ au and 14 au) is $T_{g} \sim120-170$K; hence the gas-phase chemistry to form $\mathrm{H_2O}$ molecules (e.g., O$+$H$_{2}$$\rightarrow$OH$+$H and OH$+$H$_{2}$$\rightarrow$$\mathrm{H_2O}$$+$H) is not efficient compared with the inner region at a higher temperature ($T_{g} > 170$K, $r<7-8$au).
We point out that the radial temperature profile in the midplane of the T Tauri disk is steeper than that in the midplane of the Herbig Ae disk, and this is another reason why the T Tauri disk does not have such a region with a relatively large fractional abundance of $\mathrm{H_2O}$ gas ($\sim10^{-8}$).
A similar distribution of gas-phase $\mathrm{H_2O}$ in the midplane of a Herbig Ae disk is reported in Figure 1 of \citet{Woitke2009b}.
Here we also note that \citet{Eistrup2016} calculated the chemical evolution of a disk midplane under both molecular and atomic initial conditions as initial chemical abundances.
They showed that in the latter atomic conditions, the abundance of $\mathrm{H_2O}$ gas and ice around the $\mathrm{H_2O}$ snowline ($\sim10^{-6}$) is smaller than that for molecular initial abundances ($\sim10^{-4}$). This is because $\mathrm{O_2}$ is formed in the gas-phase via O$+$OH$\rightarrow$$\mathrm{O_2}+$H and remains in the gas phase since its sublimation temperature is much lower than that of other molecules like $\mathrm{H_2O}$.
This reaction route competes with gas-phase $\mathrm{H_2O}$ formation.
 \\ \\
In the outer disk, the fractional abundance of $\mathrm{H_2O}$ gas is also relatively high ($\sim10^{-8}-10^{-7}$) in the hot surface layer and at the $\mathrm{H_2O}$ sublimation (photodesorption) front compared with the cold midplane region of the outer disk ($\lesssim10^{-12}-10^{-10}$), as also shown in the T Tauri disk model (paper I, \citealt{Notsu2016}).
\\ \\
Here we note that the region with a high $\mathrm{H_2O}$ gas abundance ($\sim10^{-4}$) in the disk midplane extends to 
a larger radius ($r\sim$10 au) at $z/r \gtrsim$ 0.1 than at $z/r\sim0$ ($r\sim 7-8$ au).
This is not seen in the T Tauri disk case (see Figure 2 of paper I, \citealt{Notsu2016}). This is because the scale height of the Herbig Ae disk (e.g., $H/r \sim 1.2$ at $r = 5$ au) is smaller than that for the T Tauri disk (e.g., $H/r \sim 1.7$ at $r = 5$ au) and the radiation from the central Herbig Ae star is stronger than that from the central T Tauri star, thus the gas temperature of the Herbig Ae disk around $z/r \sim 0.1$ is higher. In contrast, for the T Tauri disk case, since the disk scale height is larger than that of Herbig Ae disk, the values of gas temperature of the disk between $z/r \sim 0-0.1$ is constant.
\begin{figure}[htbp]
\begin{center}
\hspace{5cm}
\includegraphics[scale=0.55]{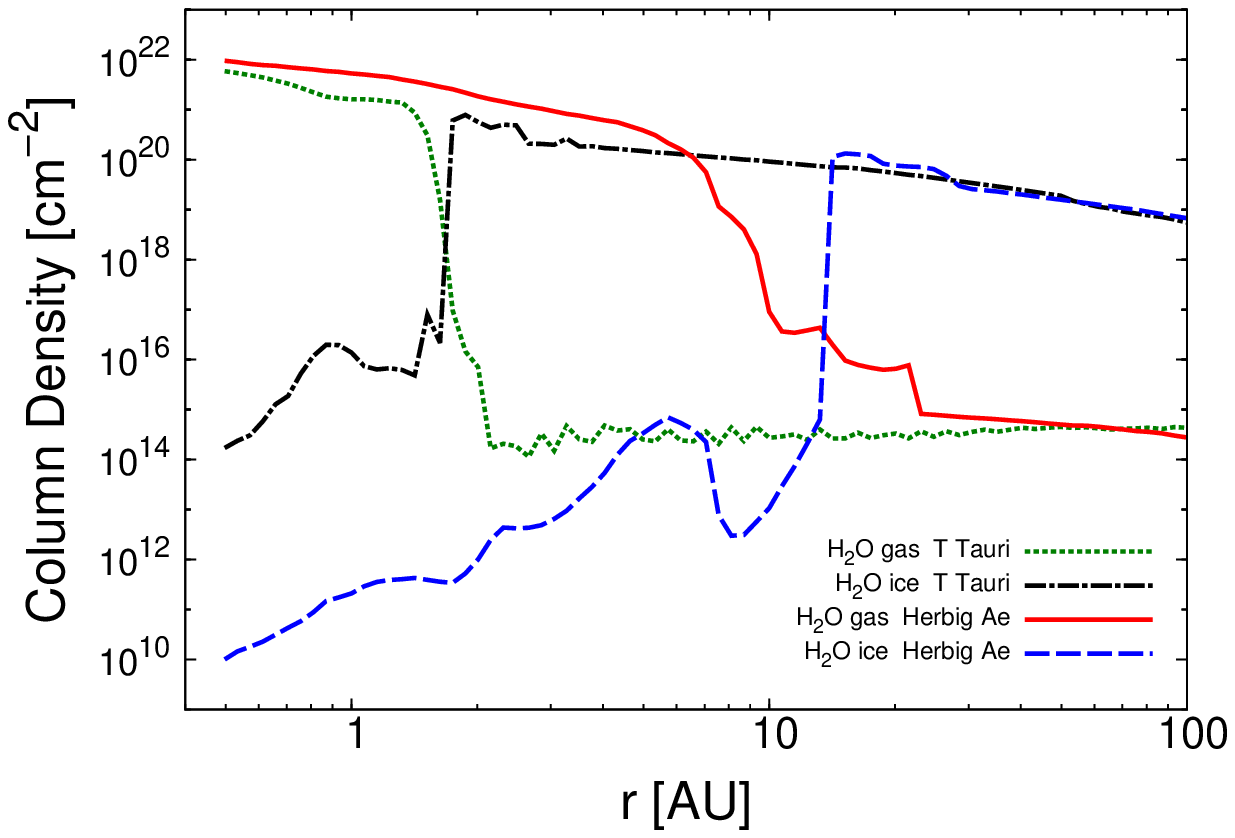}
\hspace{5cm}
\includegraphics[scale=0.55]{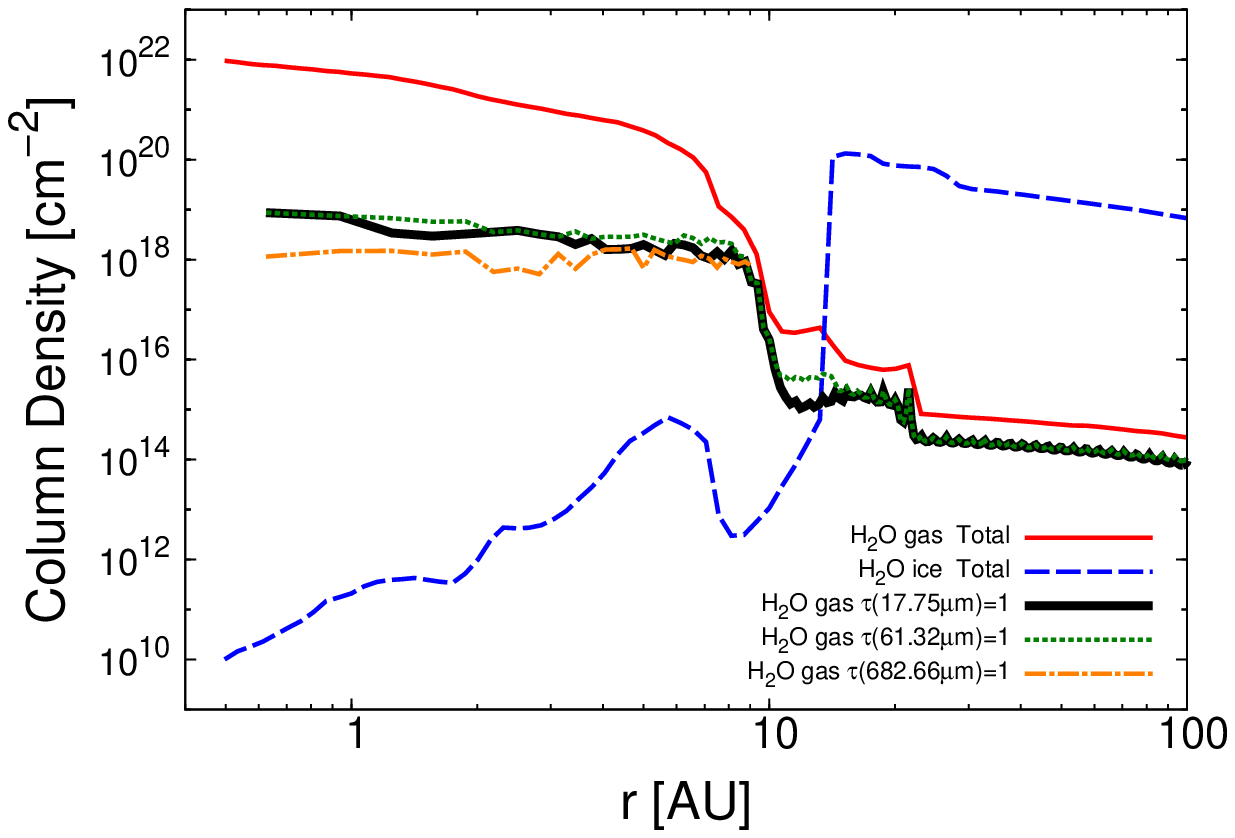}
\end{center}
\vspace{0.8cm}
\caption{
\noindent Top panel: The radial profiles of the vertically integrated column density in $\mathrm{cm}^{-2}$ of $\mathrm{H_2O}$ gas and ice in the T Tauri disk ({\it green dotted line} and {\it black dashed dotted line}) and the Herbig Ae disk ({\it red solid line} and {\it blue dashed line}).
Bottom panel: The radial profiles of the column density in $\mathrm{cm}^{-2}$ of $\mathrm{H_2O}$ ice ({\it blue dashed line}) and gas in the Herbig Ae disk, which are vertically integrated from $z=\infty$ to $-\infty$ ({\it red solid line}), to $z(\tau_{17.75\mu\mathrm{m}}=1)$ ({\it black bold solid line}), 
to $z(\tau_{61.32\mu\mathrm{m}}=1)$ ({\it green dotted line}), and to $z(\tau_{682.66\mu\mathrm{m}}=1)$ ({\it orange dashed dotted line}).
Since $\tau_{682.66\mu\mathrm{m}}$ at $z=-\infty$ is lower than unity at $r\gtrsim$10 au, the radial profile of this case is plotted only $r\lesssim$10 au.
}\label{Figure3_HerbigAe}
\end{figure} 
\\ \\
The top panel of Figure \ref{Figure3_HerbigAe} shows the radial column density profiles of $\mathrm{H_2O}$ gas and ice for both the T Tauri disk (see Figure 3 of paper I, \citealt{Notsu2016}) and the Herbig Ae disk. 
In the Herbig Ae disk case,
the column density of $\mathrm{H_2O}$ gas and ice in the disk midplane flips across the $\mathrm{H_2O}$ snowline as expected ($r\sim$ 14 au).
The column density of $\mathrm{H_2O}$ gas is high ($\sim10^{20}-10^{22}$ $\mathrm{cm}^{-2}$) in the inner high-temperature region of the disk midplane with $r<7-8$ au, relatively high ($\sim10^{16}-10^{19}$ $\mathrm{cm}^{-2}$) in the midplane between $7-8$ au and 14 au, and in contrast, low outside the $\mathrm{H_2O}$ snowline ($< 10^{16}$ $\mathrm{cm}^{-2}$). 
The column density profile of $\mathrm{H_2O}$ ice is roughly opposite. 
In the T Tauri disk case,
the column densities of $\mathrm{H_2O}$ gas and ice in the disk midplane flips across the $\mathrm{H_2O}$ snowline more steeply following the steeper temperature gradient.
The Bottom panel of Figure \ref{Figure3_HerbigAe} shows the radial profiles of the column density in $\mathrm{cm}^{-2}$ of $\mathrm{H_2O}$ ice and gas in the Herbig Ae disk, which have been vertically integrated from $z=\infty$ to (i) $-\infty$, (ii) $z(\tau_{17.75\mu\mathrm{m}}=1)$, (iii) $z(\tau_{61.32\mu\mathrm{m}}=1)$, and (iv) $z(\tau_{682.66\mu\mathrm{m}}=1)$. $\tau_{\lambda}$ is the total optical depth value at each wavelength, $\lambda$, including gas and dust components. In Section 4.3, we discuss about this panel in detail.
\\ \\
Previous analytical models and numerical simulations derived the position of the $\mathrm{H_2O}$ snowline of an optically thick disk for given parameters, such as mass ($M_{*}$) and temperature ($T_{*}$) of the central star, a viscous parameter $\alpha$, an accretion rate $\dot{M}$, a gas-to-dust mass ratio $g/d$, and the average dust grain size $a$ and opacity (e.g., \citealt{Davis2005, Garaud2007, Min2011, Oka2011, Du2014, Harsono2015, Mulders2015, Piso2015, Sato2016}), and suggested that the position of the $\mathrm{H_2O}$ snowline changes, as these parameters change.
In the case of Herbig Ae disks with $M_{\mathrm{*}} \sim 2.5M_{\bigodot}$, $\dot{M} \sim 10^{-8} M_{\bigodot}$ yr$^{-1}$, $g/d=100$, and $a \sim 0.1 \mu$m, the position of the $\mathrm{H_2O}$ snowline is $\sim 10-20$ au. 
In our calculations we use similar parameters for $M_{\mathrm{*}}$,
$\dot{M}$ and $a$, and the $\mathrm{H_2O}$ snowline appears at a radius
of around 14 au in the midplane, within the range of previous studies.
\\
\subsection{$\mathrm{H_2O}$ emission lines from a Herbig Ae disk}
\noindent In this Section, we first describe the detailed properties of seven characteristic pure rotational ortho-$\mathrm{H_2O}$ lines (see Table \ref{tab:T1} and Section 3.2.1) for the Herbig Ae disk.
These seven lines (including the ortho-$\mathrm{H_2O}$ 682.66 $\mu$m line) are candidates for tracing emission from the hot water reservoir within the $\mathrm{H_2O}$ snowline.
In Section 3.2.2, we describe the properties of the 63.32 and 538.29$\mu$m lines, which are examples of lines which are less suited to trace emission from the water reservoir within the $\mathrm{H_2O}$ snowline.
We consider these two lines to test the validity of our model calculations, since the fluxes of these two lines from protoplanetary disks have been observed with $Herschel$.
The properties of near-, and mid-infrared $\mathrm{H_2O}$ emission lines which do not trace emission from the hot water vapor within the $\mathrm{H_2O}$ snowline are also described in this subsection.
Since we investigated the profiles and properties of three lines ($\lambda$=682.66, 63.32, 538.29$\mu$m) for the T Tauri disk in paper I \citep{Notsu2016}, here we mainly focus on the differences between the line properties between the T Tauri disk and the Herbig Ae disk.
In Section 3.2.3 and Section 4.4, we show and discuss other candidate lines which trace the emission from the hot water vapor within the $\mathrm{H_2O}$ snowline from mid-infrared to sub-millimeter wavelengths, and their properties, especially the variation in line fluxes with wavelength.
In Section 3.2.4, we show and discuss normalized radial cumulative line fluxes for the lines discussed in Sections 3.2.1-3.2.3.
\\ \\
In this paper, we show and discuss only the results concerning ortho-$\mathrm{H_2O}$ lines, since the number densities and the fluxes of ortho-$\mathrm{H_2O}$ lines 
are larger than those of para-$\mathrm{H_2O}$ lines, due to the assumption, OPR=3.
The line selection process is described in detail in Section 3.2 of paper I \citep{Notsu2016}.
 \\
\subsubsection{Candidate $\mathrm{H_2O}$ emission lines which trace emission from the hot water vapor within the $\mathrm{H_2O}$ snowline}
\begin{figure*}[htbp]
\begin{center}
\includegraphics[scale=0.4]{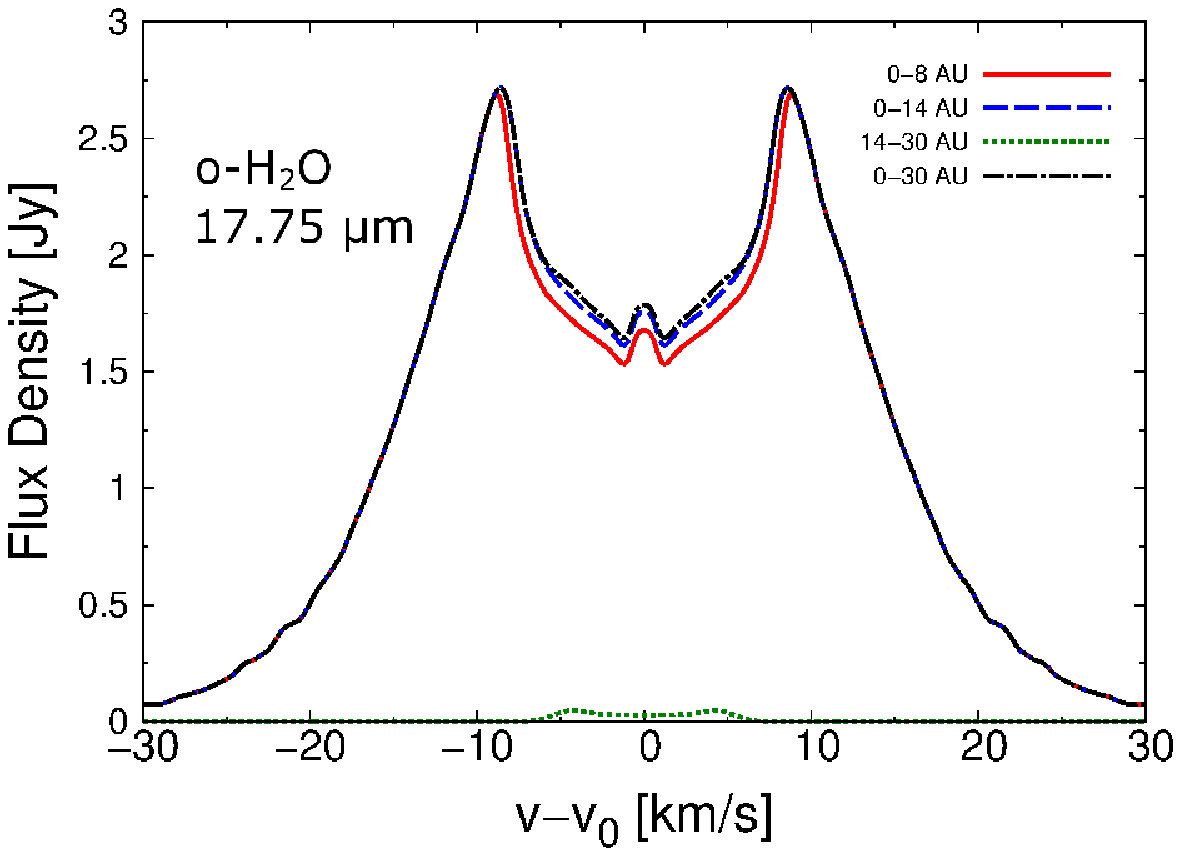}
\includegraphics[scale=0.4]{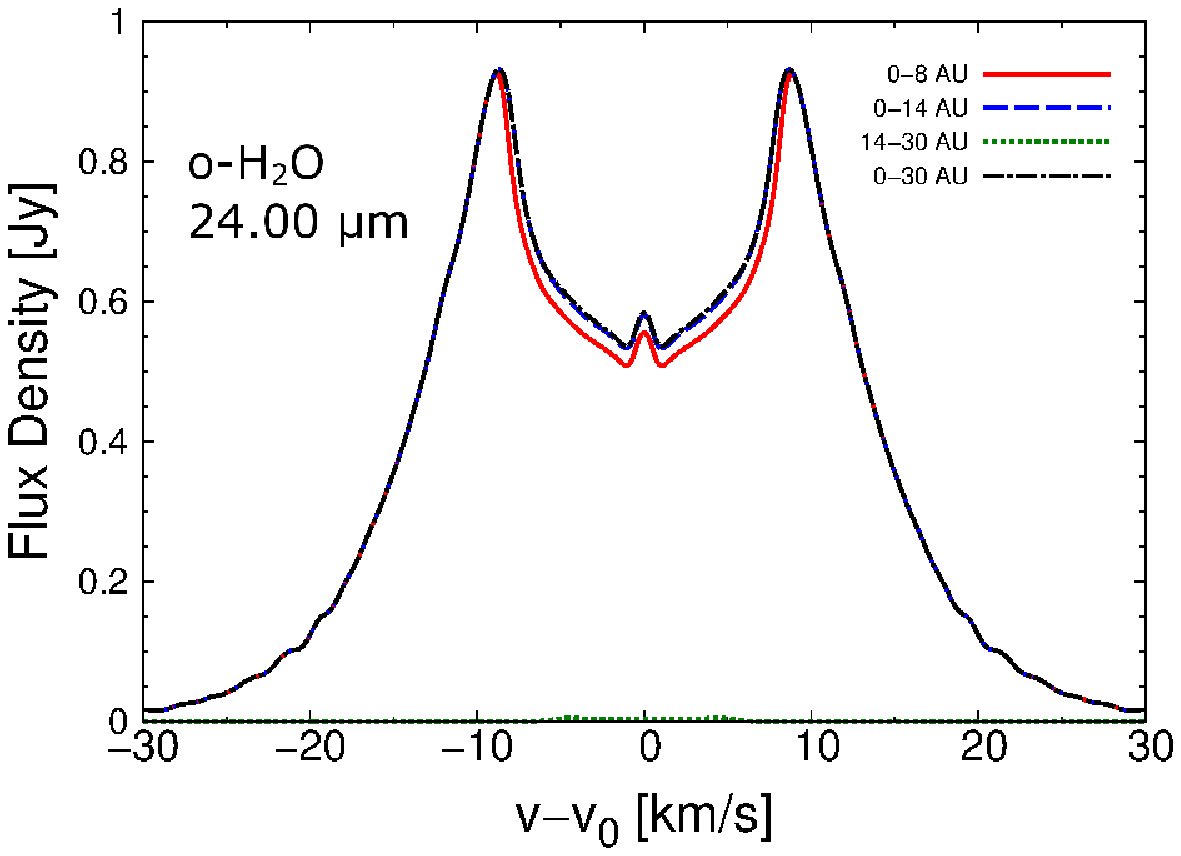}
\includegraphics[scale=0.4]{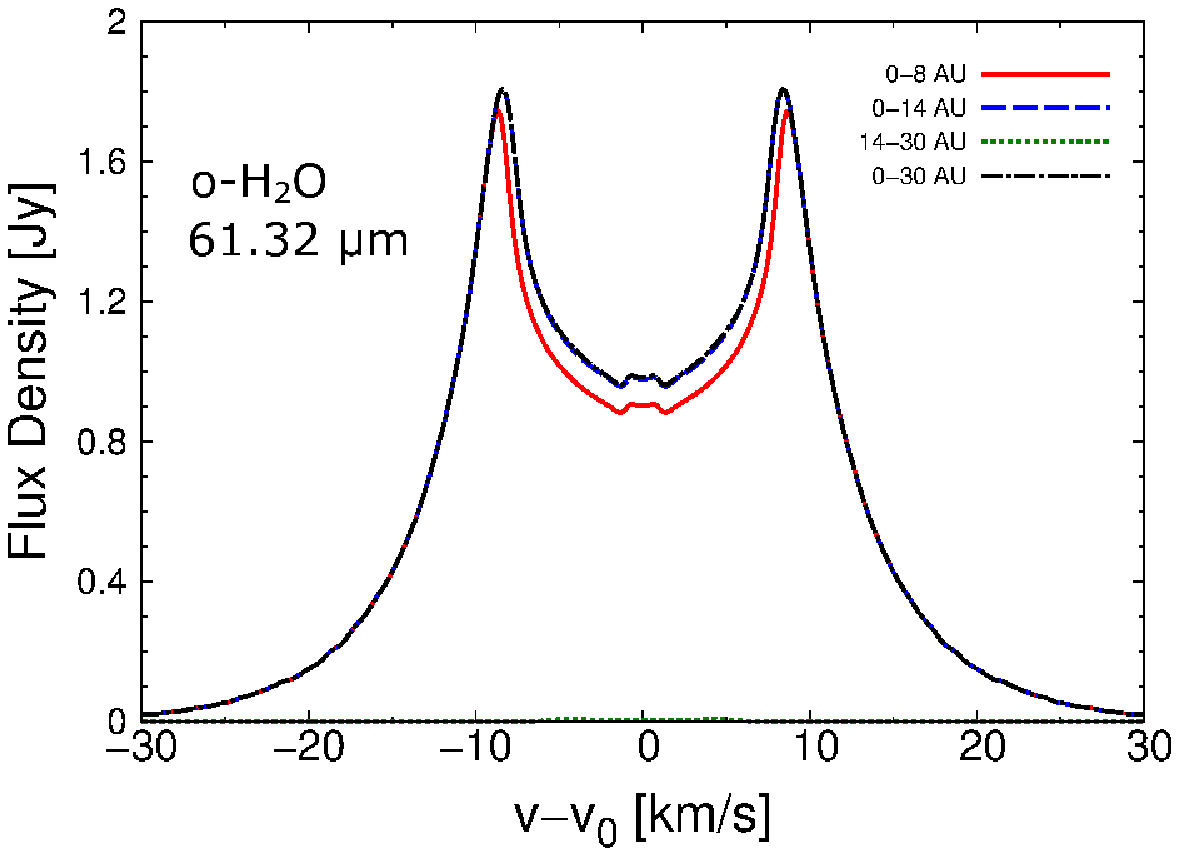}
\includegraphics[scale=0.4]{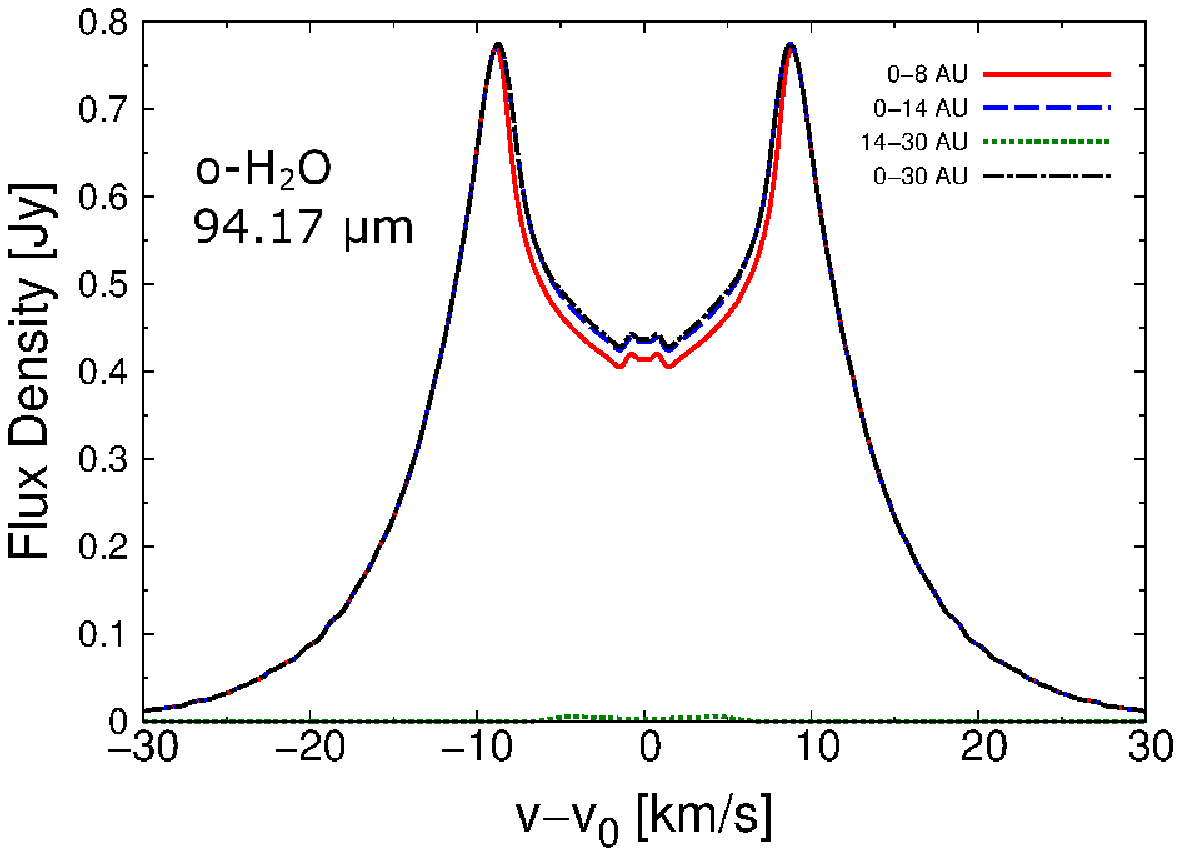}
\includegraphics[scale=0.4]{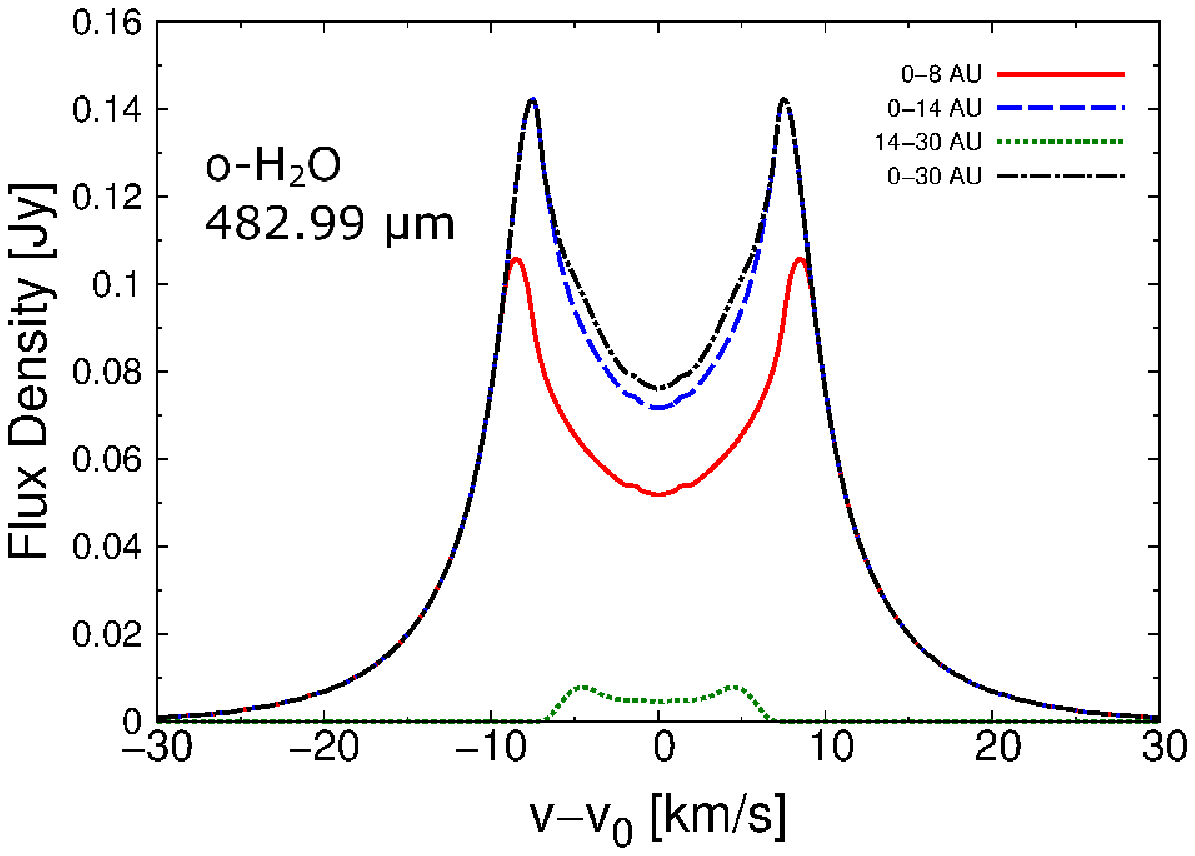}
\includegraphics[scale=0.4]{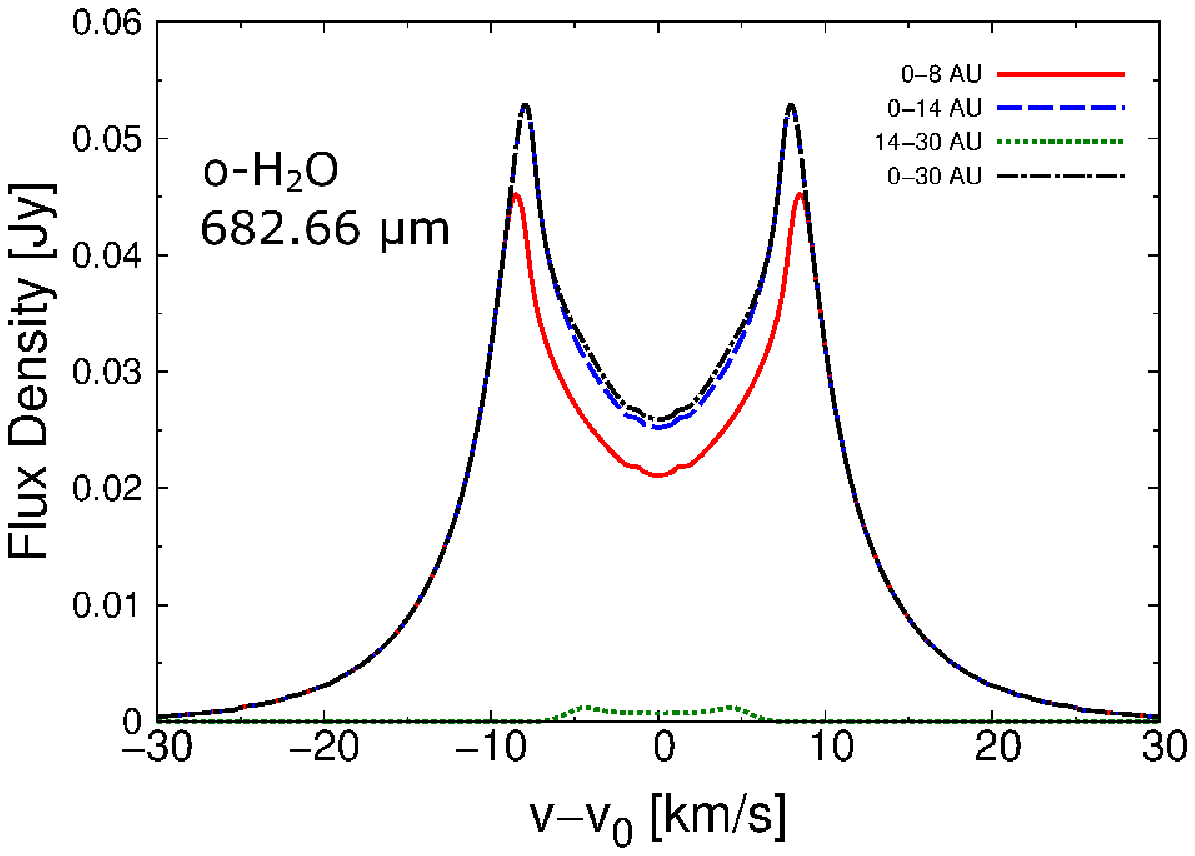}
\includegraphics[scale=0.4]{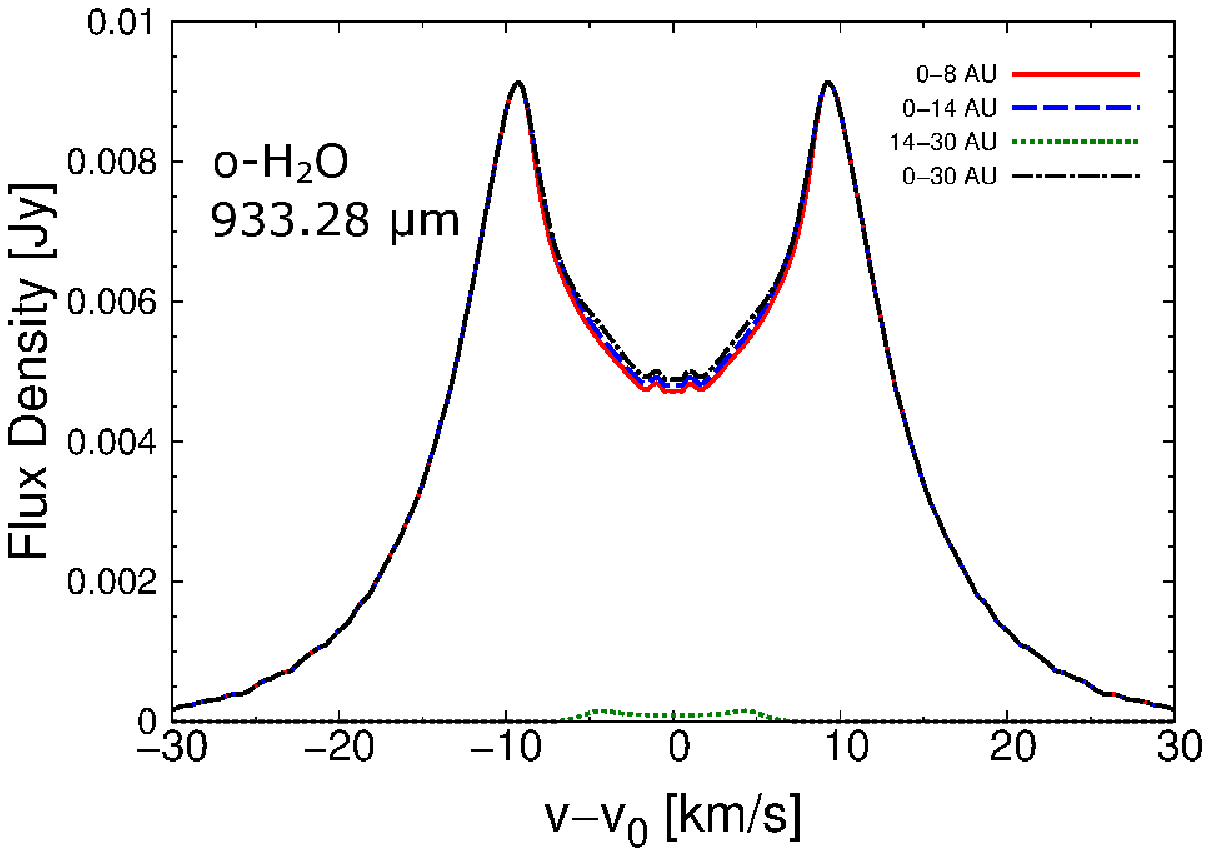}
\end{center}
\caption{\noindent The velocity profiles of seven characteristic pure rotational ortho-$\mathrm{H_2O}$ lines at $\lambda$=17.75$\mu$m (top left), 24.00$\mu$m (top center),  61.32$\mu$m (top right), 94.17$\mu$m (middle left), 482.99$\mu$m (middle center), 682.66$\mu$m (middle right), and 933.28$\mu$m (bottom), which have small $A_{ul}$ and large $E_{up}$, from the Herbig Ae disk. These are candidate $\mathrm{H_2O}$ lines to trace the hot water vapor within the $\mathrm{H_2O}$ snowline.
In calculating the line profiles in this paper (see Figures \ref{Figure4_HerbigAe}, \ref{Figure7_HerbigAe}, \ref{Figure10_HerbigAe}, and \ref{Figure14_HerbigAe}), we assume that the distance to the object $d$ is 140pc ($\sim$ the distance of Taurus molecular cloud), and the inclination angle of the disk, $i$, is 30 degree.
The parameters and total fluxes of these $\mathrm{H_2O}$ lines are listed in Table \ref{tab:T1} and \ref{tab:bT1}.1.
{\it Red solid lines} are the emission line profiles from inside 8 au ($=$the inner high temperature region), {\it blue dashed lines} are those from inside 14 au ($\sim$inside the $\mathrm{H_2O}$ snowline), {\it green dotted lines} are those from 14-30 au ($\sim$outside the $\mathrm{H_2O}$ snowline), and {\it black dashed dotted lines} are those from the total area inside 30 au. 
}\label{Figure4_HerbigAe}
\end{figure*}

\noindent Figure \ref{Figure4_HerbigAe} shows the emission profiles of seven representative characteristic pure rotational ortho-$\mathrm{H_2O}$ lines at $\lambda$=17.75$\mu$m (top left), 24.00$\mu$m (top center),  61.32$\mu$m (top right), 94.17$\mu$m (middle left), 482.99$\mu$m (middle center), 682.66$\mu$m (middle right), and 933.28$\mu$m (bottom), for the Herbig Ae disk. 
These lines have small values of $A_{ul}$ ($\sim10^{-3}-10^{-6}$ s$^{-1}$) and relatively large values of upper E$_{\mathrm{up}}$ ($\sim700-1900$K).
They are representative candidate ortho-$\mathrm{H_2O}$ lines which trace emission from the hot water gas within the $\mathrm{H_2O}$ snowline.
The $\mathrm{H_2O}$ 933.28$\mu$m, 682.66$\mu$m, and 482.99$\mu$m lines fall in ALMA band 7, 8, and 9, respectively.
The $\mathrm{H_2O}$ 17.75$\mu$m line and 24.00$\mu$m line are Q band lines at mid-infrared wavelengths, and the $\mathrm{H_2O}$ 17.75$\mu$m line falls in the wavelength coverage of SPICA/SMI-HRS (see Section 4.4).
The detailed parameters, such as transitions ($J_{K_{a}K_{c}}$), wavelength $\lambda$, frequency, $A_{ul}$, $E_{up}$, critical density $n_{\mathrm{cr}}=A_{ul}/{<\sigma v>}$\footnote[5]{$<\sigma v>$ is the collisional rates for the excitation of $\mathrm{H_2O}$ by H$_{\mathrm{2}}$ and electrons for an adopted collisional temperature of
200K from \citet{Faure2008}.}, and total line fluxes are listed in Table \ref{tab:T1}.
In Table \ref{tab:T1}, we also show the values of the total fluxes from both the Herbig Ae disk and the T Tauri disk (see also paper I, \citealt{Notsu2016}).
In calculating the values from the T Tauri disk, we use the same method as in paper I \citep{Notsu2016}.
In calculating all line profiles in this paper (see Figures \ref{Figure4_HerbigAe}, \ref{Figure7_HerbigAe}, \ref{Figure10_HerbigAe}, and \ref{Figure14_HerbigAe}), we assume that the distance $d$ to the object is 140pc ($\sim$ the distance of Taurus molecular cloud), and the inclination angle $i$ of the disk is 30 degs. 
 \begin{table*}
  \caption{{Calculated representative ortho-$\mathrm{H_2O}$ line parameters and total line fluxes}}\label{tab:T1}
 \begin{center}
    \begin{tabular}{rrrrrrrr}
     \hline 
\vspace{-0.2cm}
     $J_{K_{a}K_{c}}$& \ \ $\lambda$$^{1}$&\ Freq.&\ $A_{ul}$&$E_{up}$&$n_{\mathrm{cr}}$&HAe flux$^{2,3}$&TT flux$^{3,4}$\\
&&&&&&&\\
      &[$\mu$m]&[GHz]&[s$^{-1}$]&[K]&[$\mathrm{cm}^{-3}$]&[W $\mathrm{m}^{-2}$]&[W $\mathrm{m}^{-2}$]\\
     \hline   
      6$_{52}$-5$_{05}$ & 17.754 & 16885.840 & 2.909$\times10^{-3}$ & 1278.5 &$8.3\times10^{10}$ & $4.1\times10^{-17}$ & $2.3\times10^{-20}$\\
      5$_{50}$-5$_{05}$ & 23.996 & 12493.205 & 2.696$\times10^{-4}$ & 1067.7 &$1.9\times10^{9}$& $9.4\times10^{-18}$ & $6.4\times10^{-21}$\\
      5$_{41}$-6$_{16}$ & 61.316 & 4889.280 & 2.686$\times10^{-4}$ & 878.1 & $4.1\times10^{8}$ & $5.9\times10^{-18}$ & $3.5\times10^{-20}$\\
      6$_{52}$-7$_{25}$ & 94.172 & 3183.464 & 3.387$\times10^{-4}$ & 1278.5 & $3.1\times10^{8}$ & $1.8\times10^{-18}$ & $1.6\times10^{-20}$\\
      5$_{32}$-4$_{41}$ & 482.990 & 620.701 & 1.106$\times10^{-4}$ & 732.1 & $3.3\times10^{6}$ & $5.3\times10^{-20}$ & $1.1\times10^{-21}$\\     
      6$_{43}$-5$_{50}$ & 682.664 & 439.151 & 2.816$\times10^{-5}$ &1088.7 & $1.0\times10^{6}$ & $1.4\times10^{-20}$ & $3.1\times10^{-22}$\\
      10$_{29}$-9$_{36}$ & 933.277 & 321.226 & 6.165$\times10^{-6}$ & 1861.2 & $4.7\times10^{6}$ & $2.3\times10^{-21}$& $7.8\times10^{-23}$\\
                          &&&&&&&\\
      8$_{18}$-7$_{07}$ & 63.324 & 4734.296 & 1.772 & 1070.6 & $1.5\times10^{10}$ & $1.1\times10^{-16}$ & $5.7\times10^{-18}$\\
      1$_{10}$-1$_{01}$ & 538.289 & 556.936 & 3.497$\times 10^{-3}$& 61.0 & $2.9\times10^{7}$ & $7.2\times10^{-20}$ & $1.1\times10^{-20}$\\
                          &&&&&&&\\
      17$_{4 13}$-16$_{3 14}$ & 12.396 & 24184.126 & 7.728 & 5780.8 & $1.1\times10^{11}$  & $6.7\times10^{-17}$ & $5.3\times10^{-19}$\\
      13$_{7 6}$-12$_{4 9}$ & 12.453 & 24073.032 & 1.053 & 4212.6 & $1.1\times10^{13}$ & $6.9\times10^{-17}$ & $2.5\times10^{-19}$\\
      7$_{6 1}$-6$_{5 2}$ & 4.958 & 60463.186 & 3.260 & 4180.4 & $1.6\times10^{13}$ & $2.2\times10^{-16}$ & $1.1\times10^{-18}$\\
      7$_{6 1}$-6$_{3 4}$ & 4.432 & 67646.817 & 2.080$\times10^{-4}$ & 4180.4 & $6.5\times10^{11}$ & $5.0\times10^{-20}$ & $8.1\times10^{-23}$\\
\hline
  \multicolumn{4}{l}{\hbox to 0pt{\parbox{120mm}{
  \footnotesize
     \footnotemark[1] In calculating the value of line wavelength from the value of line frequency, we use the value of speed of light $c$$=$$2.99792458\times 10^{8}$ m s$^{-1}$. \\
     \footnotemark[2] The total flux of each emission line from the Herbig Ae disk. \\
     \footnotemark[3] In calculating the total fluxes of these $\mathrm{H_2O}$ lines, we use a distance of $d=140$pc and an inclination
     angle of $i=$30 degree. \\
     \footnotemark[4] The total flux of each emission line from the T Tauri disk (see also paper I, \citealt{Notsu2016}).}}}
   \end{tabular}
\end{center}
    \end{table*}
\\ \\
As shown in all panels in Figure \ref{Figure4_HerbigAe},
the contributions from the optically thin surface layer of the outer disk ($r=$14-30 au) are very small compared with those from the optically thick region near the midplane of the inner disk ($r <$14 au), and they show the characteristic double-peaked profile due to Keplerian rotation.
This is because these lines, which have small Einstein $A$ coefficients ($A_{ul}\sim10^{-3}-10^{-6}$ s$^{-1}$) and relatively large upper state energies (E$_{\mathrm{up}}\sim$1000K), mainly trace the hot $\mathrm{H_2O}$ vapor inside the $\mathrm{H_2O}$ snowline.
In Section 2.3 and 3.2.1 of paper I \citep{Notsu2016}, we explained the reason why these lines have such properties.
\\ \\
In the cases of candidate $\mathrm{H_2O}$ lines except the 482.99$\mu$m and 682.66$\mu$m lines (see Figure \ref{Figure4_HerbigAe}), almost all of the emission fluxes ($> 95\%$) come from the region with a high $\mathrm{H_2O}$ gas abundance ($\sim$$10^{-4}$, $r<$8 au), and the position of the two peaks and the rapid drop in flux density between the peaks contains information on the position of the outer edge of this region.
In contrast, in the cases of the 482.99$\mu$m and 682.66$\mu$m lines (see Figure \ref{Figure4_HerbigAe}), most of the emission fluxes ($\sim 80-90\%$) are emitted from the region with a high $\mathrm{H_2O}$ gas abundance ($\sim$$10^{-4}$, $r<$8 au), and some fluxes ($\sim 10-20\%$) are emitted from the region with a relatively high $\mathrm{H_2O}$ gas abundance ($\sim$$10^{-8}$, $r=$8-14 au).
The position of the two peaks and the rapid drop in flux density between the peaks contains information on the distribution of hot $\mathrm{H_2O}$ gas within the $\mathrm{H_2O}$ snowline.
\\ \\
Figures \ref{Figure5_HerbigAe} and \ref{Figure6_HerbigAe} show the line-of-sight emissivity (emissivity times extinction, $\eta_{ul} e^{-\tau_{ul}}$;
see Equation (14) of Paper I, \citealt{Notsu2016}) and the total optical depth, $\tau_{ul}$ (gas emission and dust) distributions of these seven $\mathrm{H_2O}$ lines, respectively.
We assume that the inclination angle, $i$, of the disk is 0 deg in making these figures (see Figures \ref{Figure5_HerbigAe}, \ref{Figure6_HerbigAe}, \ref{Figure8_HerbigAe}, \ref{Figure9_HerbigAe}, and \ref{Figure13_HerbigAe}), and thus the line-of-sight direction is from $z=+\infty$ to $-\infty$ at each disk radius.
In the left panels of Figure \ref{Figure5_HerbigAe}, we overplot the total optical depth contours ($\tau_{ul}=$0.1, 1, and 10) on top of these line emissivity panels (see also Figure \ref{Figure6_HerbigAe}). In the right panels, we overplot the gas temperature $T_{g}$ contours ($T_{g}=$120, 170, and 300K, see also Figure \ref{Figure1_HerbigAe}).
Figure \ref{Figure13_HerbigAe} in Appendix A shows the vertical distributions of the normalized cumulative line emissivity at $r=$5 au (top two panels), $r=$10 au (middle two panels), and $r=$30 au (bottom two panels), and of the gas temperature $T_{g}$.
The left three panels show the distributions for these seven $\mathrm{H_2O}$ lines.
We normalize the cumulative emissivity of each line using the values at $z=-\infty$. 
According to Figures \ref{Figure5_HerbigAe}, \ref{Figure6_HerbigAe}, and \ref{Figure13_HerbigAe}, the values of emissivity at $r\lesssim$ 14 au ($=$ the position of the $\mathrm{H_2O}$ snowline), $T_{g}\gtrsim$120K, and $z/r \sim 0.05-0.12$ are larger than those of the optically thin hot surface layer and the photodesorbed layer of the outer disk, and in particular those
in the region with a higher $\mathrm{H_2O}$ gas abundance ($\sim$$10^{-4}$, $r<7-8$ au, and $T_{g}\gtrsim$170K) and $z/r \sim 0.05-0.12$ are much larger.
Emission from $z \sim 0$ at $r\lesssim$ 7 au is not possible to detect, because the optical depth of the inner disk midplane is high due to absorption by dust grains and excited $\mathrm{H_2O}$ molecules in the upper disk layer. Nevertheless, we can extract information on the distribution of hot $\mathrm{H_2O}$ vapor inside the $\mathrm{H_2O}$ snowline. This is because within $r<$14 au (= the position of the $\mathrm{H_2O}$ snowline), the $\mathrm{H_2O}$ gas fractional abundance is relatively constant over $z/r \sim$ 0-0.1 for the same disk radius $r$ (see also Figure \ref{Figure2_HerbigAe}). 
Strictly speaking, as we described in Section 3.1, the region with a high $\mathrm{H_2O}$ gas abundance ($\sim$$10^{-4}$) extends to a radius of $r\sim$10 au at $z/r \sim$ 0.1 compared with that at $z \sim 0$ ($r\sim$7-8 au), and is reflected in the emissivity distribution. However, since the radial shift in the distribution is small (a few au), its influence is not so serious for obtaining information on the general trend of the $\mathrm{H_2O}$ distribution in the inner disk within the $\mathrm{H_2O}$ snowline.
\begin{figure*}[htbp]
\begin{center}
\includegraphics[scale=0.55]{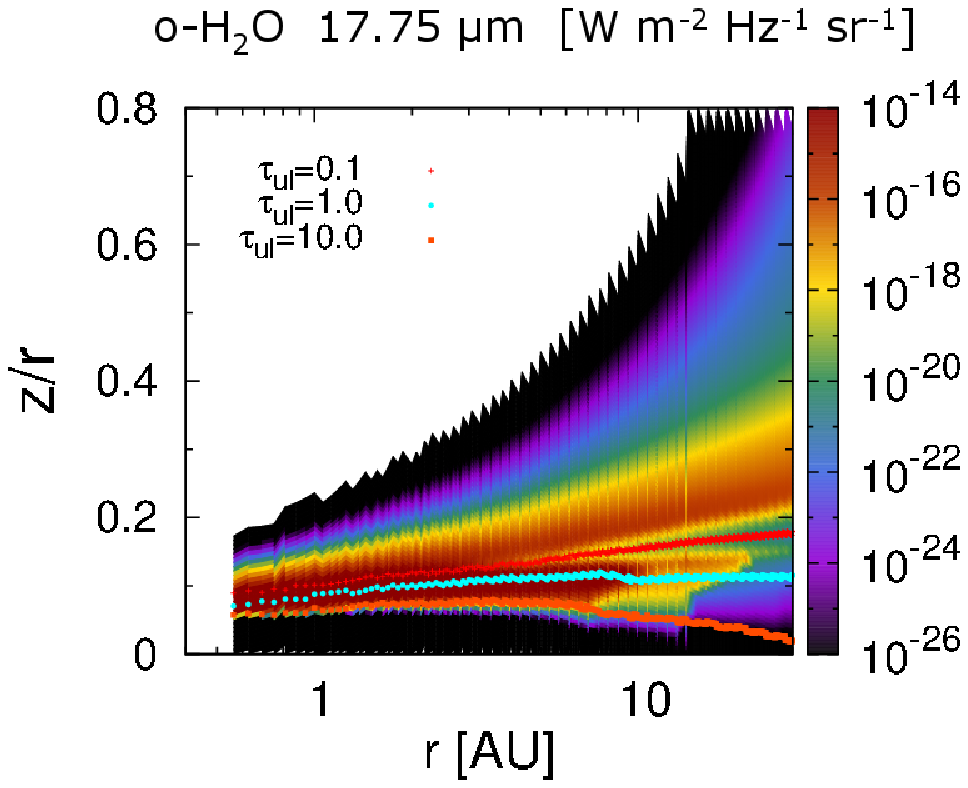}
\includegraphics[scale=0.55]{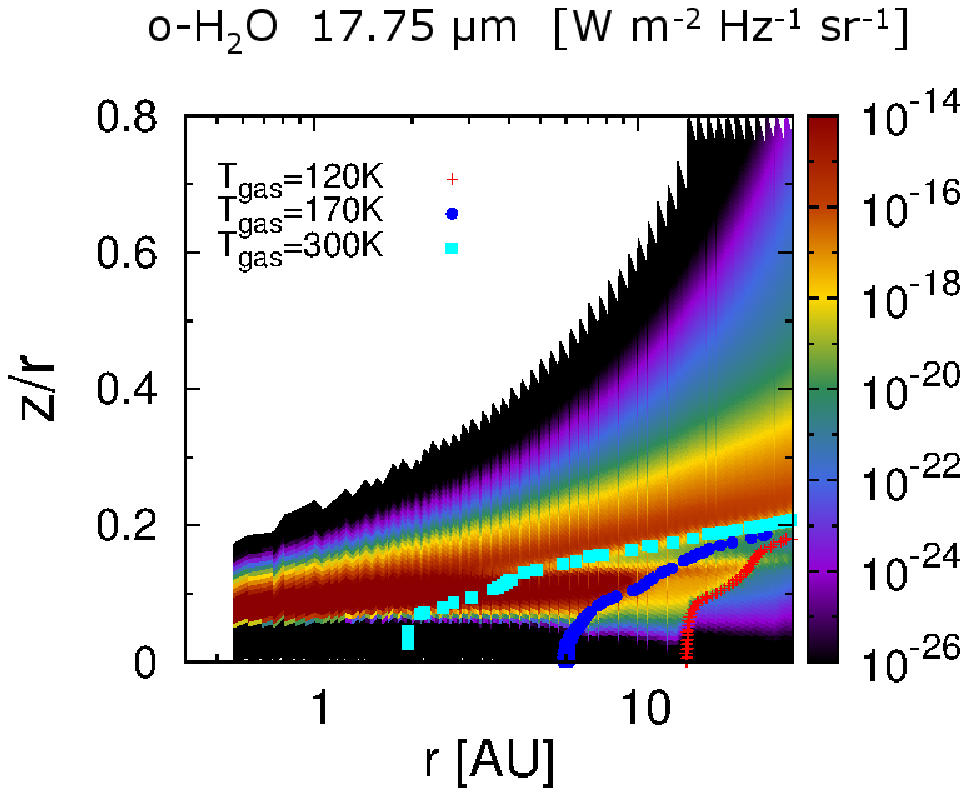}
\includegraphics[scale=0.55]{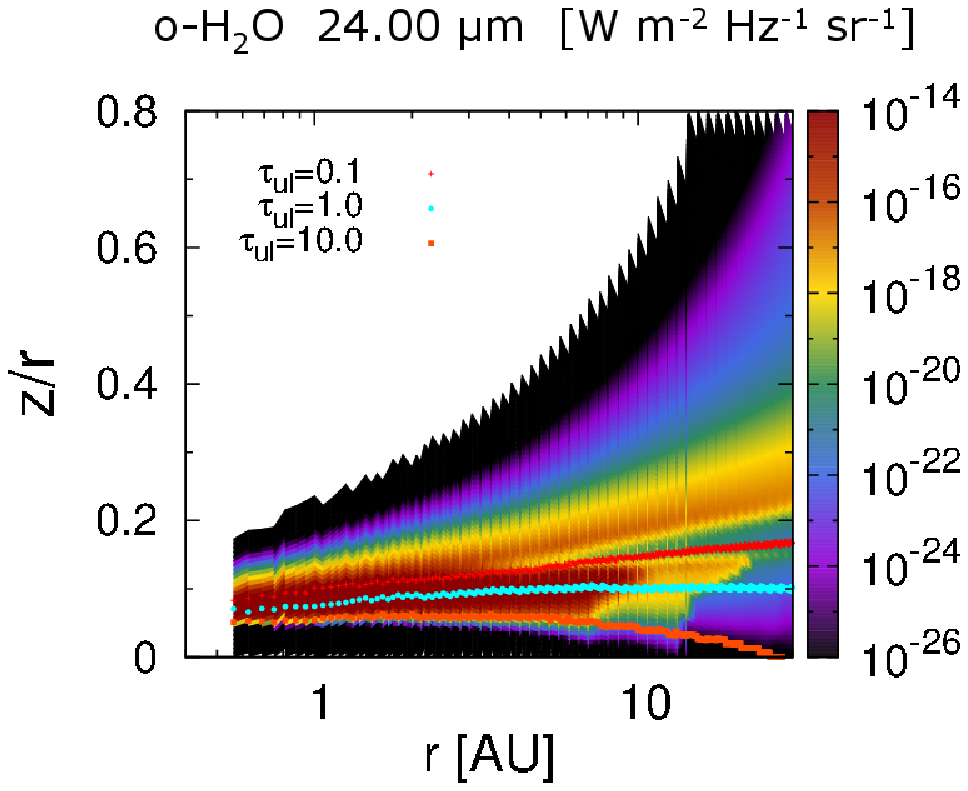}
\includegraphics[scale=0.55]{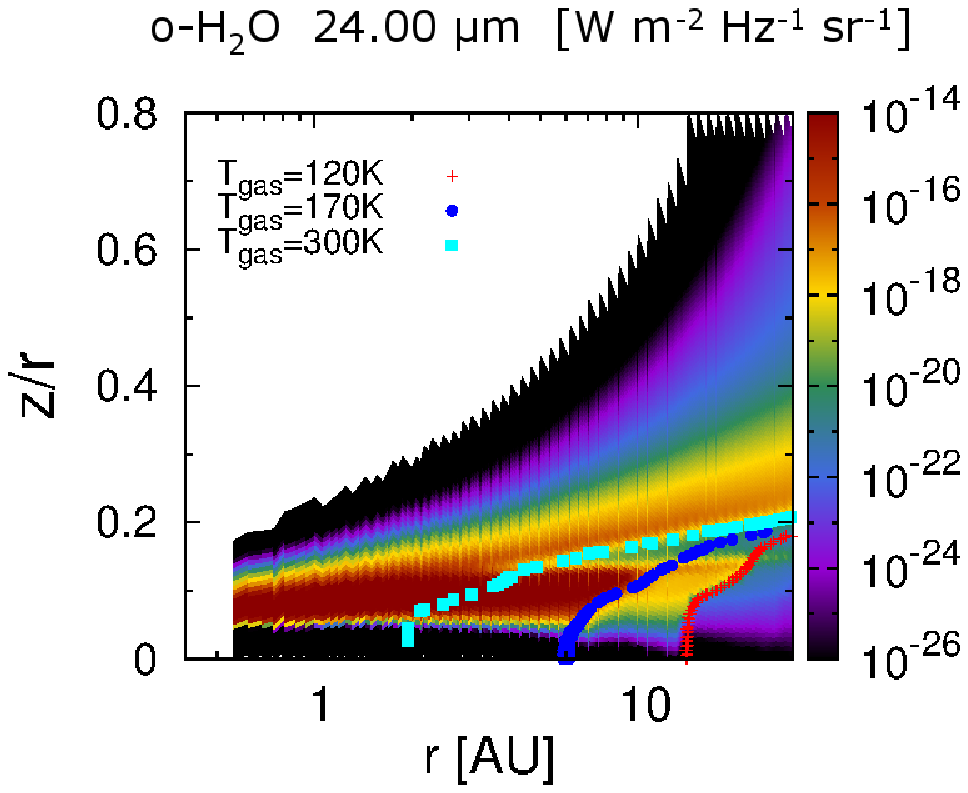}
\includegraphics[scale=0.55]{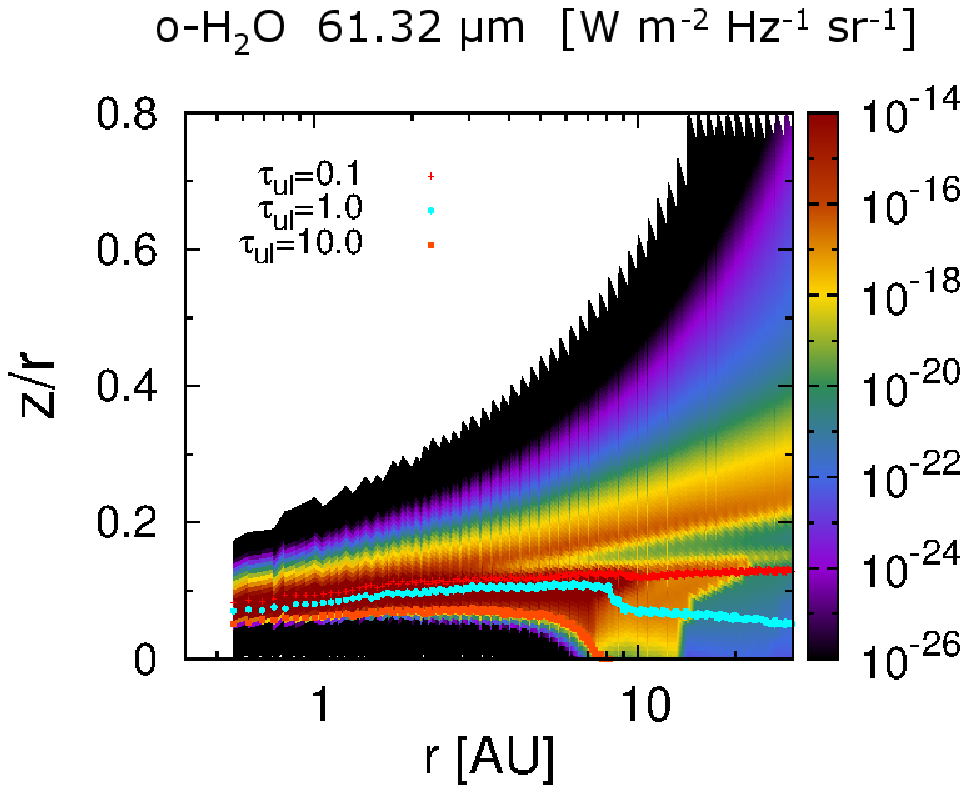}
\includegraphics[scale=0.55]{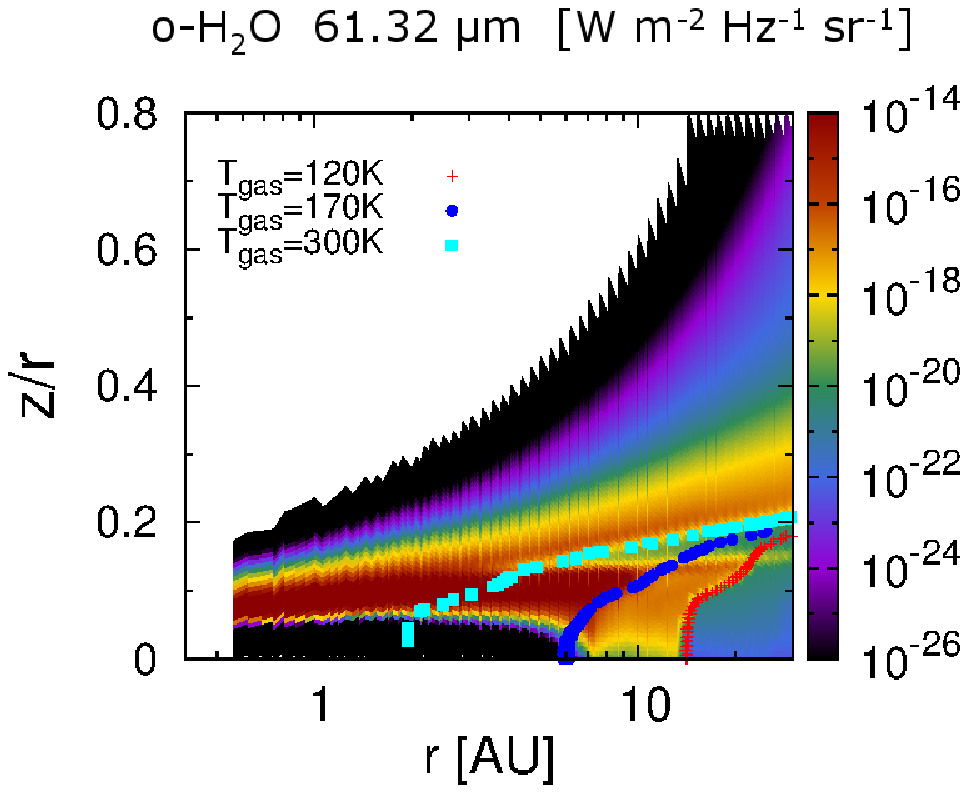}
\end{center}
\caption{\noindent The line-of-sight emissivity distributions of the $\mathrm{H_2O}$ lines at $\lambda$=17.75$\mu$m (top left and right), 24.00$\mu$m (middle left and right),  61.32$\mu$m (bottom left and right), which have small $A_{ul}$ and large $E_{up}$, from the Herbig Ae disk. 
In the left panels, we overplot the total optical depth contours $\bigl($$\tau_{ul}=$0.1 (red cross points), 1 (cyan circle points), and 10 (orange square points)$\bigr)$ on top of these line emissivity panels (see also Figure \ref{Figure6_HerbigAe}). In the right panels, we overplot the gas temperature $T_{g}$ contours $\bigl($$T_{g}=$120K (red cross points), 170K (blue circle points), and 300K (cyan square points), see also Figure \ref{Figure1_HerbigAe}$\bigr)$.
We assume that the inclination angle, $i$, of the disk is 0 deg in making these figures in this paper (see Figures \ref{Figure5_HerbigAe}, and \ref{Figure8_HerbigAe}), and the emissivity is calculated along the line from z=+$\infty$ to -$\infty$ at each disk radius.
The units are W $\mathrm{m}^{-2}$ $\mathrm{Hz}^{-1}$ ${\mathrm{sr}}^{-1}$.
}\label{Figure5_HerbigAe}
\end{figure*}   
\setcounter{figure}{4}
\begin{figure*}[htbp]
\begin{center}
\includegraphics[scale=0.55]{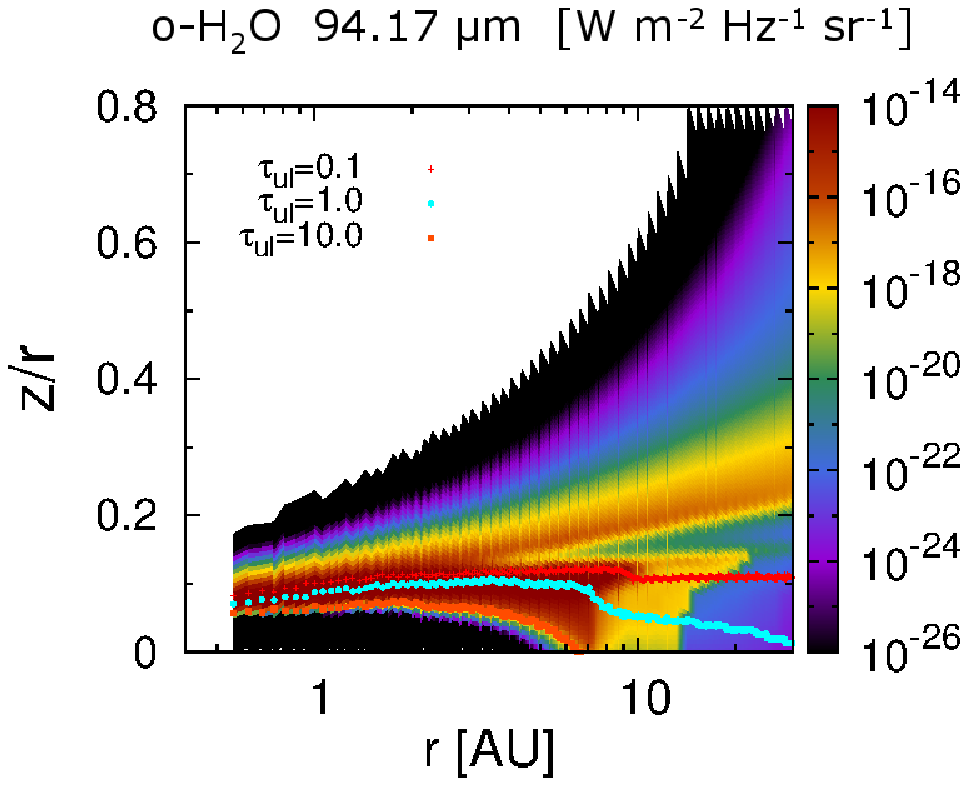}
\includegraphics[scale=0.55]{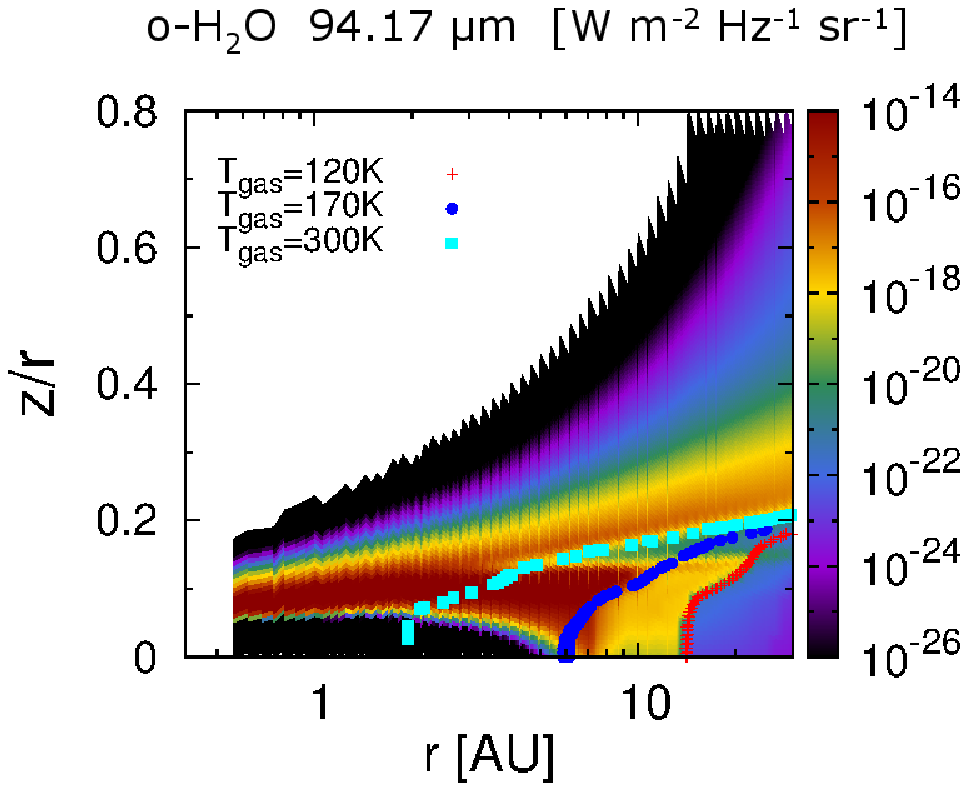}
\includegraphics[scale=0.55]{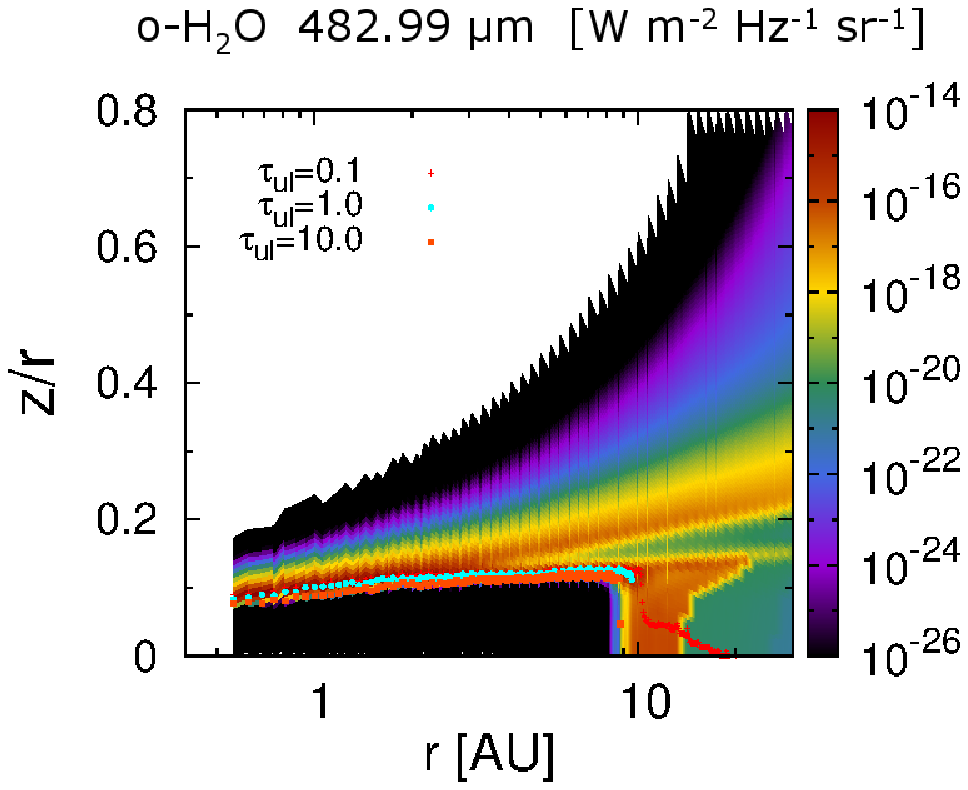}
\includegraphics[scale=0.55]{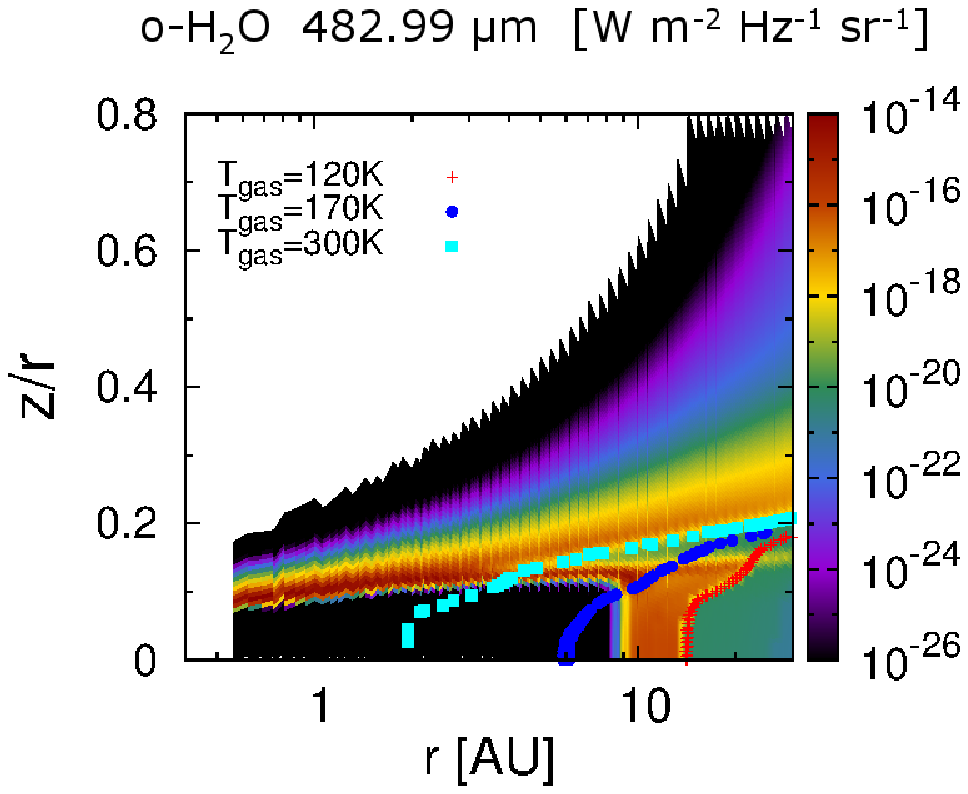}
\includegraphics[scale=0.55]{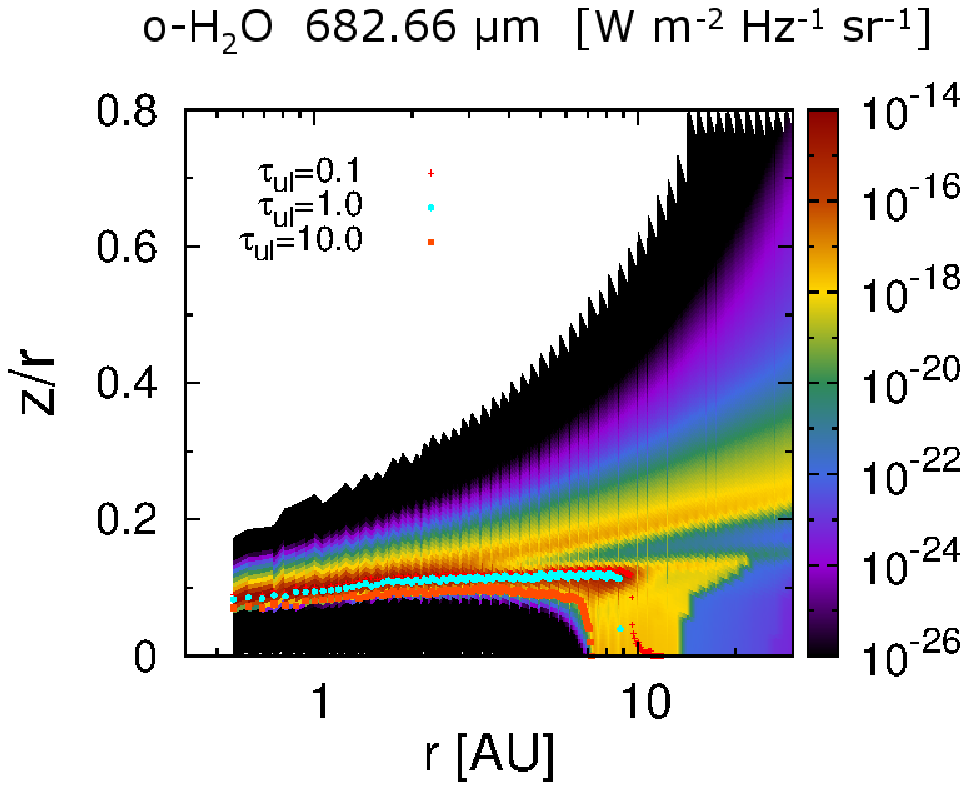}
\includegraphics[scale=0.55]{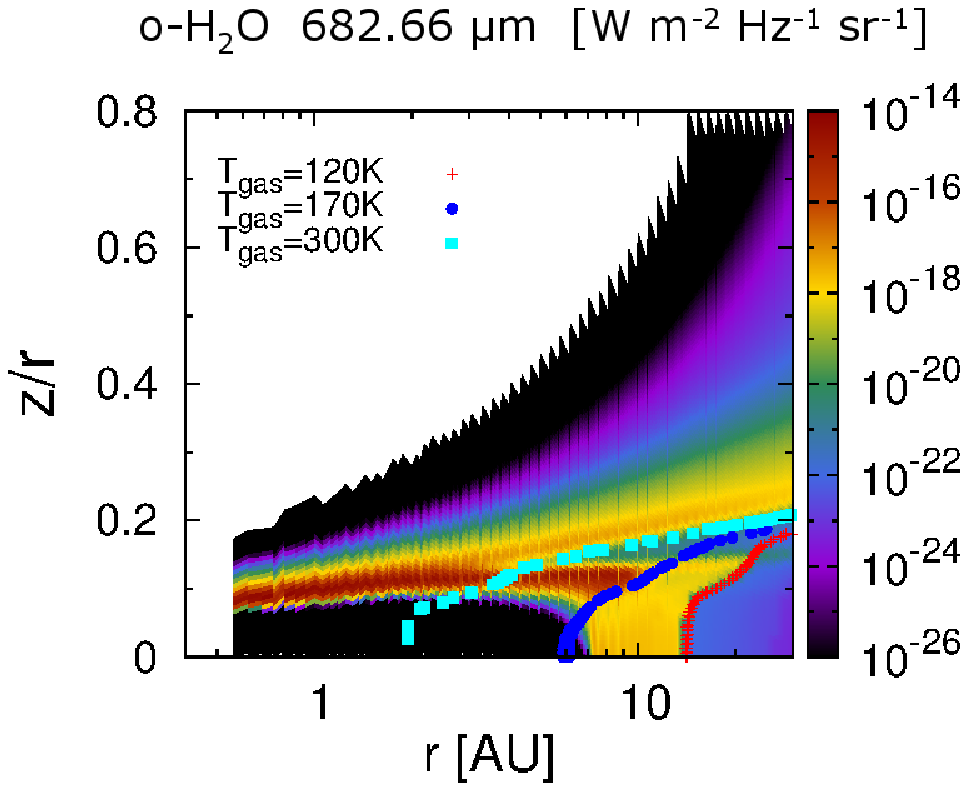}
\includegraphics[scale=0.55]{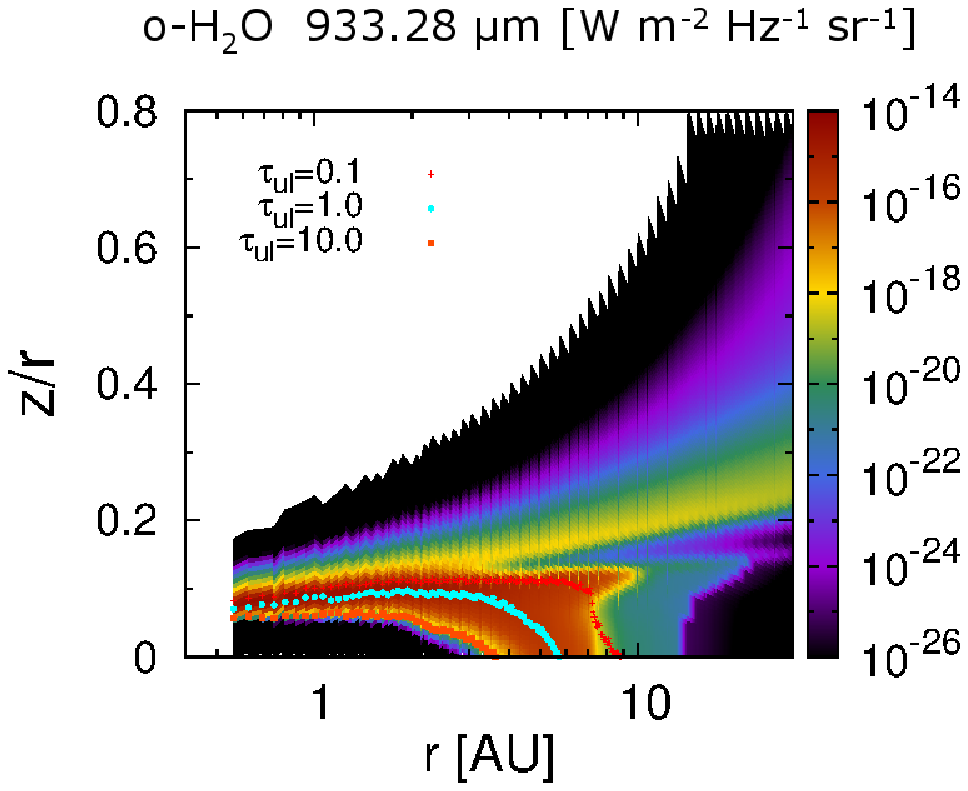}
\includegraphics[scale=0.55]{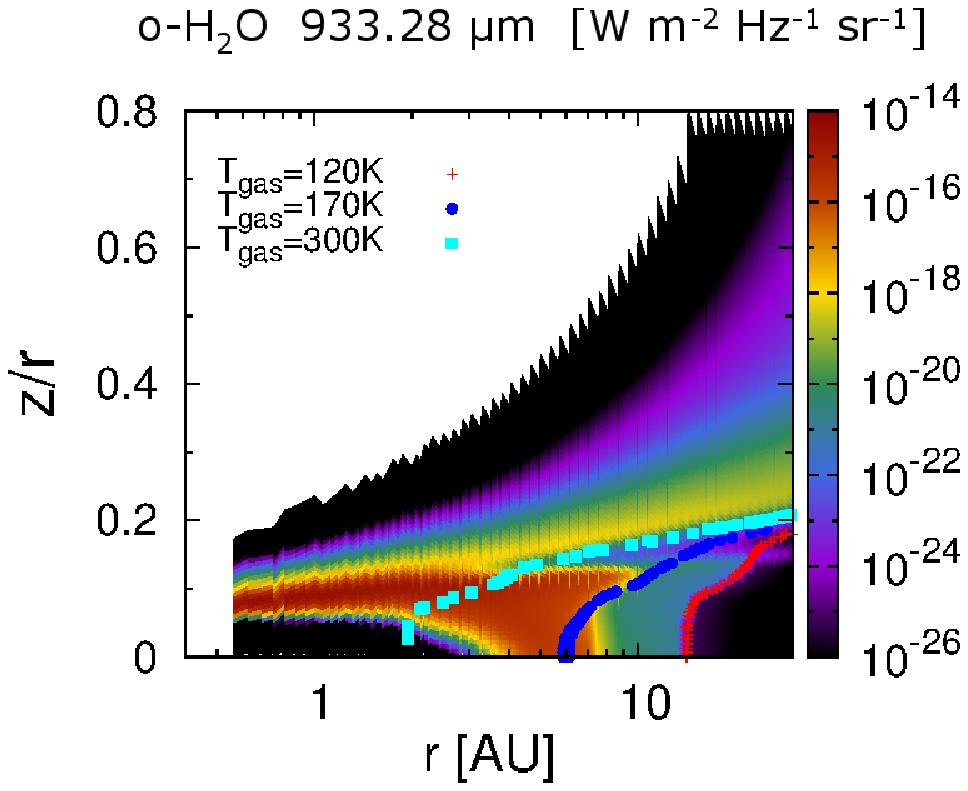}
\end{center}
\caption{\noindent (Continued.) The line-of-sight emissivity distributions of the $\mathrm{H_2O}$ lines at $\lambda$=94.17$\mu$m (top left and right), 482.99$\mu$m (second line left and right), 682.66$\mu$m (third line left and right), and 933.28$\mu$m (bottom left and right), which have small $A_{ul}$ and large $E_{up}$, from the Herbig Ae disk.
}\label{Figure5_HerbigAe}
\end{figure*}
\begin{figure*}[htbp]
\begin{center}
\includegraphics[scale=0.5]{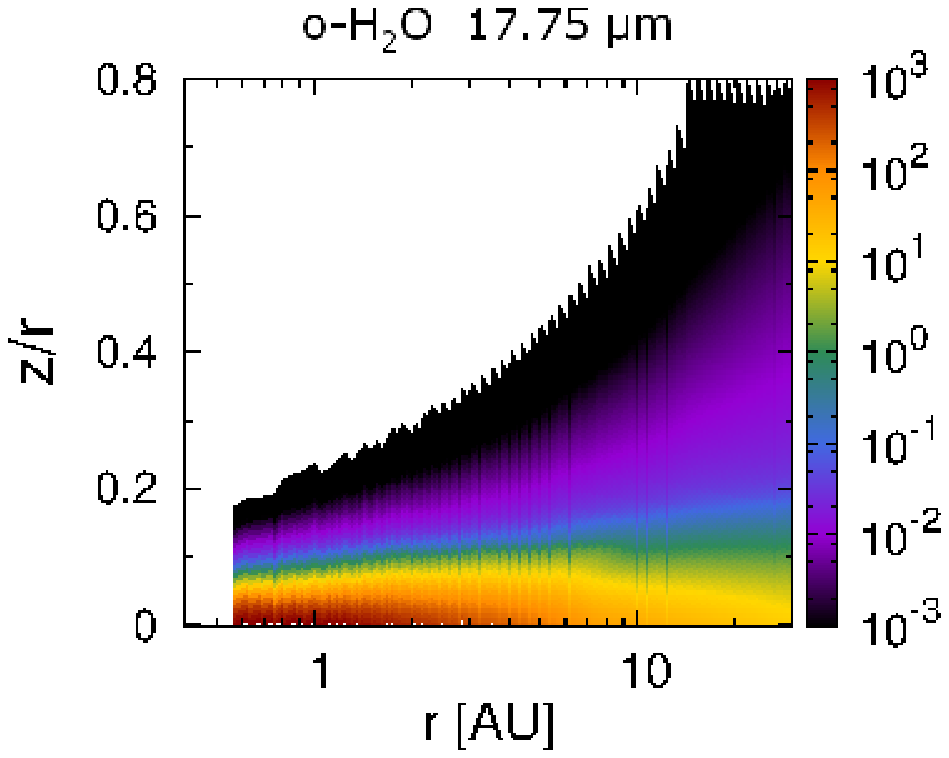}
\includegraphics[scale=0.5]{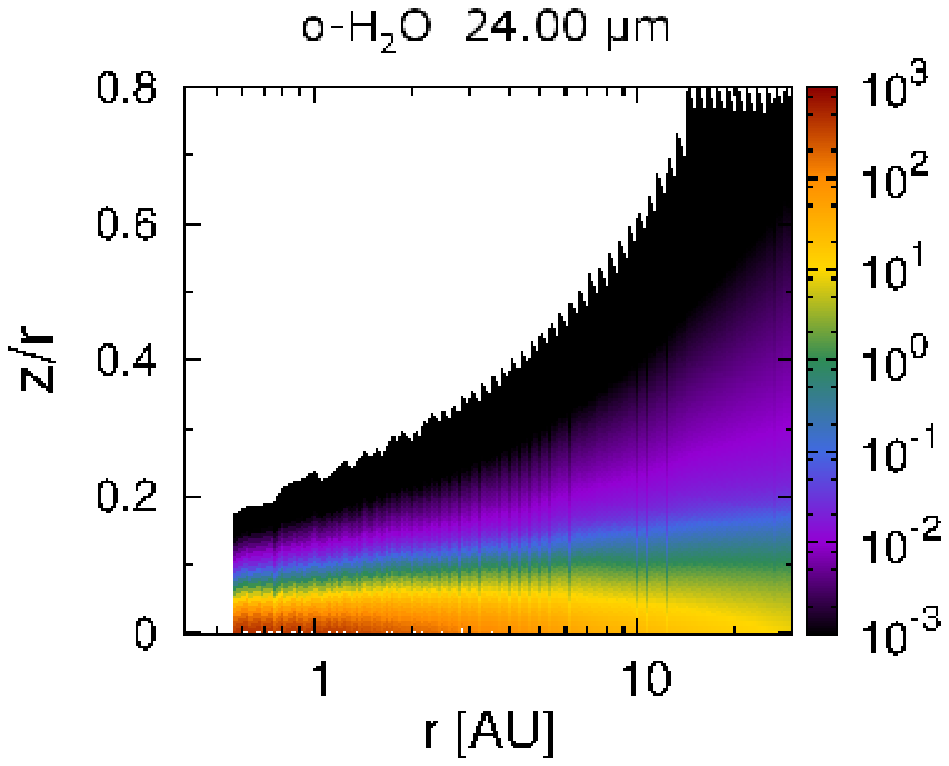}
\includegraphics[scale=0.5]{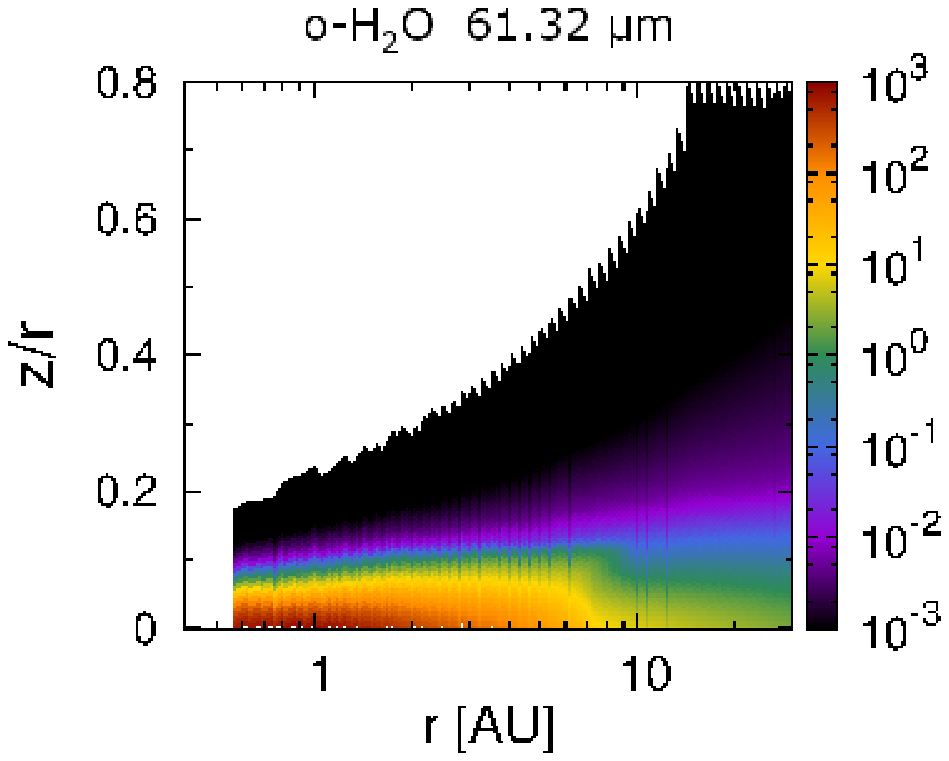}
\includegraphics[scale=0.5]{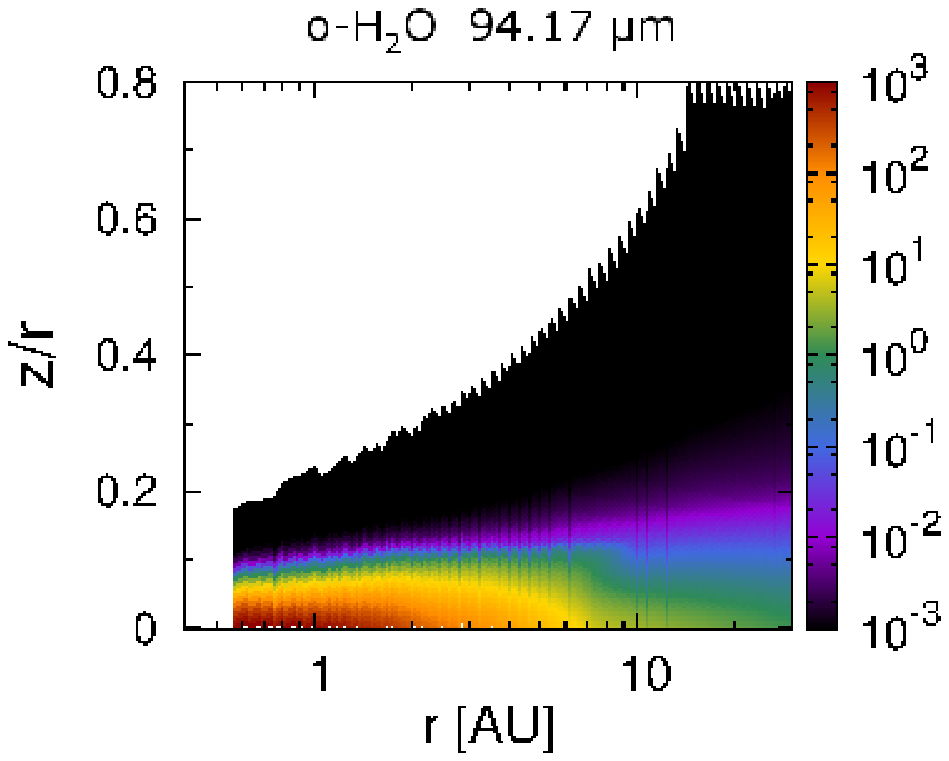}
\includegraphics[scale=0.5]{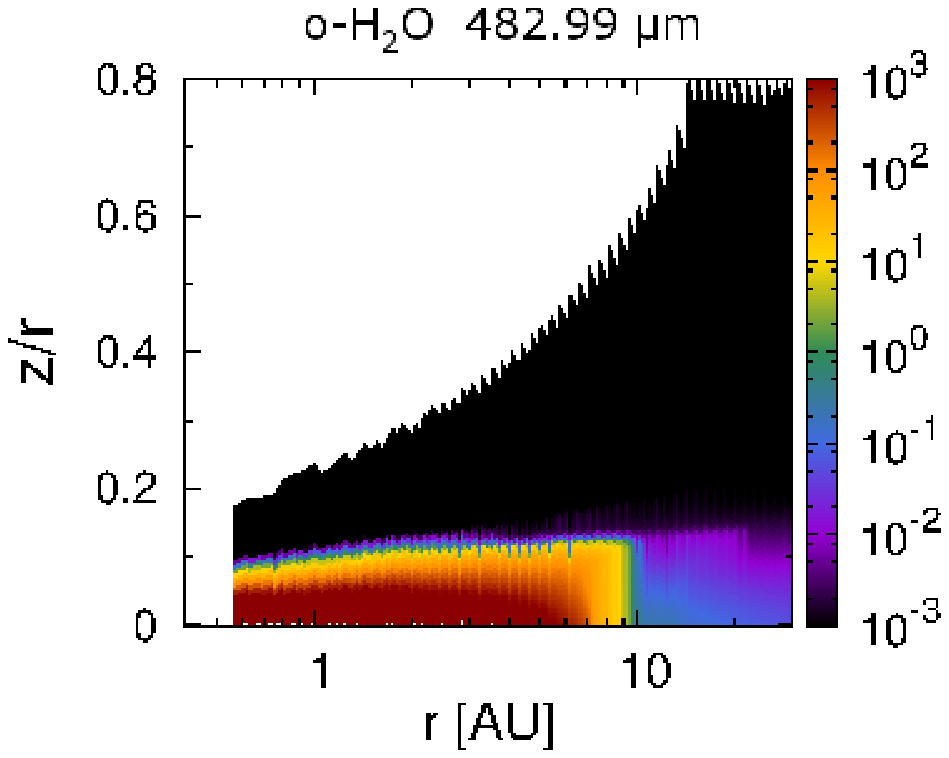}
\includegraphics[scale=0.5]{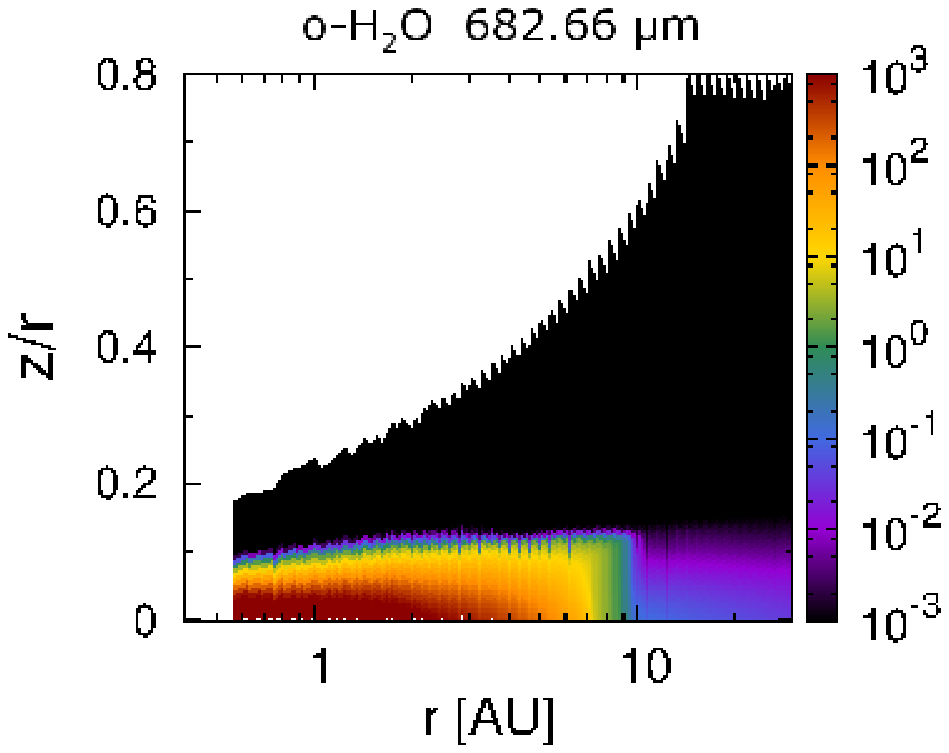}
\includegraphics[scale=0.5]{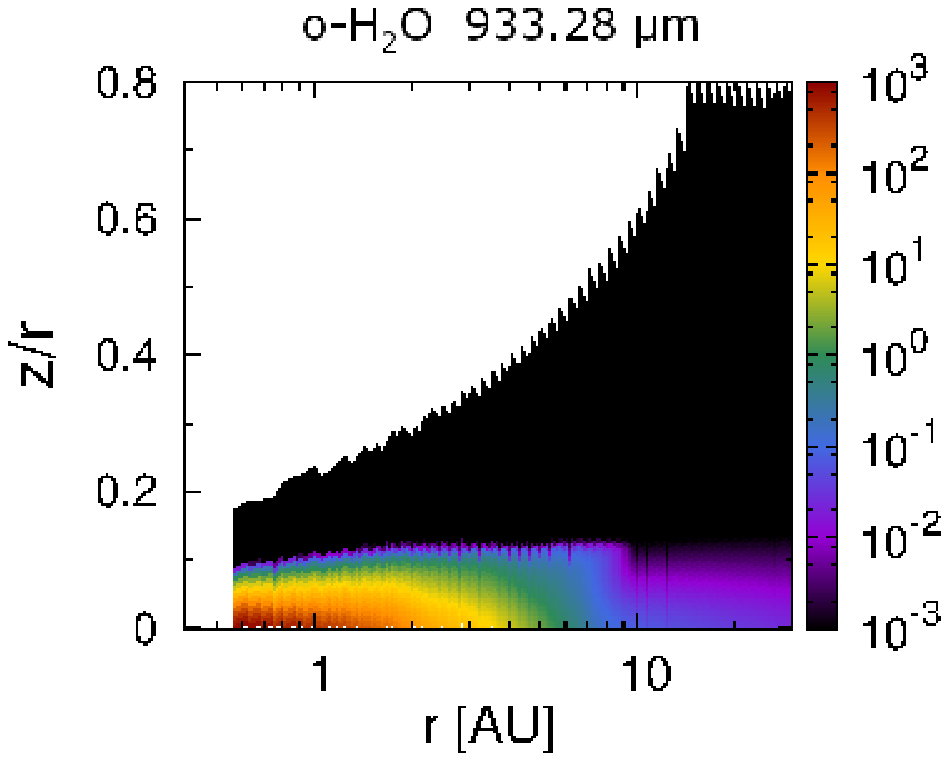}
\end{center}
\caption{\noindent The line-of-sight total optical depth $\tau_{ul}(s,x,y,\nu)$ (gas emission and dust) distributions of the $\mathrm{H_2O}$ lines at $\lambda$=17.75$\mu$m (top left), 24.00$\mu$m (top center),  61.32$\mu$m (top right), 94.17$\mu$m (middle left), 482.99$\mu$m (middle center), 682.66$\mu$m (middle right), and 933.28$\mu$m (bottom), which have small $A_{ul}$ and large $E_{up}$, from the Herbig Ae disk. 
We assume that the inclination angle, $i$, of the disk is 0 deg in making these figures in this paper (see Figure \ref{Figure6_HerbigAe} and \ref{Figure9_HerbigAe}), and thus the optical depth is calculated along the line from z=+$\infty$ to -$\infty$ at each disk radius.
In calculating the values of $\tau_{ul}(s,x,y,\nu)$, we consider the contributions of both absorption by dust grains and the line absorption by the $\mathrm{H_2O}$ gas.
}\label{Figure6_HerbigAe}
\end{figure*}
\\ \\
The differences in the properties of the line profiles (see Figures \ref{Figure4_HerbigAe}, \ref{Figure5_HerbigAe}, \ref{Figure6_HerbigAe}, and  \ref{Figure13_HerbigAe}) come from the differences in $A_{ul}$, $E_{up}$, and wavelengths among lines.
For the lines with similar wavelengths, the values of optical depth tend to be larger as values of $A_{ul}$ of the lines become larger, since the absorption by excited $\mathrm{H_2O}$ molecules increases. In addition, the values of optical depth become larger as values of $E_{up}$ become smaller, since the line absorption become stronger even in the colder region of the disk.
\\ \\
For the lines at shorter wavelengths, the opacity of the dust grains becomes larger (e.g., \citealt{NomuraMillar2005}).
In the cases of shorter wavelength (mid- and far-infrared) lines, the absorption by the dust grains mainly determines total optical depth profiles (including gas and dust components) and line emitting regions in the both inner and outer disk. 
In contrast, for the lines with longer wavelengths, 
the values of the line absorption by excited molecules become larger (see also Equation (10) of Paper I \citep{Notsu2016}) even if the values of $A_{ul}$ and $E_{up}$ are similar.
In the case of longer wavelength lines (sub-millimeter lines),
the line absorption by excited molecules mainly determines the total optical depth profiles and line emitting regions in the inner disk midplane with a high $\mathrm{H_2O}$ gas abundance ($\sim$$10^{-4}$, $r<7-8$ au), although the absorption by dust grains mainly determines those in the disk surface and outer colder disk midplane.
\\ \\
The $\mathrm{H_2O}$ 482.99$\mu$m and 682.66$\mu$m lines have relatively smaller values of $E_{up}$ ($<$1100K), and thus they can also trace outer colder region compared to lines with larger values of $E_{up}$. In addition, they have longer wavelengths ($>$ 400$\mu$m) compared with other lines, thus the dust opacity is smaller and they can trace the regions closer to the midplane in the outer disk.
These are the reasons why emission fluxes from the region with a relatively high $\mathrm{H_2O}$ gas abundance ($\sim$$10^{-8}$, $r=$8-14 au) are not so small ($\sim 10-20\%$ of total emission fluxes) compared to the region with a high $\mathrm{H_2O}$ gas abundance ($\sim$$10^{-4}$, $r<$8 au). 
Although the $\mathrm{H_2O}$ 933.28 $\mu$m line resides in the sub-millimeter region, this line has a larger $E_{up}$($=$1861.2K) than other lines, and thus most of the emission flux is emitted from the region with high temperature and a high $\mathrm{H_2O}$ gas abundance ($\sim$$10^{-4}$, $r<$8 au).
\\ \\
Since the radial difference of the positions between the exact $\mathrm{H_2O}$ snowline location and the outer edge of the hot $\mathrm{H_2O}$ gas region is not so large (several au), the influence is not so serious when we want to get information on the overall $\mathrm{H_2O}$ distribution of the inner disk and roughly estimate the position of the $\mathrm{H_2O}$ snowline.
However, if we observe several candidate $\mathrm{H_2O}$ lines with small $A_{ul}$ ($\sim 10^{-6} - 10^{-3}$), various E$_{up}$ (e.g., $\sim700-2100$K), and at various wavelengths between mid-infrared and sub-millimeter, there is the possibility to confine the detailed distribution, not only the position of the $\mathrm{H_2O}$ snowline, but also the $\mathrm{H_2O}$ gas abundance and the gas temperature in the disk midplane.
In addition, we could trace the water reservoir within the $\mathrm{H_2O}$ snowline from the Keplerian line profiles independently regardless of the assumption of the relation between disk gas temperature and radius, as adopted in previous works to get the $\mathrm{H_2O}$ distributions (e.g., \citealt{Zhang2013, Blevins2016}).
We note that the previous observations of $\mathrm{H_2O}$ lines with large $A_{ul}$ ($\sim 10^{-1}-10^{0}$ s$^{-1}$) and very high $E_{up}$ ($>$3000K) in many cases mainly trace the emission from the disk surface (e.g., \citealt{Salyk2008, Pontoppidan2010a, Pontoppidan2010b, Fedele2011, Mandell2012, vanDishoeck2014, Banzatti2016, Blevins2016}, see also Section 3.2.2).
 \\
\subsubsection{$\mathrm{H_2O}$ lines which are less suited to trace emission from water reservoir within the $\mathrm{H_2O}$ snowline}
\begin{figure*}[htbp]
\begin{center}
\includegraphics[scale=0.45]{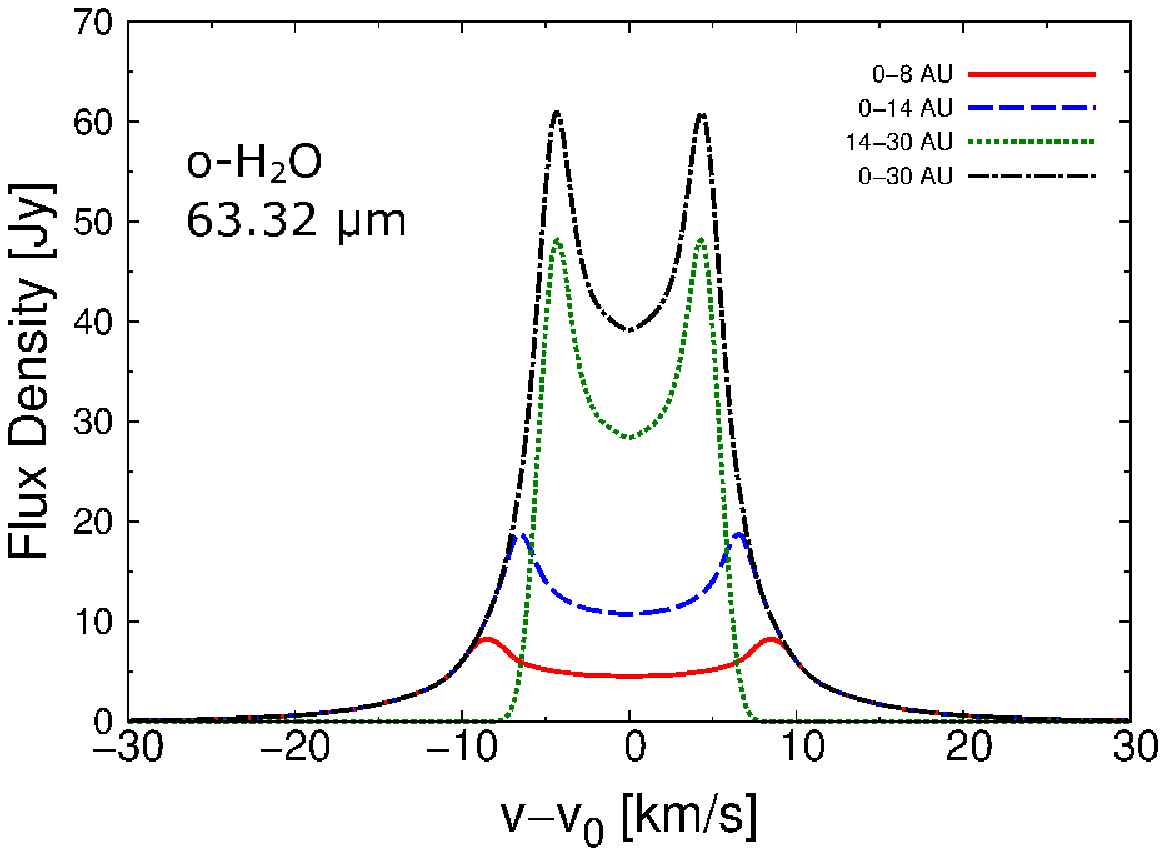}
\includegraphics[scale=0.45]{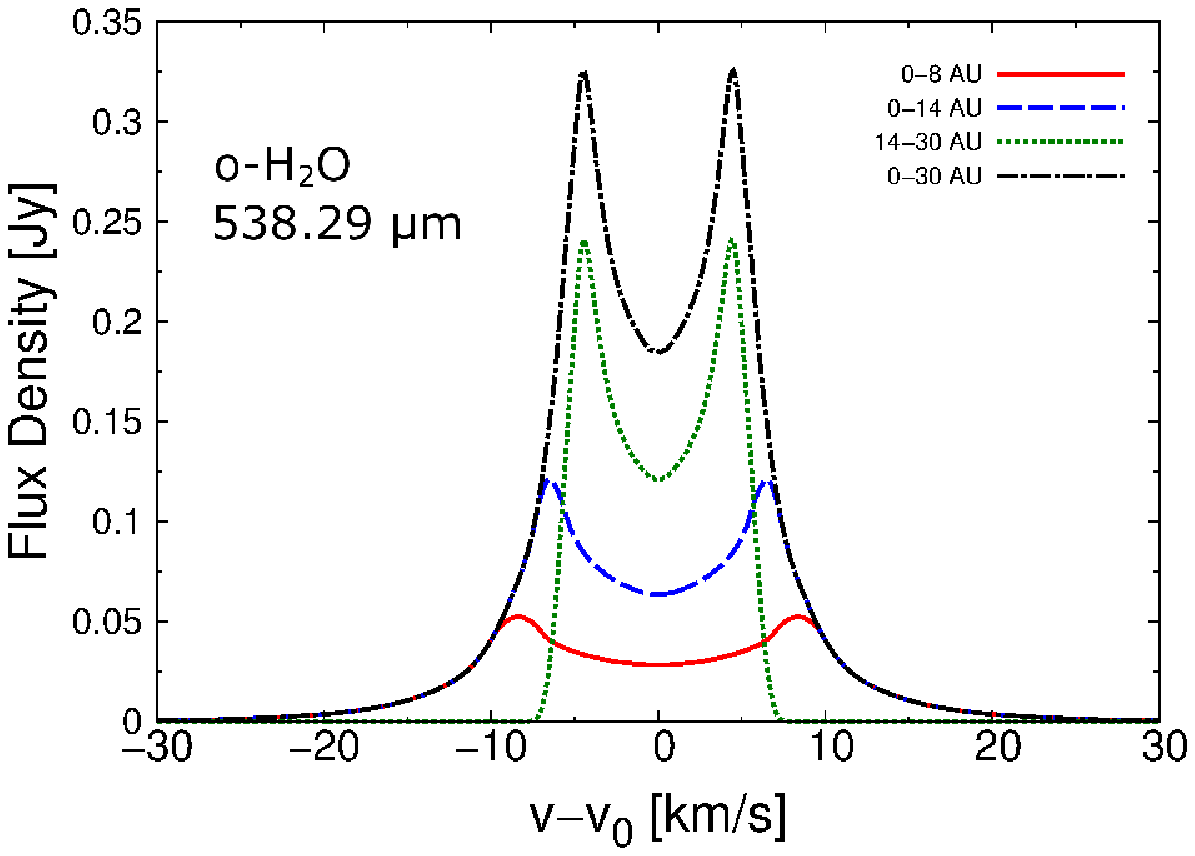}
\includegraphics[scale=0.45]{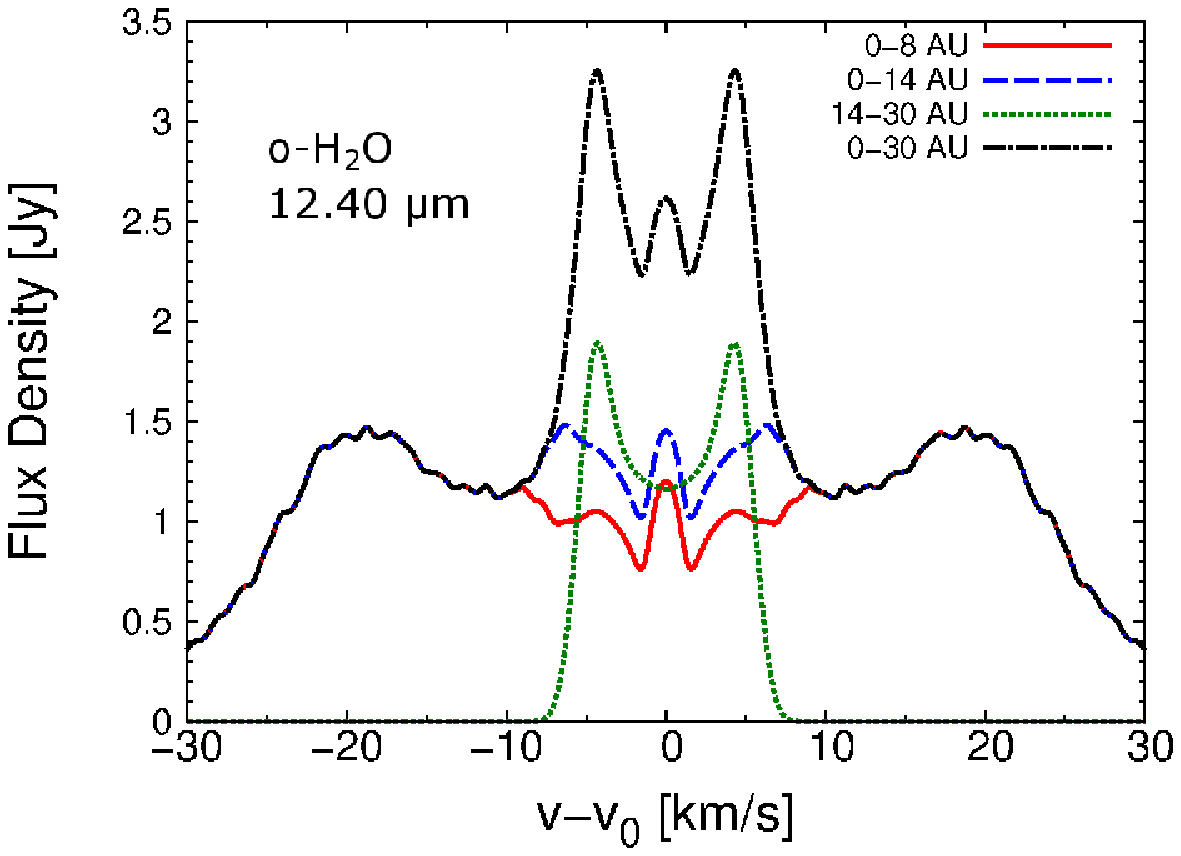}
\includegraphics[scale=0.45]{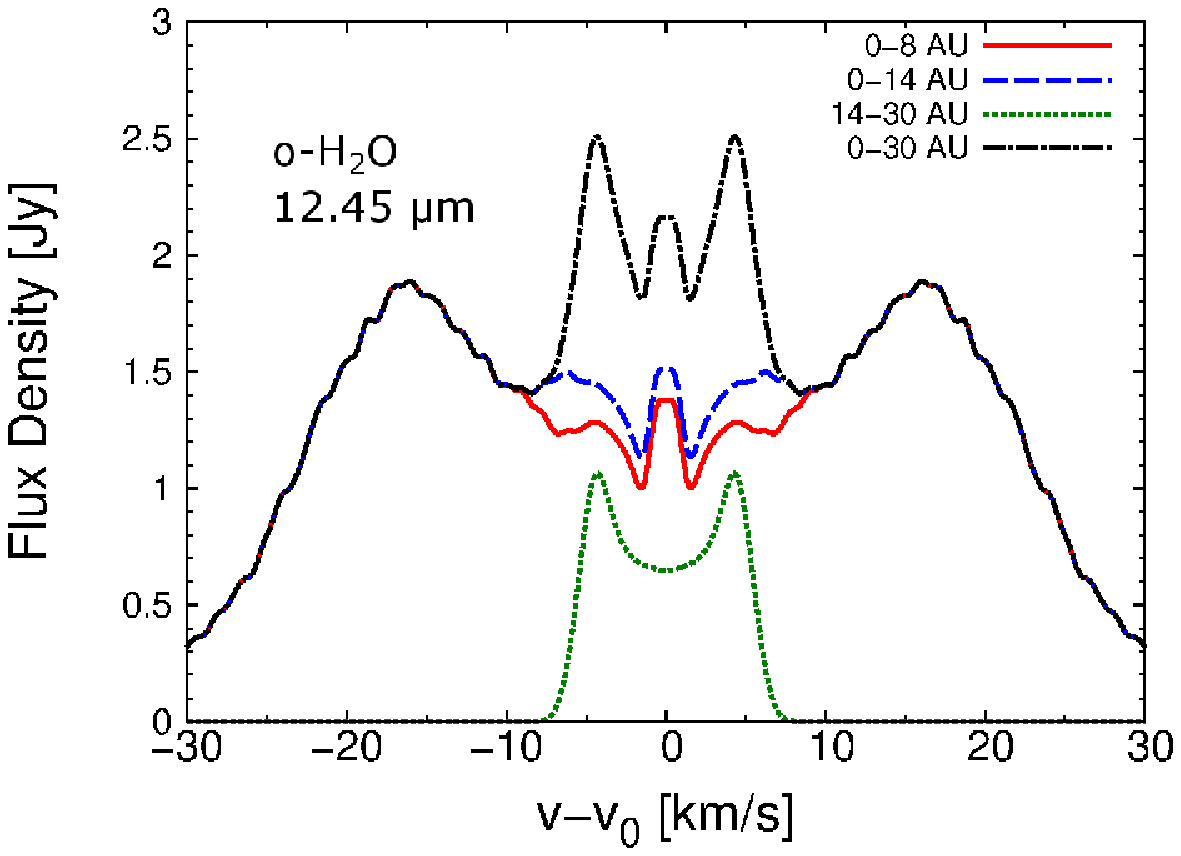}
\includegraphics[scale=0.45]{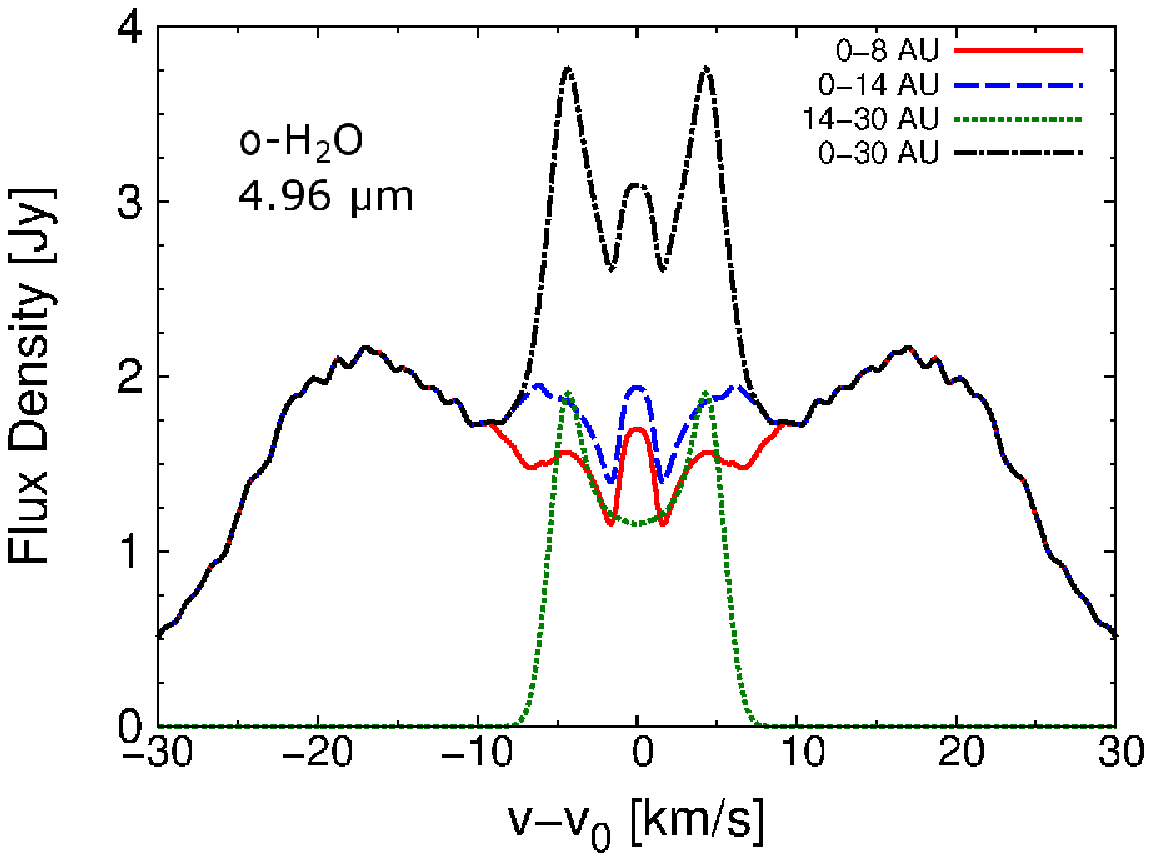}
\includegraphics[scale=0.45]{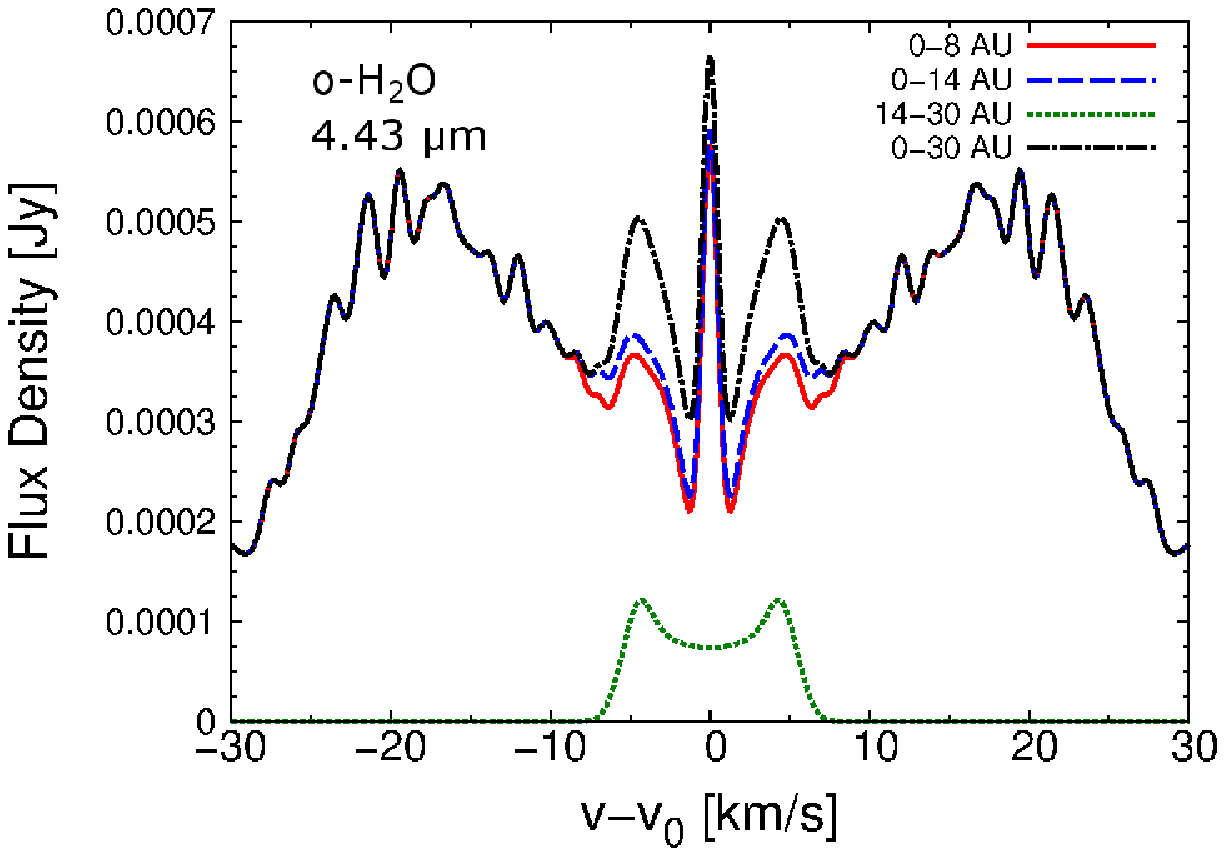}
\end{center}
\caption{\noindent (Top two panels): The velocity profiles of two characteristic pure rotational ortho-$\mathrm{H_2O}$ lines at $\lambda=$ 63.32$\mu$m (top left) and 538.29$\mu$m (top right) from the Herbig Ae disk. 
They are examples of lines which are less suited to trace emission from water vapor within the $\mathrm{H_2O}$ snowline.
(Middle two panels): The velocity profiles of mid-infrared ortho-$\mathrm{H_2O}$ lines at $\lambda$=12.40$\mu$m (middle left), 12.45$\mu$m (middle right) from the Herbig Ae disk. Both lines have much larger values of $A_{ul}$ and $E_{up}$ than those of the candidate mid-infared $\mathrm{H_2O}$ lines to trace the emission from the $\mathrm{H_2O}$ vapor within the $\mathrm{H_2O}$ snowline.
(Bottom two panels): The velocity profiles of near-infrared ortho-$\mathrm{H_2O}$ lines at $\lambda$=4.96$\mu$m (bottom left), 4.43$\mu$m (bottom right) from the Herbig Ae disk. Both lines have the same values of $E_{up}$ ($=$4180.4K), the former line has a larger value of $A_{ul}$ and the latter line has a smaller value of $A_{ul}$.
{\it Red solid lines} are the emission line profiles from inside 8 au ($=$the inner high temperature region), {\it blue dashed lines} are those from inside 14 au ($\sim$ inside the $\mathrm{H_2O}$ snowline), {\it green dotted lines} are those from 14-30 au ($\sim$ outside the $\mathrm{H_2O}$ snowline), and {\it black dashed dotted lines} are those from the total area inside 30au. 
}\label{Figure7_HerbigAe}
\end{figure*}

\noindent In the top left panel of Figure \ref{Figure7_HerbigAe}, we show the line profile for the $\mathrm{H_2O}$ 63.32$\mu$m line. The contribution from the optically thin surface layer of the outer disk ($r=$14-30 au) is large (three times larger in flux density) compared with that of the optically thick region near the midplane of inner disk ($r <$14 au), and the shape of the line profile is a narrower by around a factor of two compared to those of candidate $\mathrm{H_2O}$ lines which trace the emission from the hot water vapor within the $\mathrm{H_2O}$ snowline.
This is because the $\mathrm{H_2O}$ 63.32$\mu$m line has a large $A_{ul}$ (=1.772 s$^{-1}$), although $E_{up}$ (=1070.6K) is similar to that of the candidate ortho-$\mathrm{H_2O}$ lines (e.g., $E_{up}=$1088.7K for the 682.66$\mu$m line).
The detailed parameters, such as transitions ($J_{K_{a}K_{c}}$), wavelength, frequency, $A_{ul}$, $E_{up}$, critical density $n_{\mathrm{cr}}$, and total line fluxes of the ortho-$\mathrm{H_2O}$ lines discussed in this subsection are listed in Table \ref{tab:T1}. 
\\ \\
According to Figures \ref{Figure8_HerbigAe}, \ref{Figure9_HerbigAe} and \ref{Figure13_HerbigAe}, the values of emissivity of the $\mathrm{H_2O}$ 63.32$\mu$m line in the optically thin hot surface layer of outer disk
are as strong as that of the optically thick region inside the $\mathrm{H_2O}$ snowline. 
The area of the outer $\mathrm{H_2O}$ line emitting region is larger than that of the inner region for this line, and thus emission from the outer part dominates.
Therefore, we propose that this line is not optimal to detect emission from the hot water vapor within the $\mathrm{H_2O}$ snowline in Herbig Ae disk case, as we also found for the T Tauri disk \citep{Notsu2016}.
\begin{figure*}[htbp]
\begin{center}
\includegraphics[scale=0.55]{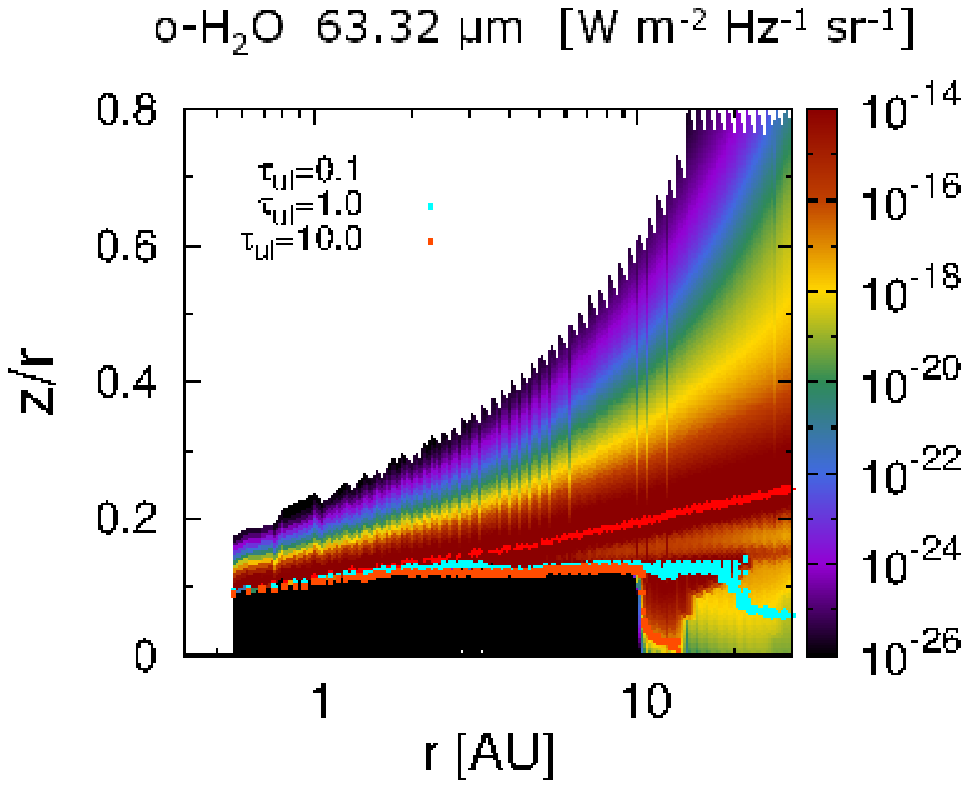}
\includegraphics[scale=0.55]{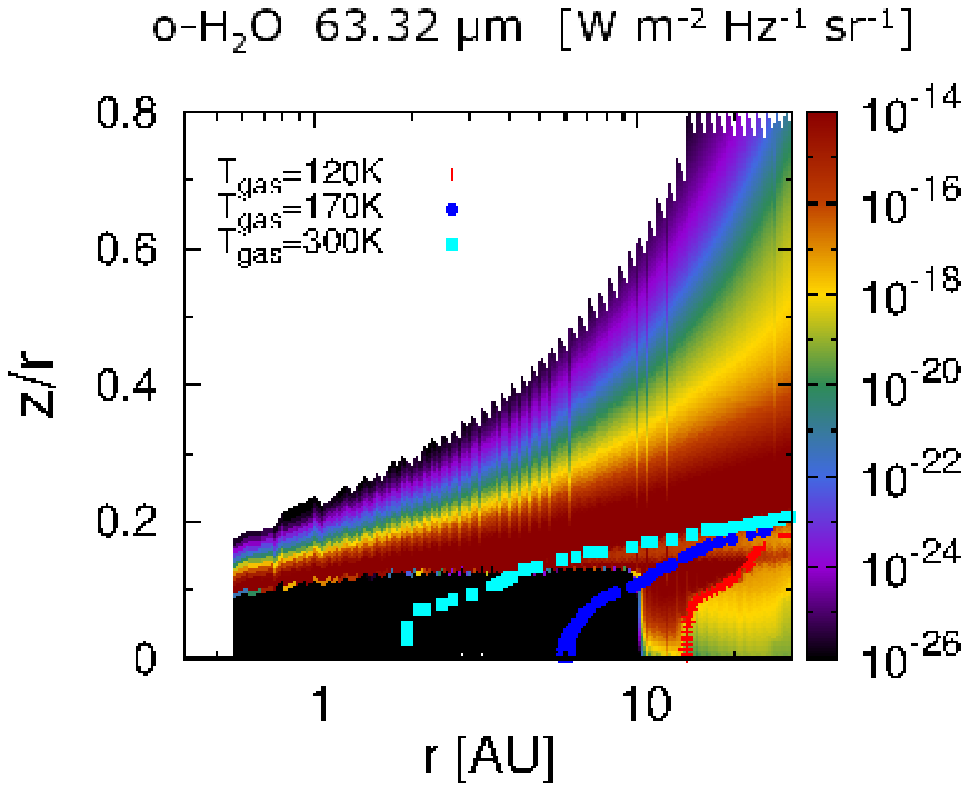}
\includegraphics[scale=0.55]{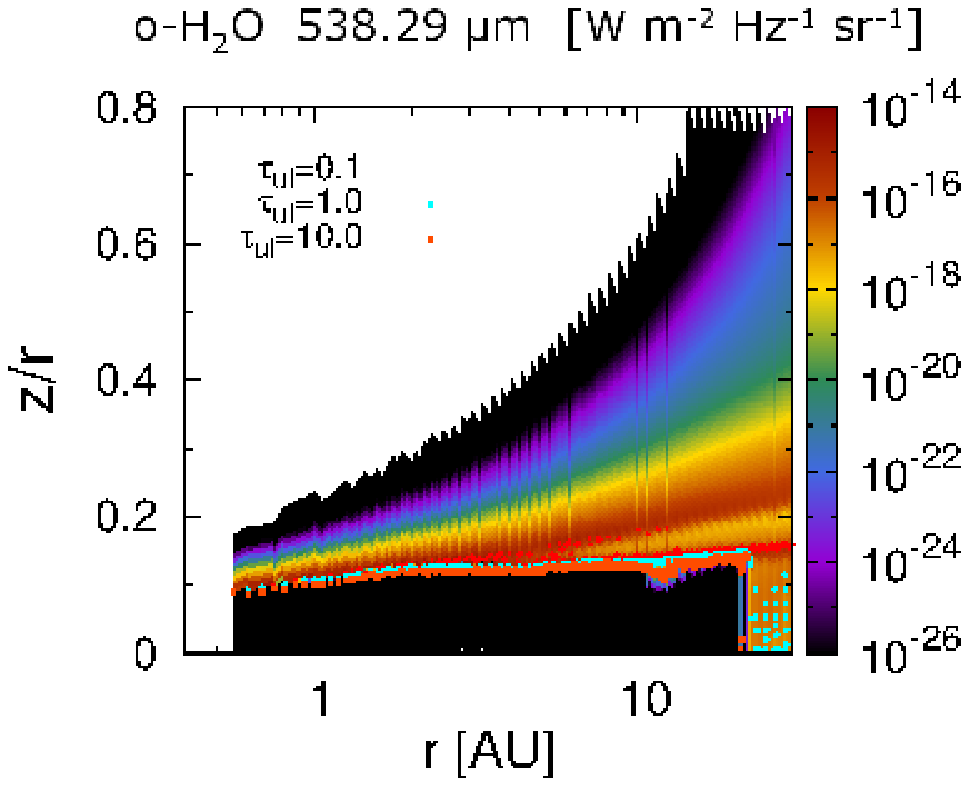}
\includegraphics[scale=0.55]{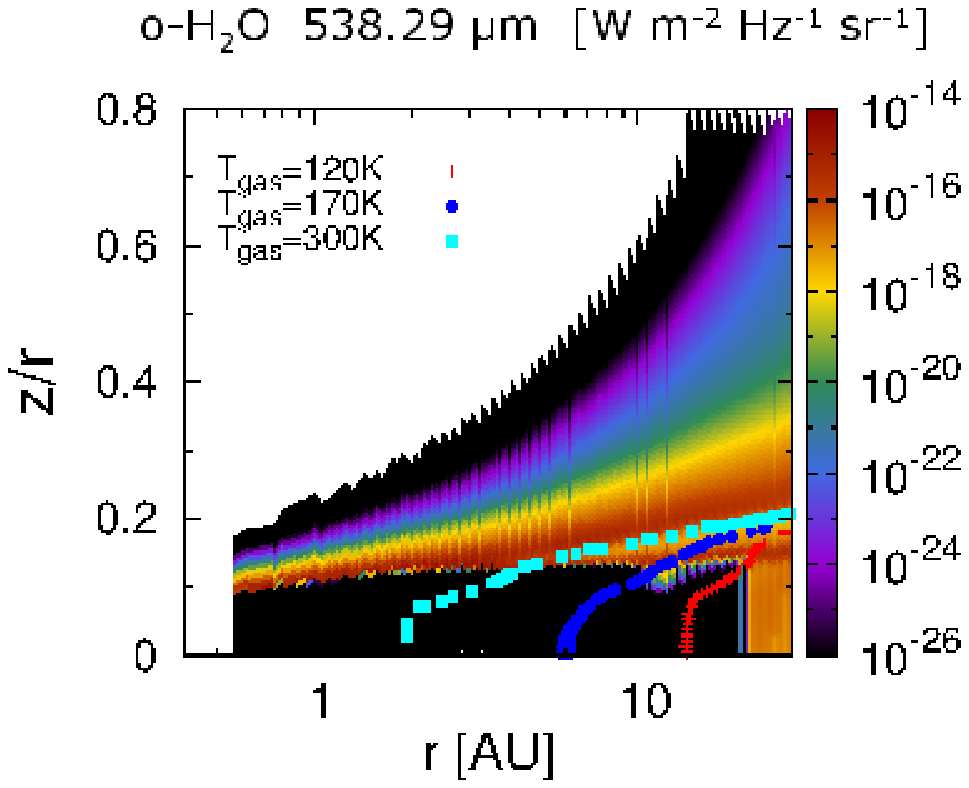}
\end{center}
\caption{\noindent The line-of-sight emissivity distributions of the two characteristic $\mathrm{H_2O}$ lines at $\lambda$=63.32$\mu$m (top left and right) and 538.29$\mu$m (bottom left and right), which have various $A_{ul}$ and $E_{up}$, from the Herbig Ae disk. 
In the left panels, we overplot the total optical depth contours $\bigl($$\tau_{ul}=$0.1 (red cross points), 1 (cyan circle points), and 10 (orange square points)$\bigr)$ on top of these line emissivity panels (see also Figure \ref{Figure9_HerbigAe}). In the right panels, we overplot the gas temperature $T_{g}$ contours $\bigl($$T_{g}=$120K (red cross points), 170K (blue circle points), and 300K (cyan square points), see also Figure \ref{Figure1_HerbigAe}$\bigr)$.
}\label{Figure8_HerbigAe}
\end{figure*}    
\setcounter{figure}{7}
\begin{figure*}[htbp]
\begin{center}
\includegraphics[scale=0.5]{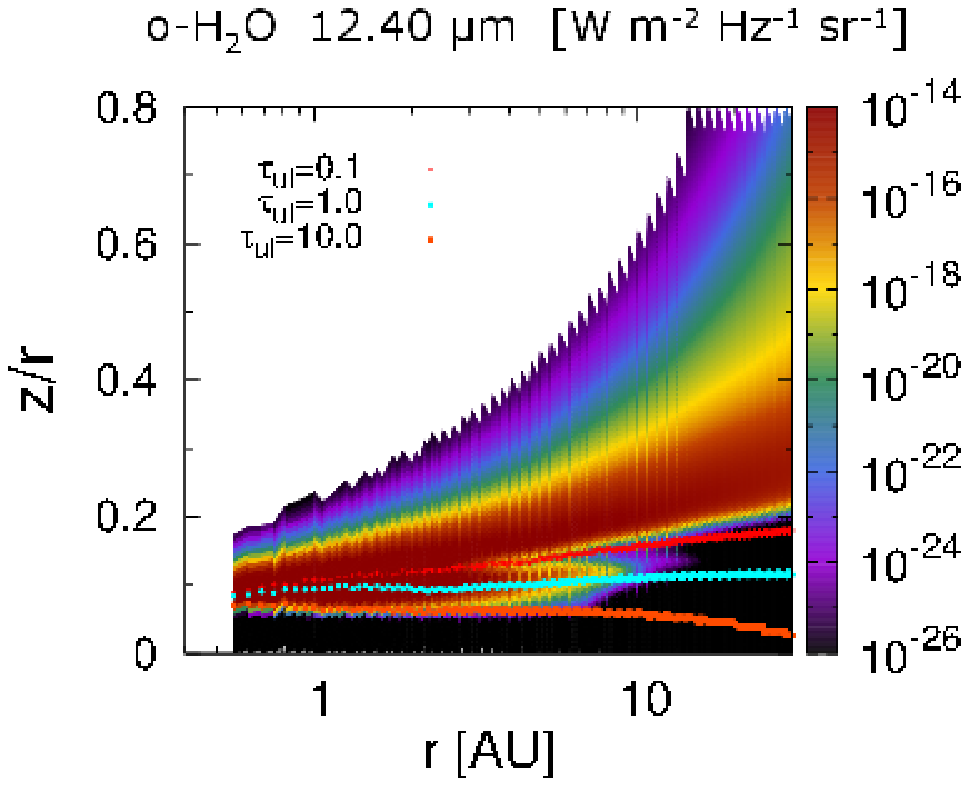}
\includegraphics[scale=0.5]{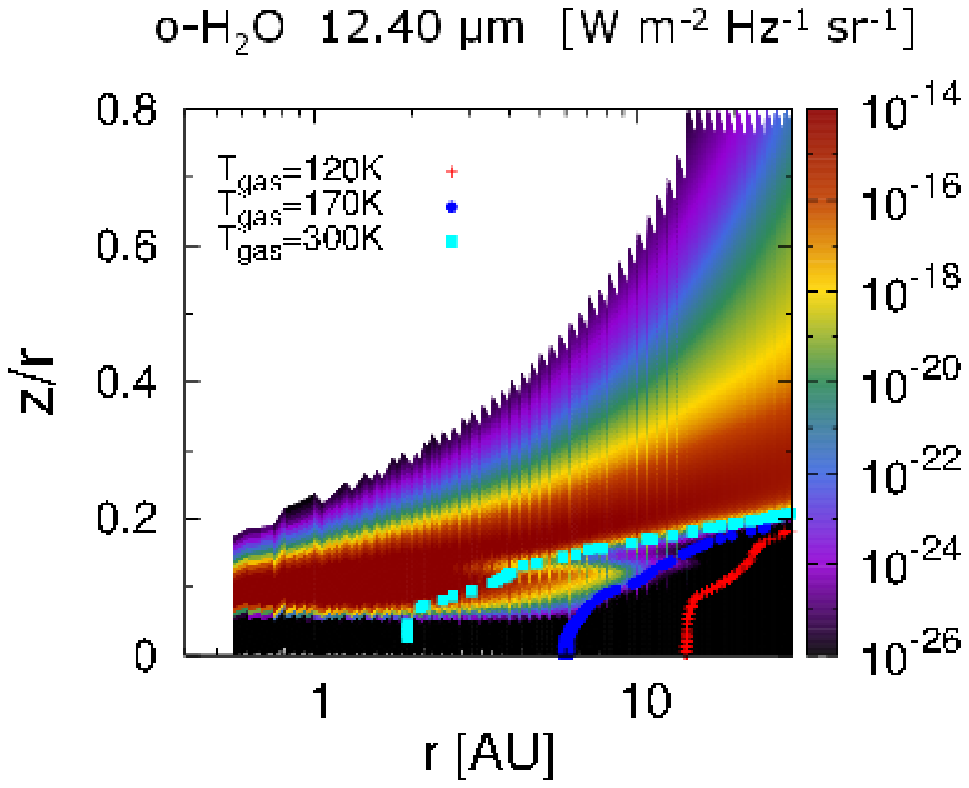}
\includegraphics[scale=0.5]{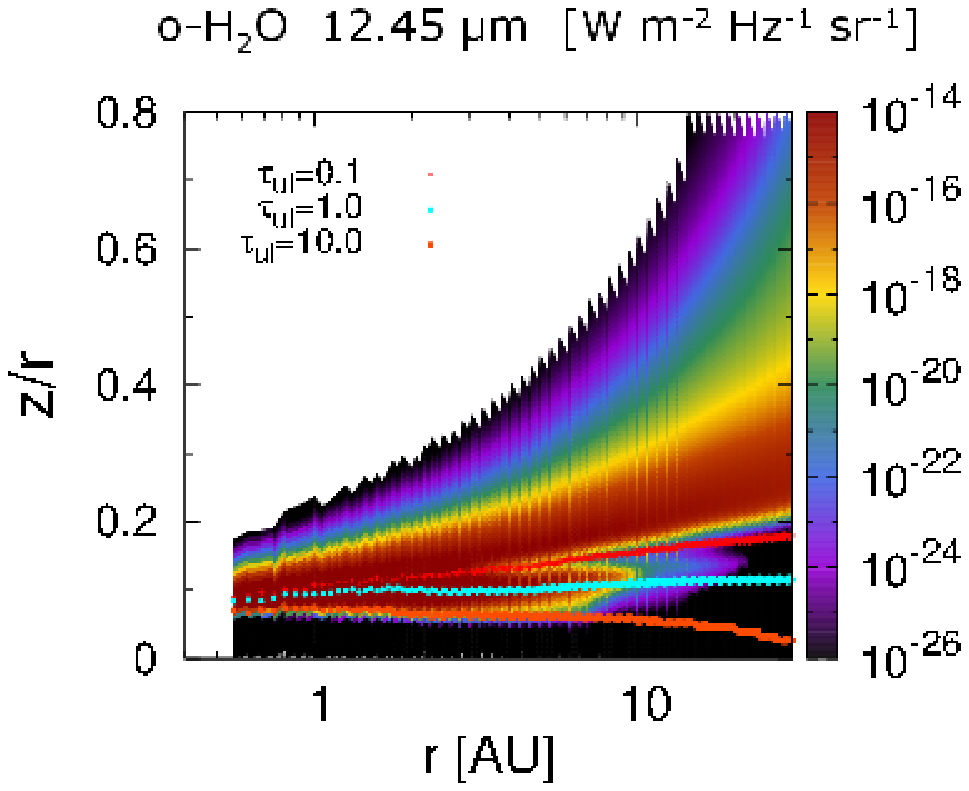}
\includegraphics[scale=0.5]{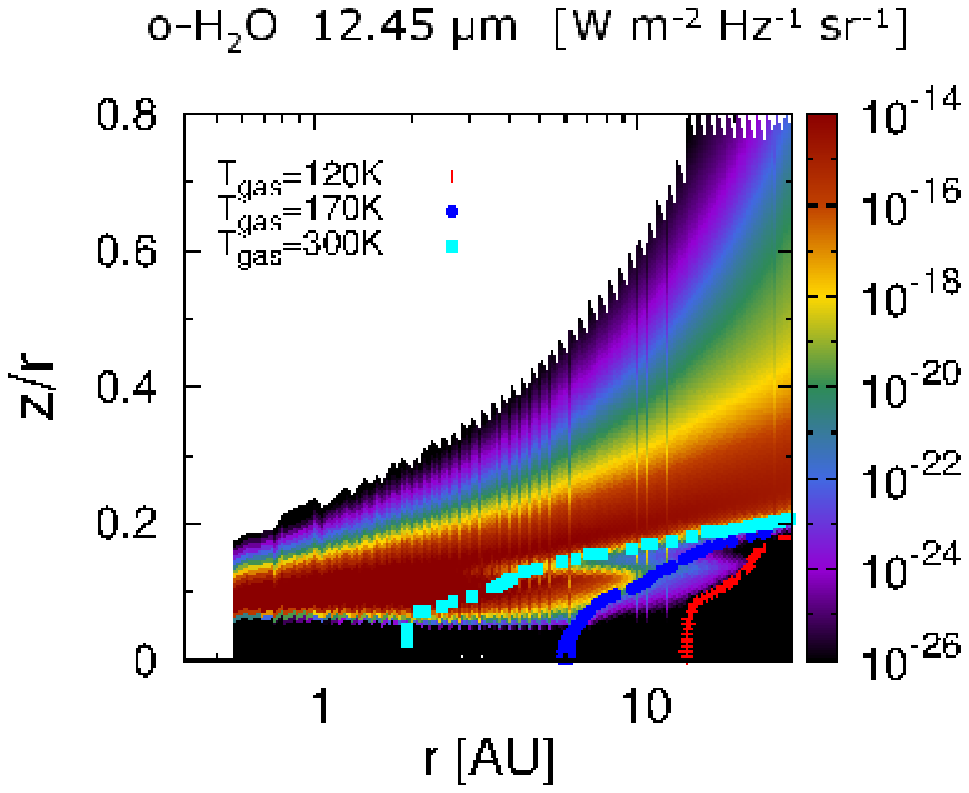}
\includegraphics[scale=0.5]{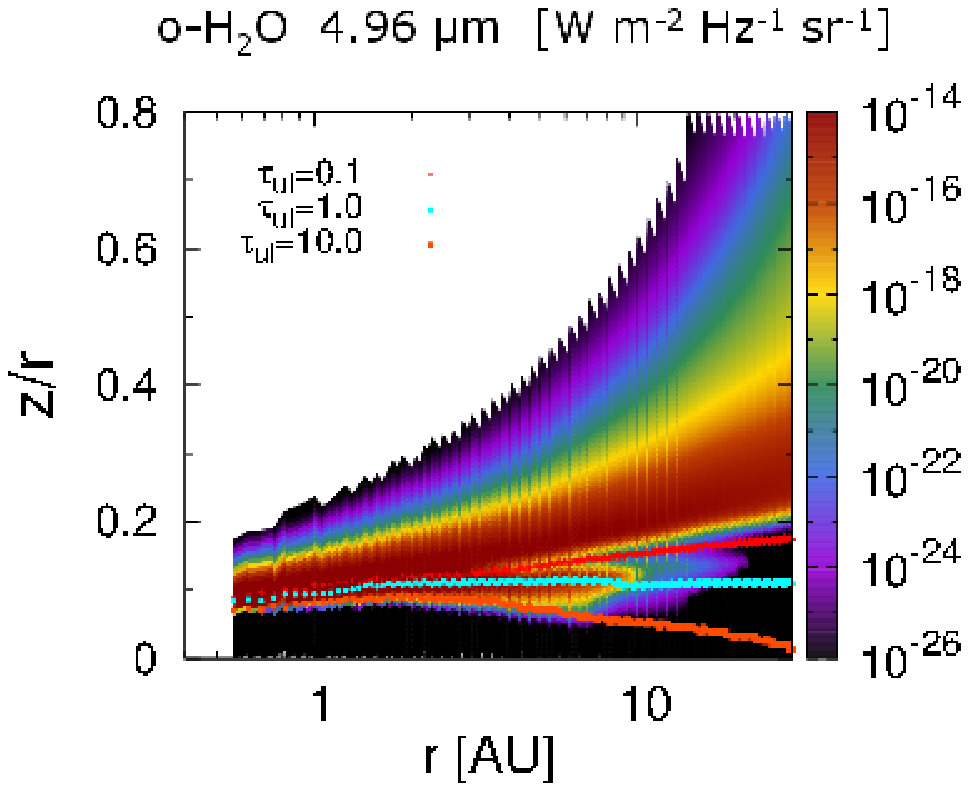}
\includegraphics[scale=0.5]{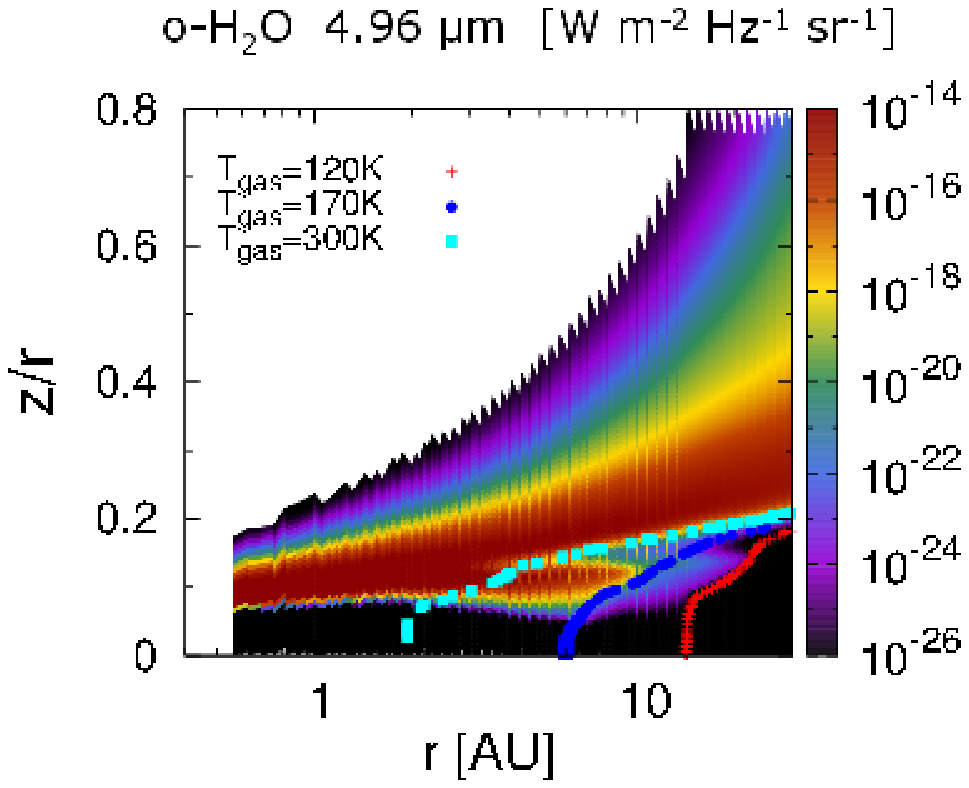}
\includegraphics[scale=0.5]{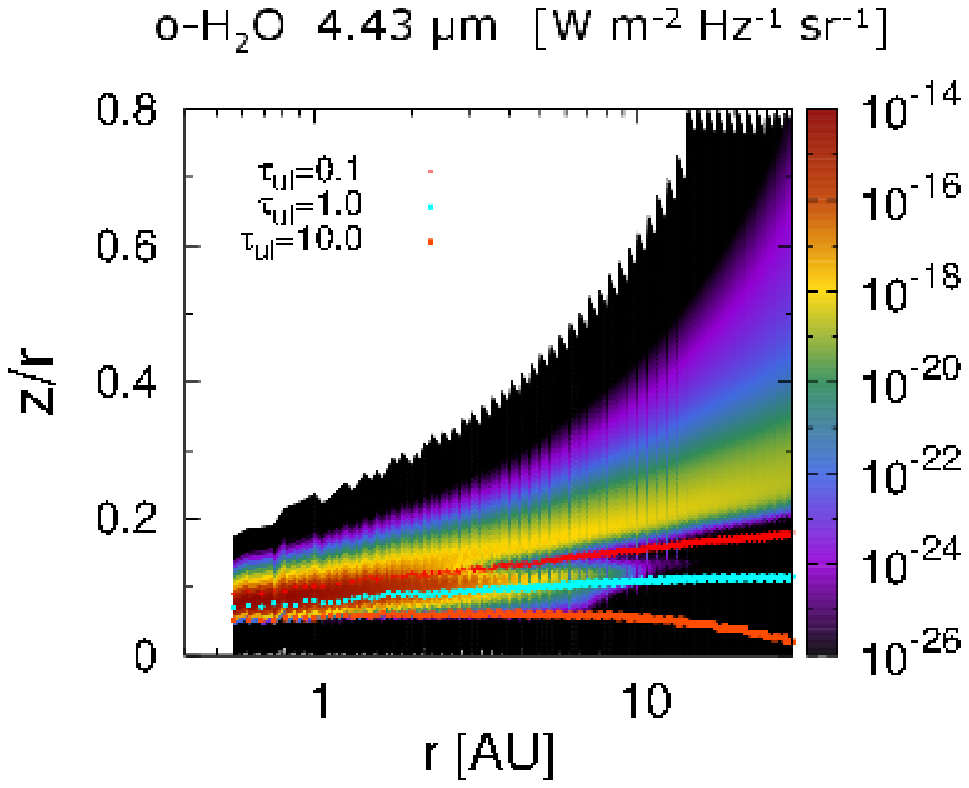}
\includegraphics[scale=0.5]{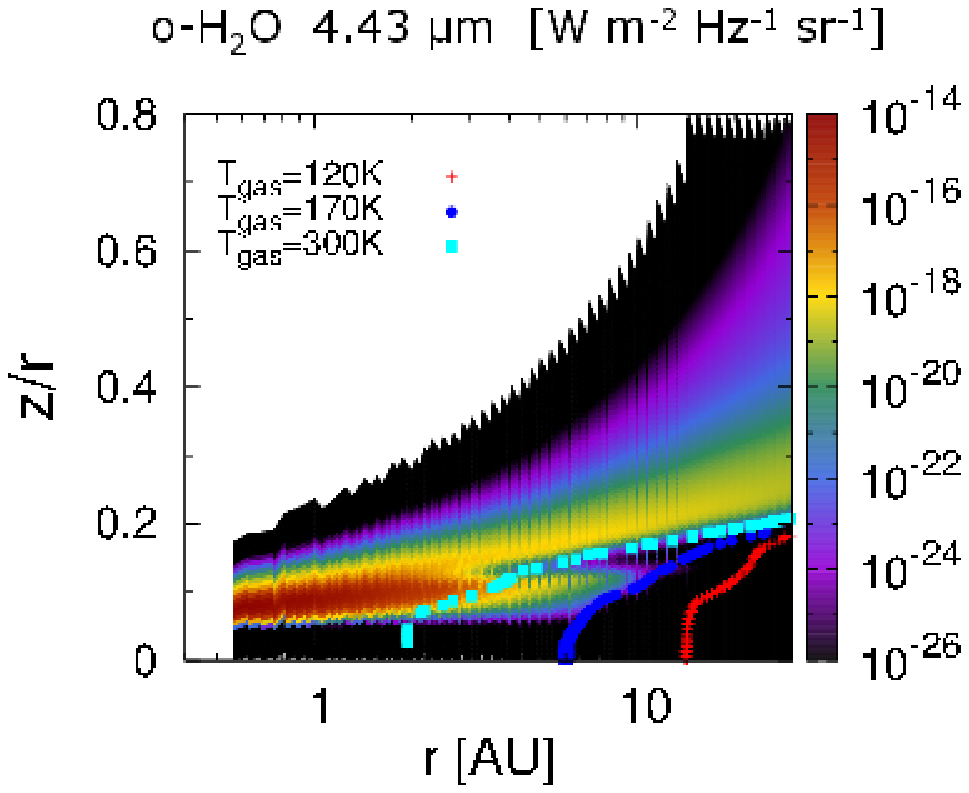}
\end{center}
\caption{\noindent (Continued.) 
(Top and second line panels): The line-of-sight emissivity distributions of mid-infrared $\mathrm{H_2O}$ lines at $\lambda$=12.40$\mu$m (top left and right), 12.45$\mu$m (second line left and right) from the Herbig Ae disk. Both lines have much larger values of $A_{ul}$ and $E_{up}$ than those of the candidate mid-infared $\mathrm{H_2O}$ lines which trace emission from the hot water vapor within $\mathrm{H_2O}$ snowline.
(Third line and bottom panels): The line-of-sight emissivity distributions of near-infrared $\mathrm{H_2O}$ lines at $\lambda$=4.96$\mu$m (third left and right), 4.43$\mu$m (bottom left and right) from the Herbig Ae disk. Both lines have the same values of $E_{up}$ ($=$4180.4K), the former line has a larger value of $A_{ul}$ and the latter line has a smaller value of $A_{ul}$.
}\label{Figure8_HerbigAe}
\end{figure*} 
\begin{figure*}[htbp]
\begin{center}
\includegraphics[scale=0.5]{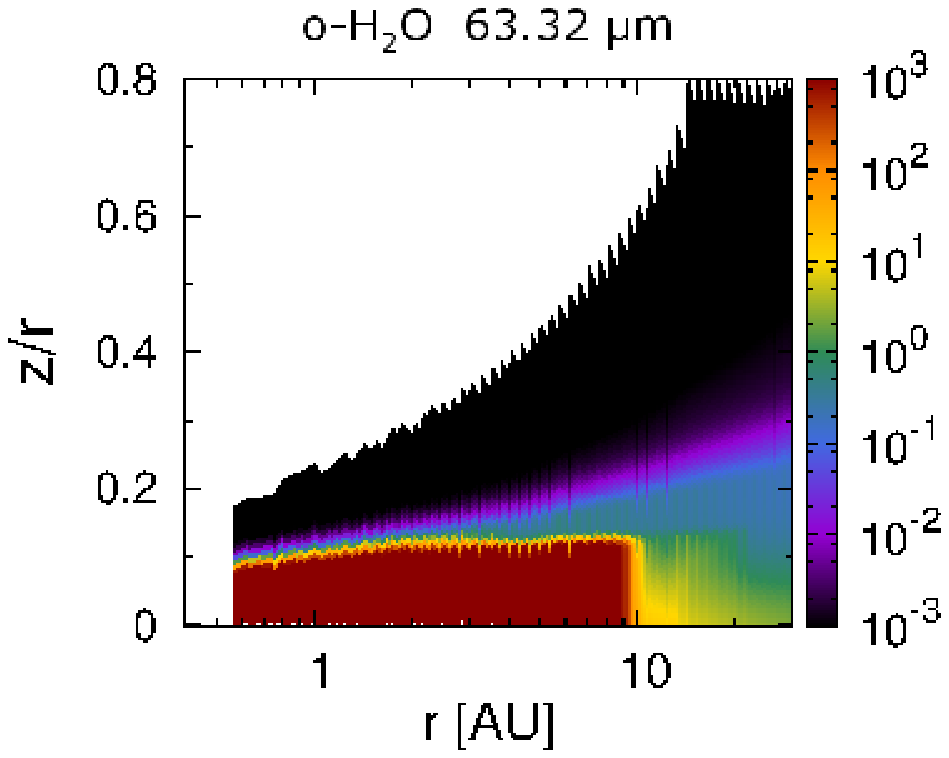}
\includegraphics[scale=0.5]{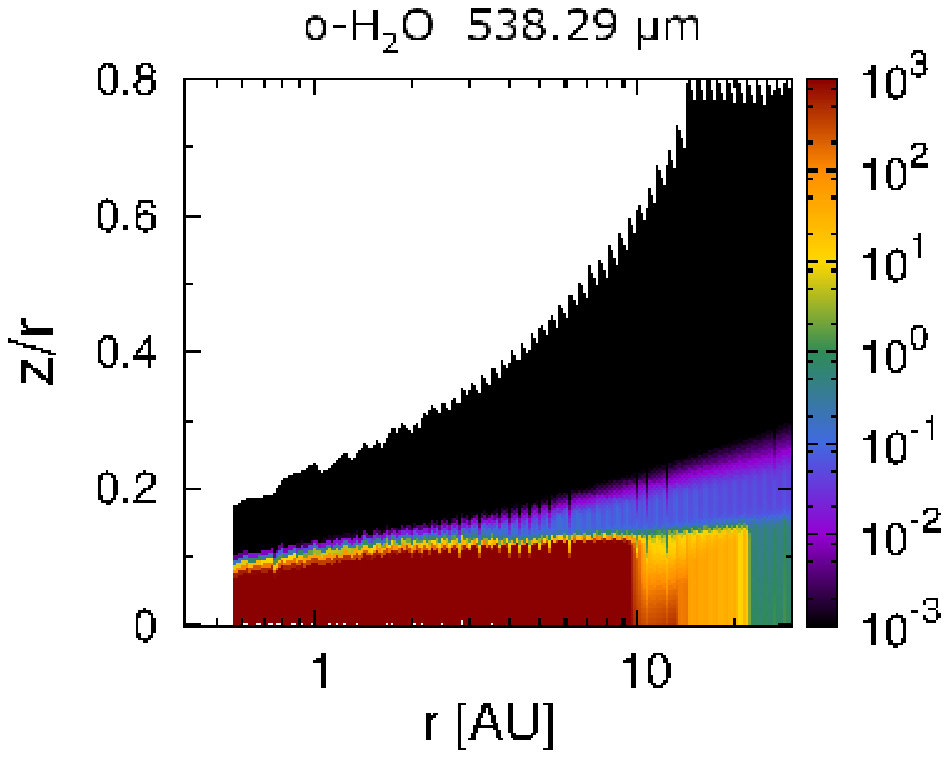}
\includegraphics[scale=0.5]{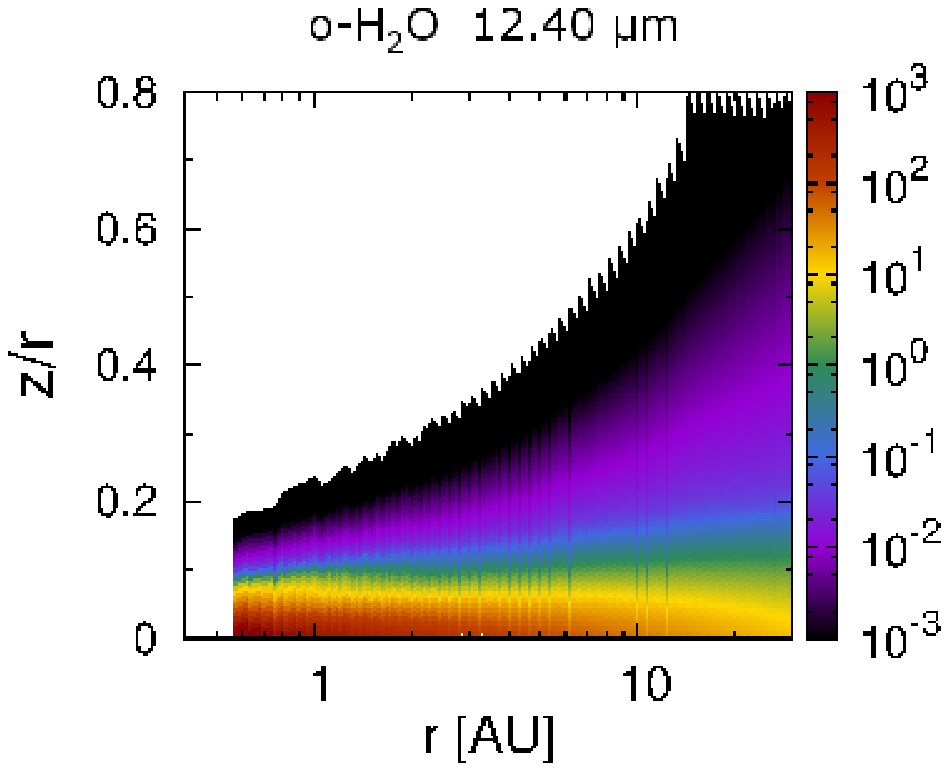}
\includegraphics[scale=0.5]{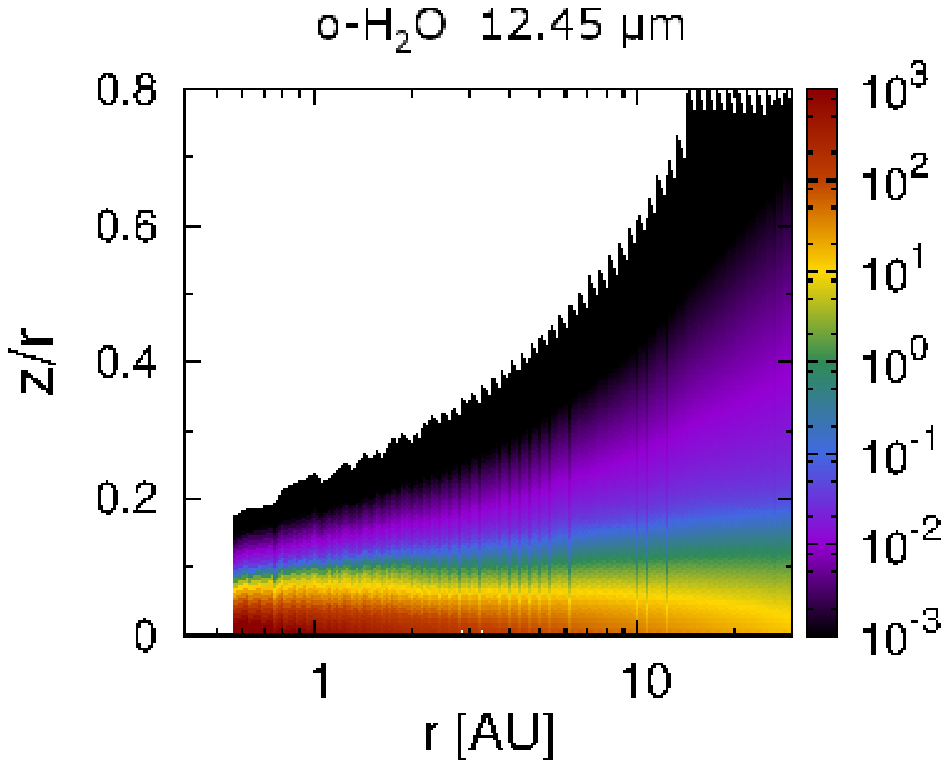}
\includegraphics[scale=0.5]{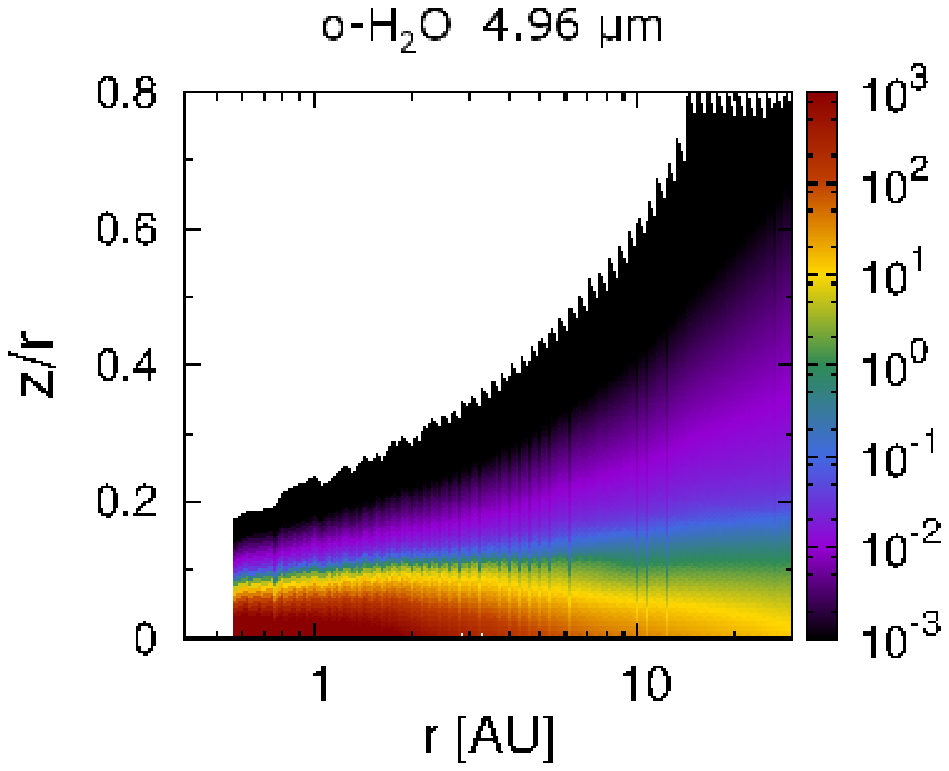}
\includegraphics[scale=0.5]{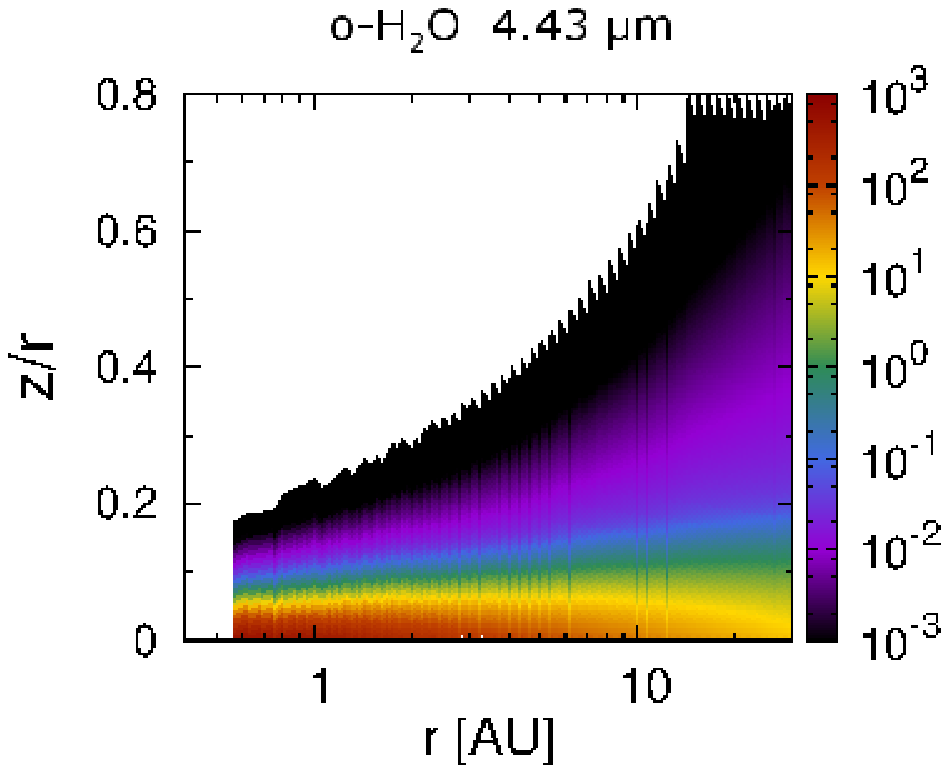}
\end{center}
\caption{\noindent (Top two panels):  The line-of-sight total optical depth $\tau_{ul}(s,x,y,\nu)$ (gas emission and dust) distributions of two characteristic $\mathrm{H_2O}$ lines at 63.32$\mu$m (top left) and 538.29$\mu$m (top right) from the Herbig Ae disk. 
(Middle two panels): The line-of-sight optical depth distributions of mid-infrared ortho-$\mathrm{H_2O}$ lines at $\lambda$=12.40$\mu$m (middle left), 12.45$\mu$m (middle right) from the Herbig Ae disk. Both lines have much larger values of $A_{ul}$ and $E_{up}$ than those of the candidate mid-infared $\mathrm{H_2O}$ lines which trace emission from the hot water vapor within $\mathrm{H_2O}$ snowline. 
(Bottom two panels): The line-of-sight optical depth distributions of near-infrared ortho-$\mathrm{H_2O}$ lines at $\lambda$=4.96$\mu$m (bottom left), 4.43$\mu$m (bottom right) from the Herbig Ae disk. Both lines have the same values of $E_{up}$ ($=$4180.4K), the former line has a larger value of $A_{ul}$ and the latter line has a smaller value of $A_{ul}$.
}\label{Figure9_HerbigAe}
\end{figure*}
\\ \\
We note that previous space far-infrared low-dispersion spectroscopic observations with \\ $Herschel$/PACS ($R\sim$1500) detected some far-infrared $\mathrm{H_2O}$ emission lines with large $A_{ul}$ ($\sim 10^{-1} -10^{0}$ s$^{-1}$) and relatively large $E_{up}$ ($\sim$ 1000K) including this 63.32$\mu$m line from the gas-rich Herbig Ae disk around HD163296, although the detections of these lines are only slightly above 3$\sigma$ (e.g., \citealt{Fedele2012, Fedele2013, Meeus2012, Tilling2012, Dent2013}). 
Although the profiles of these lines are spectrally unresolved, a comparison of line strength with models indicates that the emitting region of these observations originates in the hot surface layer of the outer disk ($r$$>$$15$ au, e.g., \citealt{Fedele2012, Fedele2013}).
The total integrated line flux of this $\mathrm{H_2O}$ 63.32$\mu$m line from the disk around HD163296 (at a distance $d$ of $\sim$ 122 pc and inclination $i$ of 44 deg) is observed to be $(2.0 \pm 0.6)\times 10^{-17}$ W $\mathrm{m}^{-2}$, and the values of the other lines (e.g., ortho-$\mathrm{H_2O}$  56.82$\mu$m and 71.95$\mu$m lines) are roughly similar (e.g., \citealt{Fedele2012, Meeus2012}).
\citet{Meeus2012} and \citet{Fedele2013} determined that the upper limits of the total fluxes of such $\mathrm{H_2O}$ emission lines, including the $\mathrm{H_2O}$ 63.32$\mu$m from other Herbig Ae disks ($d\sim 100-150$ pc), are between a few $10^{-18}$ and a few $10^{-17}$ W $\mathrm{m}^{-2}$.
These values are roughly several tens of times smaller than the value which we calculate in this paper for this particular Herbig Ae disk model (see also Table \ref{tab:T1}, $d=$140 pc) if we consider the difference in the distances from the solar system. We note that our model disk is not intended to be representative of any particular source.
We discuss this issue further in Section 4.3.
\\ \\
\noindent In the top right panel of Figure \ref{Figure7_HerbigAe} we show the line profile for the $\mathrm{H_2O}$ line at 538.29$\mu$m. The contribution from the outer disk ($r=$14-30 au) is large compared to that from the optically thick region near the midplane of the inner disk ($r <$14 au) and the width between the double peaks of the line profile is around two times narrower than those of candidate $\mathrm{H_2O}$ lines which trace the emission from the hot water vapor within the $\mathrm{H_2O}$ snowline, although the $A_{ul}$ is not so high (=3.497$\times10^{-3}$s$^{-1}$). This is because this $\mathrm{H_2O}$ 538.29$\mu$m line is the ground-state rotational transition and has a low $E_{up}$ (=61.0K) compared with the other lines discussed in this paper.
The flux of this line comes mainly from the outer cold water vapor in the photodesorbed layer.
\\ \\
According to Figures \ref{Figure8_HerbigAe} and \ref{Figure13_HerbigAe}, the value of the emissivity of the $\mathrm{H_2O}$ 538.29$\mu$m at each $(r,z)$ in the photodesorbed layer is comparable inside and outside the $\mathrm{H_2O}$ snowline.
However, because of the larger surface area of the outer disk most disk-integrated emission from this line arises from the outer disk.
In addition, in the outer disk midplane opacity of this line (see Figure \ref{Figure9_HerbigAe}) is around $10^{3}-10^{4}$ times higher than those of the ortho-$\mathrm{H_2O}$ 482.99$\mu$m and 682.66$\mu$m lines, although the wavelength and thus the dust opacity are similar. 
This is because this line has a small value of $E_{up}$ and the level population of $\mathrm{H_2O}$ for this line is very high near the midplane of cold outer disk.
On the basis of these properties, we propose that this line is not optimal to detect emission from the hot water vapor within the $\mathrm{H_2O}$ snowline in the Herbig Ae disk case, as also concluded for the T Tauri disk \citep{Notsu2016}.
\\ \\
We mention that previous space high-dispersion spectroscopic observations with $Herschel$/HIFI detected this line from disks around one Herbig Ae star HD100546 and two T Tauri stars TW Hya and DG Tau (e.g., \citealt{Hogerheijde2011, Podio2013, vanDishoeck2014, Salinas2016}). The number of detections is small because the line flux is low compared with the sensitivity of that instrument \citep{Antonellini2015}.
The detected line profile and other line modeling work (e.g., \citealt{Meijerink2008, Woitke2009b, Antonellini2015}) suggested that the emitting region arises in the cold outer disk, consistent with the results of our model calculations.
\\ \\
Here we note that since we define OPR $=$3 (=the value in the high temperature region) throughout the disk, we may be overestimating the line flux of the ortho-$\mathrm{H_2O}$ 538.29$\mu$m line (for more details, see Section 2.3 of paper I, \citealt{Notsu2016}).
In addition, since the flux of this line is controlled by the outer cold $\mathrm{H_2O}$ gas which is desorbed from the cold dust-grain surfaces, it also may be necessary to include grain-surface reactions (e.g., \citealt{Hasegawa1992}) 
for accurate determination of the gas phase $\mathrm{H_2O}$ abundance in this region.
\\ \\
The middle two panels of Figure \ref{Figure7_HerbigAe} show the profiles for the pure rotational mid-infrared ortho-$\mathrm{H_2O}$ lines at $\lambda$=12.40$\mu$m (middle left), 12.45$\mu$m (middle right) from the Herbig Ae disk. 
Figures \ref{Figure8_HerbigAe}, \ref{Figure9_HerbigAe}, and \ref{Figure13_HerbigAe} also show the line-of-sight emissivity, the total optical depth (gas emission and dust), and the vertical normalized cumulative emissivity distributions of these two mid-infrared $\mathrm{H_2O}$ lines from the Herbig Ae disk, respectively.
Both lines have much larger values of $A_{ul}$ ($>$$1$ s$^{-1}$) and $E_{up}$ ($>$4000K) than those of the candidate mid-infared $\mathrm{H_2O}$ lines that trace emission from the $\mathrm{H_2O}$ vapor within the $\mathrm{H_2O}$ snowline in the disk midplane (see Table \ref{tab:T1} and \ref{tab:bT1}.1), and thus they mainly trace emission from the hot surface of the inner and outer disk.
The velocity profiles of these two lines were obtained by previous ground-based mid-infrared spectroscopic observations using VLT/VISIR \citep{Pontoppidan2010b} from bright T Tauri disks (AS 205N and RNO 90). They show the Keplerian double-peaked or flat-topped (for low inclination objects) profiles, and the line emitting region is the hot disk surface.
\\ \\
The bottom two panels of Figure \ref{Figure7_HerbigAe} show the profiles of pure rotational near-infrared ortho-$\mathrm{H_2O}$ lines at $\lambda$=4.96$\mu$m (bottom left), 4.43$\mu$m (bottom right) from the Herbig Ae disk. 
Figures \ref{Figure8_HerbigAe}, \ref{Figure9_HerbigAe}, and \ref{Figure13_HerbigAe} also show the line-of-sight emissivity, the total optical depth (gas emission and dust), and the vertical normalized cumulative emissivity distributions of these near-infrared lines for the Herbig Ae disk, respectively.
Both lines have the same much larger values of $E_{up}$ ($=$4180.4K), the former line has a larger value of $A_{ul}$ ($=$3.260 s$^{-1}$) and the latter has a smaller value of $A_{ul}$ ($=2.080\times 10^{-4}$ s$^{-1}$).
For the former case, since it has much larger values of $A_{ul}$ and $E_{up}$ than those of the candidate $\mathrm{H_2O}$ lines (see Table \ref{tab:T1} and \ref{tab:bT1}.1), it mainly traces the emission from the hot surface of the inner and outer disk.
This line has similar values of $A_{ul}$ and $E_{up}$ with the observed near-infrared rovibrational $\mathrm{H_2O}$ lines in L band \citep{Mandell2012}.
For the latter smaller $A_{ul}$ line case, since the value of $E_{up}$ in this near-infrared line is much larger ($>$4000K) than those of the candidate $\mathrm{H_2O}$ lines which trace the emission from the hot water vapor within the $\mathrm{H_2O}$ snowline from mid-infrared to sub-millimeter wavelengths, it only traces the very innermost region ($r<$ 3 au for the 4.43$\mu$m line case).
In addition, the widths between the two peaks of the Keplerian profiles of these near- and mid-infrared lines with large $E_{up}$ are larger than those of candidate $\mathrm{H_2O}$ lines which trace the emission from the hot water within the $\mathrm{H_2O}$ snowline (see Figures \ref{Figure4_HerbigAe}, \ref{Figure7_HerbigAe}, and \ref{Figure14_HerbigAe}). These are because they trace the innermost hot region compared with the region around the $\mathrm{H_2O}$ snowline.
\\ \\
Here we note that previous near- and mid-infrared spectroscopic observations using VLT/CRIRES and $Spitzer$/IRS for Herbig Ae disks (e.g., \citealt{Pontoppidan2010a, Fedele2011, Salyk2011}) have not detected any $\mathrm{H_2O}$ lines regardless of the value of $A_{ul}$. We discuss this issue in Section 4.3.
Moreover, as we described in Section 2.3, the level populations of the water molecule are calculated under LTE, as opposed to non-LTE. However, as we discuss further in Section 4.2, in our LTE calculations there is a possibility that we have overestimated the emission fluxes of strong $\mathrm{H_2O}$ lines with large $A_{ul}$ which trace the hot surface layer, as found in previous studies (e.g., \citealt{Meijerink2009, Woitke2009b, Banzatti2012, Antonellini2015}).
\\
\subsubsection{The total fluxes of candidate $\mathrm{H_2O}$ emission lines which trace emission from the hot water vapor within $\mathrm{H_2O}$ snowline}
\begin{figure}[htbp]
\begin{center}
\includegraphics[scale=0.6]{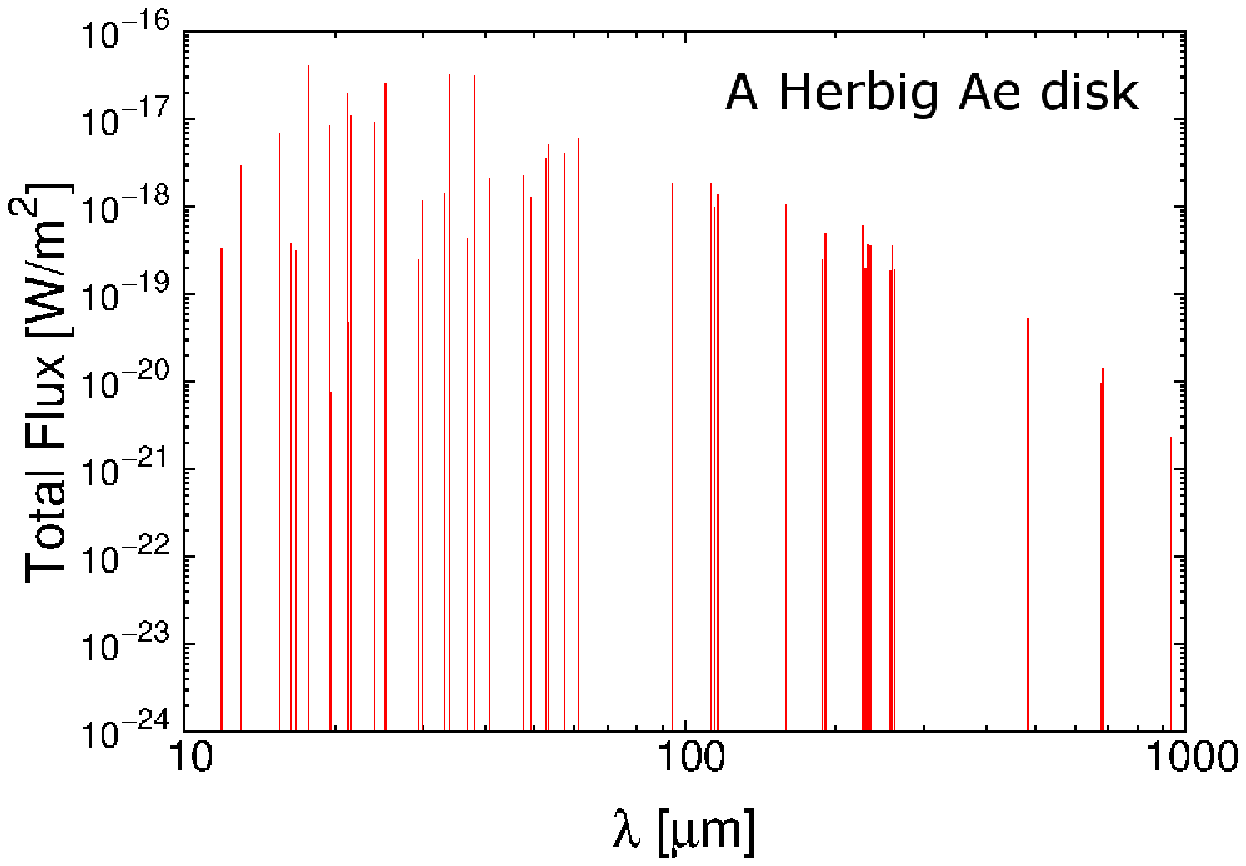}
\includegraphics[scale=0.6]{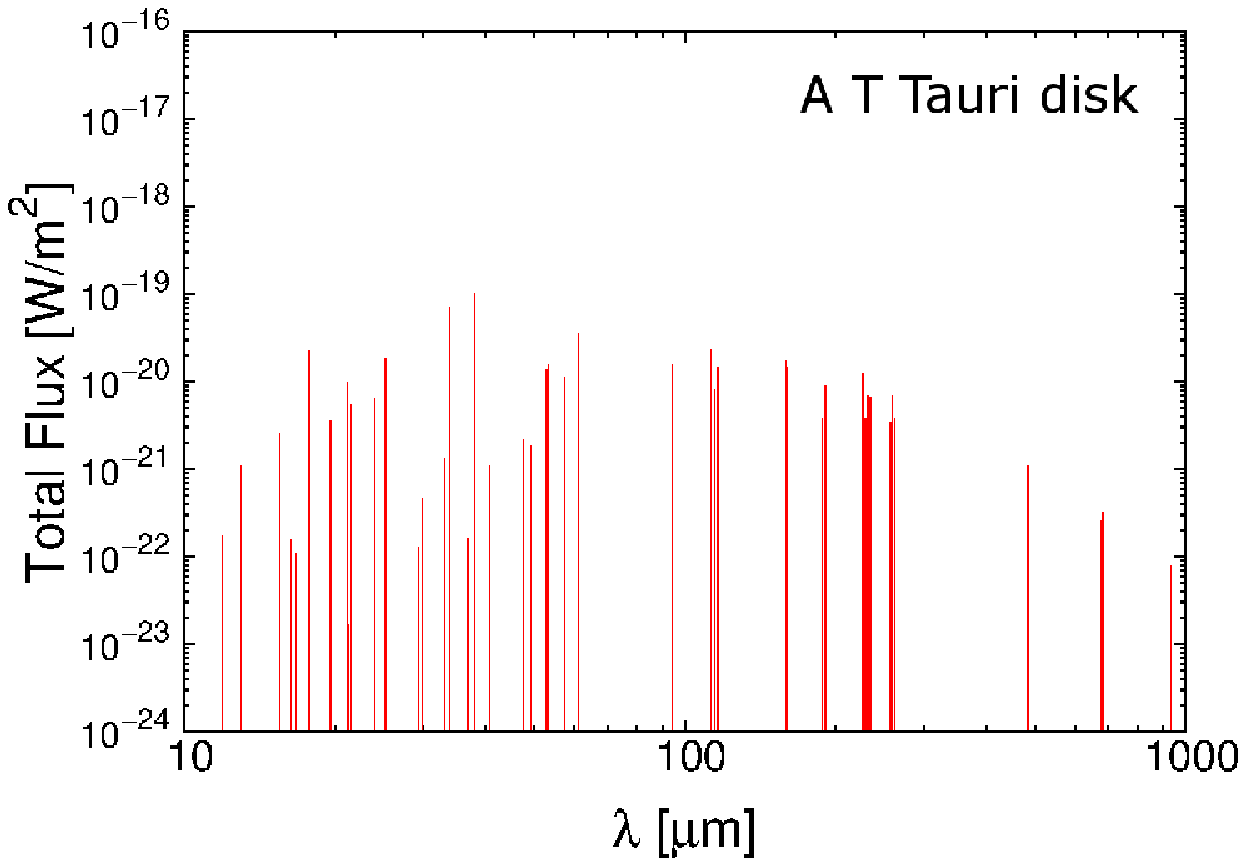}
\end{center}
\caption{\noindent The total fluxes of the ortho-$\mathrm{H_2O}$ lines which are best candidates to trace the emission from the water vapor within the $\mathrm{H_2O}$ snowline, 
for a Herbig Ae disk (top panel) and a T Tauri disk (bottom panel).
We select these lines based on their small Einstein $A$ coefficients of $10^{-6} < A_{ul} < 10^{-2}$ s$^{-1}$ and relatively large excitation energies of $700 < E_{up} < 2100$K. The wavelengths of these lines range from mid-infrared to sub-millimeter, $\lambda \sim 11-1000 $$\mu$m. 
}\label{Figure10_HerbigAe}
\end{figure}

\noindent
Figure \ref{Figure10_HerbigAe} shows the total fluxes of the various ortho-$\mathrm{H_2O}$ lines which are the candidates for tracing emission from hot water vapor within the $\mathrm{H_2O}$ snowline
for a Herbig Ae disk (top panel) and a T Tauri disk (bottom panel).
We select those lines from the LAMDA database (see Section 2.3 of paper I, \citealt{Notsu2016}) which have both small values of $A_{ul}$ ($10^{-6} < A_{ul} < 10^{-2}$ s$^{-1}$) and relatively large values of $E_{up}$ ($700 < E_{up} < 2100$K). 
The wavelengths of these lines range from mid-infrared to sub-millimeter, $\lambda \sim 11-1000 $$\mu$m, because we do not have candidate lines which trace emission from the hot water vapor within $\mathrm{H_2O}$ snowline with wavelengths $\lambda < 10\mu$m on the basis of our criteria for $A_{ul}$ and $E_{up}$.
The values of $E_{up}$ of lines for wavelengths $\lambda <$ 10$\mu$m are too large ($\gtrsim$3000K), and the opacity of the dust grains for wavelengths $\lambda < 10\mu$m is expected to be too large to trace the emission from the midplane of the disk (see Sections 3.2.2 and 4.3).
The detailed parameters, such as transitions ($J_{K_{a}K_{c}}$), wavelength, frequency, $A_{ul}$, $E_{up}$, and total line fluxes of these candidate ortho-$\mathrm{H_2O}$ lines shown in Figure \ref{Figure10_HerbigAe} are listed in Table \ref{tab:bT1}.1 of Appendix B. 
In Figure \ref{Figure10_HerbigAe} and Table \ref{tab:bT1}.1, we show both values for the total fluxes from the Herbig Ae disk and the T Tauri disk.
In addition, Figure \ref{Figure14_HerbigAe} in Appendix C shows the profiles of mid-infrared candidate lines ($\lambda \sim 11-25 $$\mu$m) for the Herbig Ae disk. All lines in this Figure are listed in Table \ref{tab:bT1}.1.
\\ \\
On the basis of Figure \ref{Figure10_HerbigAe}, Figure \ref{Figure14_HerbigAe}, Table \ref{tab:T1}, and Table \ref{tab:bT1}.1, the values of fluxes of these  lines from the Herbig Ae disk are around $10^{2}-10^{3}$ larger than those of the T Tauri disk.
This is because the position of the $\mathrm{H_2O}$ snowline in the Herbig Ae disk exists at a larger radius from the central star than that in the T Tauri disk.
In addition, the peak fluxes of these lines become larger as the values of A$_{ul}$ become larger and E$_{up}$ become smaller.
Moreover, the values of total flux tend to be larger as the wavelengths of these $\mathrm{H_2O}$ lines become shorter.
This is because the peak wavelengths of the Planck function at the gas temperatures around the $\mathrm{H_2O}$ snowline ($T_{g}$$\sim$ 100-200 K) are at mid-infrared wavelengths. 
In the cases of the Herbig Ae disk and the T Tauri disk, the values of total fluxes of the mid-infrared $\mathrm{H_2O}$ lines which trace emission from the hot water vapor within $\mathrm{H_2O}$ snowline are $\sim 10^{2}-10^{4}$ and $\sim 10^{1}-10^{2}$ times larger than those of sub-millimeter $\mathrm{H_2O}$ lines, respectively, and there are differences in the line flux ratio of mid-infrared lines to sub-millimeter lines between the Herbig Ae and the T Tauri disks. 
These are because the amount of hot $\mathrm{H_2O}$ gas around the region at $\tau_{ul} \lesssim 1$ in mid-infrared lines and inside the $\mathrm{H_2O}$ snowline are higher in the Herbig Ae disk model than that in the T Tauri disk.
\\ \\
On the basis of Figure \ref{Figure14_HerbigAe}, most of the emission flux from these mid-infrared lines comes from the region with a high $\mathrm{H_2O}$ gas abundance ($\sim$$10^{-4}$, $r<$8 au), and the position of the two peaks and the rapid drop in flux density between the peaks contains information on the position of the outer edge of this region. This is because they have shorter wavelengths ($\lambda \sim 11-25 $$\mu$m) and relatively larger values of $E_{up}$ ($\sim$$1500-2000$K except for the $\mathrm{H_2O}$ 17.75 and 24.00 $\mu$m lines) among all candidate lines which trace the emission from the hot $\mathrm{H_2O}$ vapor within the $\mathrm{H_2O}$ snowline (see Table \ref{tab:bT1}.1).
\\
\subsubsection{The radial distributions of normalized cumulative line fluxes}
\begin{figure}[htbp]
\begin{center}
\includegraphics[scale=0.6]{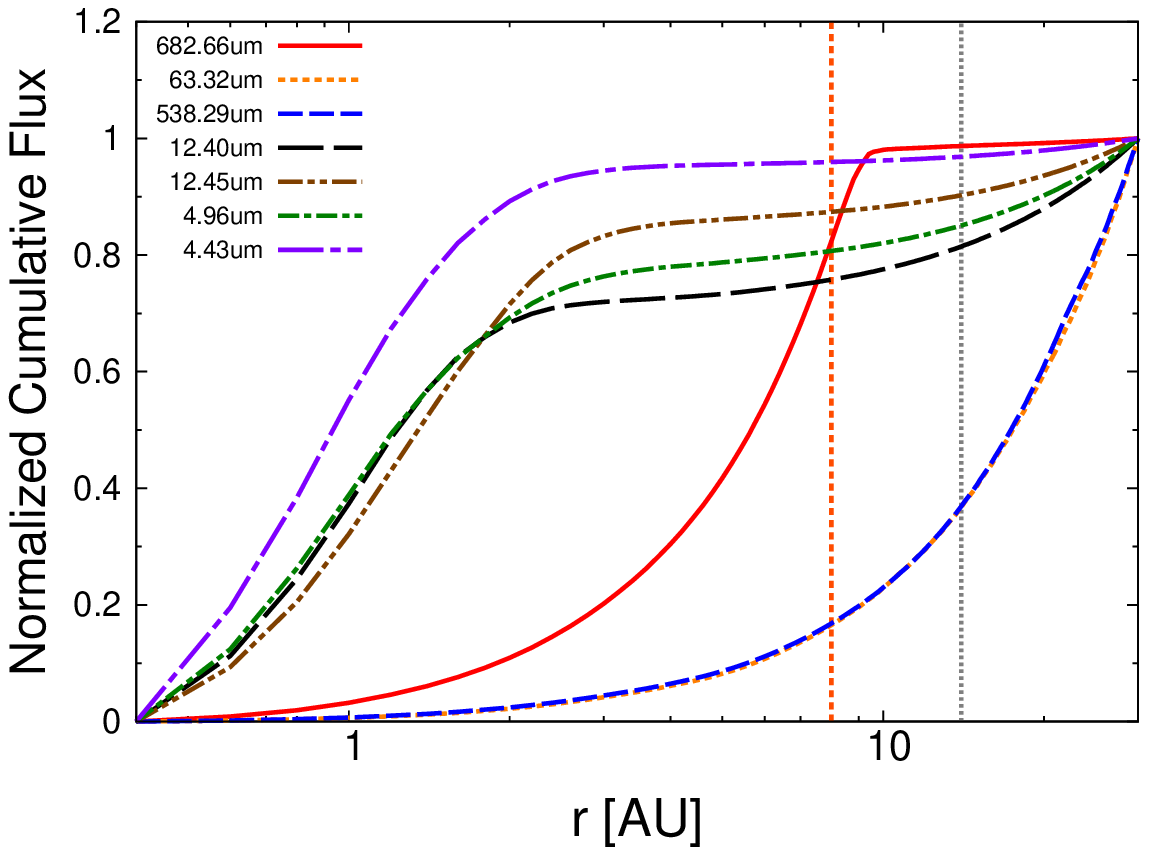}
\vspace{0.7cm}
\includegraphics[scale=0.6]{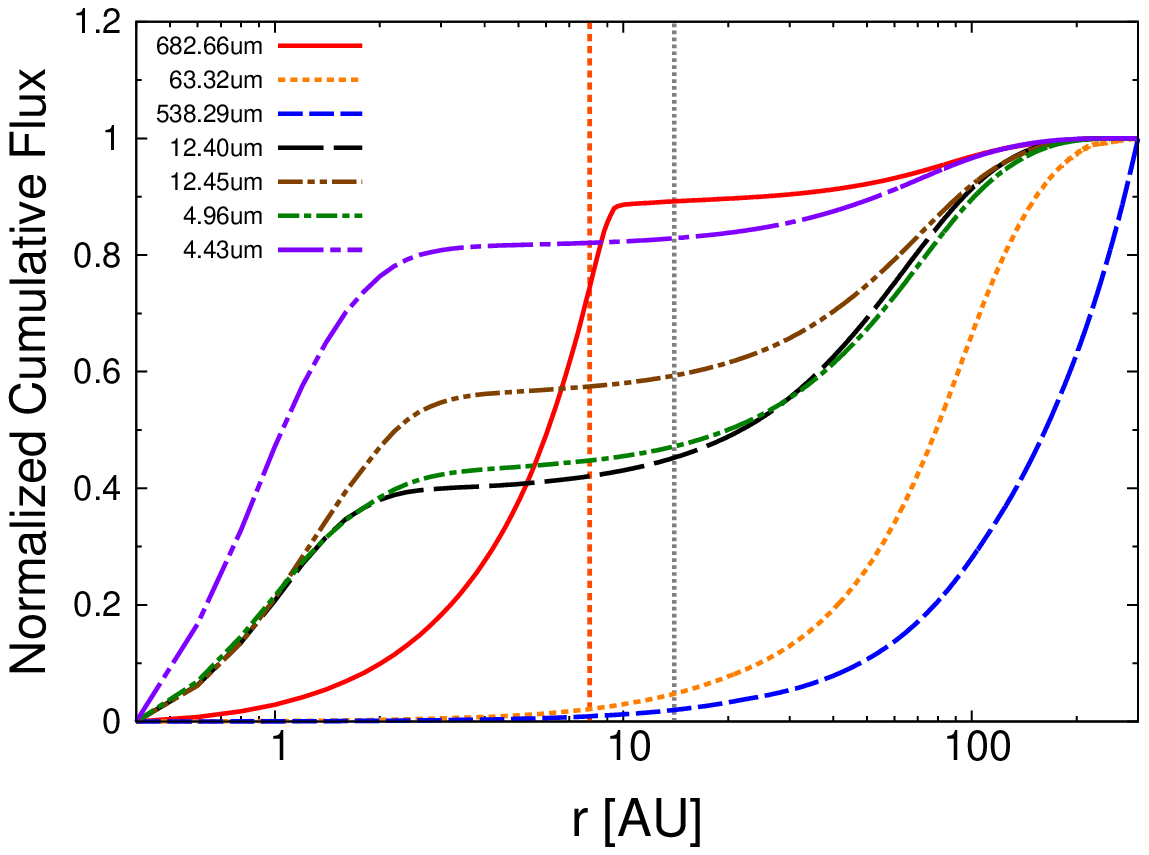}
\end{center}
\vspace{0.4cm}
\caption{\noindent The radial distributions of the normalized cumulative flux for seven pure rotational ortho-$\mathrm{H_2O}$ lines at $\lambda$=682.66$\mu$m ({\it red solid line}), 63.32$\mu$m ({\it orange dotted line}), 538.29$\mu$m ({\it blue dashed line}), 12.40$\mu$m ({\it black long dashed line}), 12.45$\mu$m ({\it brown dashed two dotted line}), 4.96$\mu$m ({\it green dashed dotted line}), and 4.43$\mu$m ({\it violet long dashed dotted line}).
The vertical straight lines display the positions of $r$=8 au ({\it orange dotted line}) and 14 au ({\it grey dotted line}), respectively.
We normalized the cumulative flux of each line using the values at $r=30$ au (top panel) and at $r=300$ au (bottom panel).
We assume that the inclination angle of the disk $i$ is 0 degree in making these figures.}\label{Figure11_HerbigAe}
\end{figure}

\noindent Figure \ref{Figure11_HerbigAe} shows the normalized radial cumulative fluxes for seven $\mathrm{H_2O}$ lines at $\lambda$=682.66$\mu$m, 63.32$\mu$m, 538.29$\mu$m, 12.40$\mu$m, 12.45$\mu$m, 4.96$\mu$m, and 4.43$\mu$m.
We discussed the properties of these seven lines in Sections 3.2.1 and 3.2.2.
According to these figures, around 90$\%$ of the flux of the 682.66$\mu$m line is emitted from the region inside the $\mathrm{H_2O}$ snowline ($r<$14 au). In contrast, emission from the 63.32$\mu$m and the 538.29$\mu$m lines is emitted mostly from the region outside the $\mathrm{H_2O}$ snowline. 
In addition, although the 63.32$\mu$m line is mainly emitted from the region between $r\sim25-200$ au, the 538.29$\mu$m line is mainly emitted from a region much further out ($r\sim 50-300$ au).
The properties of these three lines for the Herbig Ae disk are similar to those for the T Tauri disk which we showed and discussed in Figure 7 and Section 3.2.5 in paper I \citep{Notsu2016}.
\\ \\
The 12.40$\mu$m, 12.45$\mu$m, and 4.96$\mu$m lines are mainly emitted both from the regions within 3 au and outside the $\mathrm{H_2O}$ snowline ($r>$14 au).
The emission from the region between 3-14au ($=$ the $\mathrm{H_2O}$ snowline) is much smaller.
The 4.43$\mu$m line is mainly emitted from the innermost region of the disk ($r<3$ au).
These are because these four lines have much larger values of $E_{up}$ ($>$4000K) than both the candidate $\mathrm{H_2O}$ lines that trace the hot water vapor within the $\mathrm{H_2O}$ snowline and the 63.32$\mu$m line. Thus, the values of flux densities from the hot surface layer of the inner disk are larger.
The 4.43$\mu$m line has a smaller value of $A_{ul}$ ($=2.080\times 10^{-4}$ s$^{-1}$) and thus the values of flux densities from the hot surface layer of the outer disk are much smaller, but the other three lines 
have larger values of $A_{ul}$ ($>$1 s$^{-1}$) and thus the values of flux densities from the hot surface layer of the outer disk are larger. 
\begin{figure}[htbp]
\begin{center}
\includegraphics[scale=0.6]{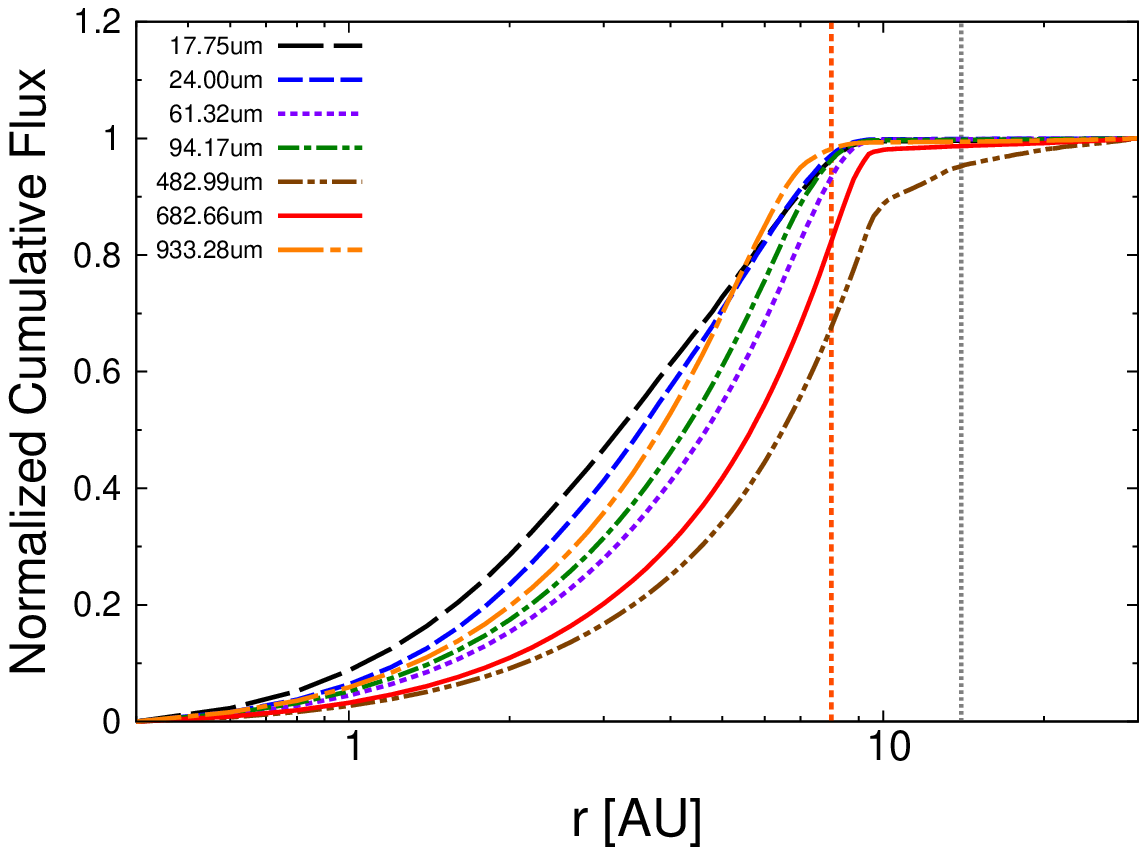}
\vspace{0.7cm}
\includegraphics[scale=0.6]{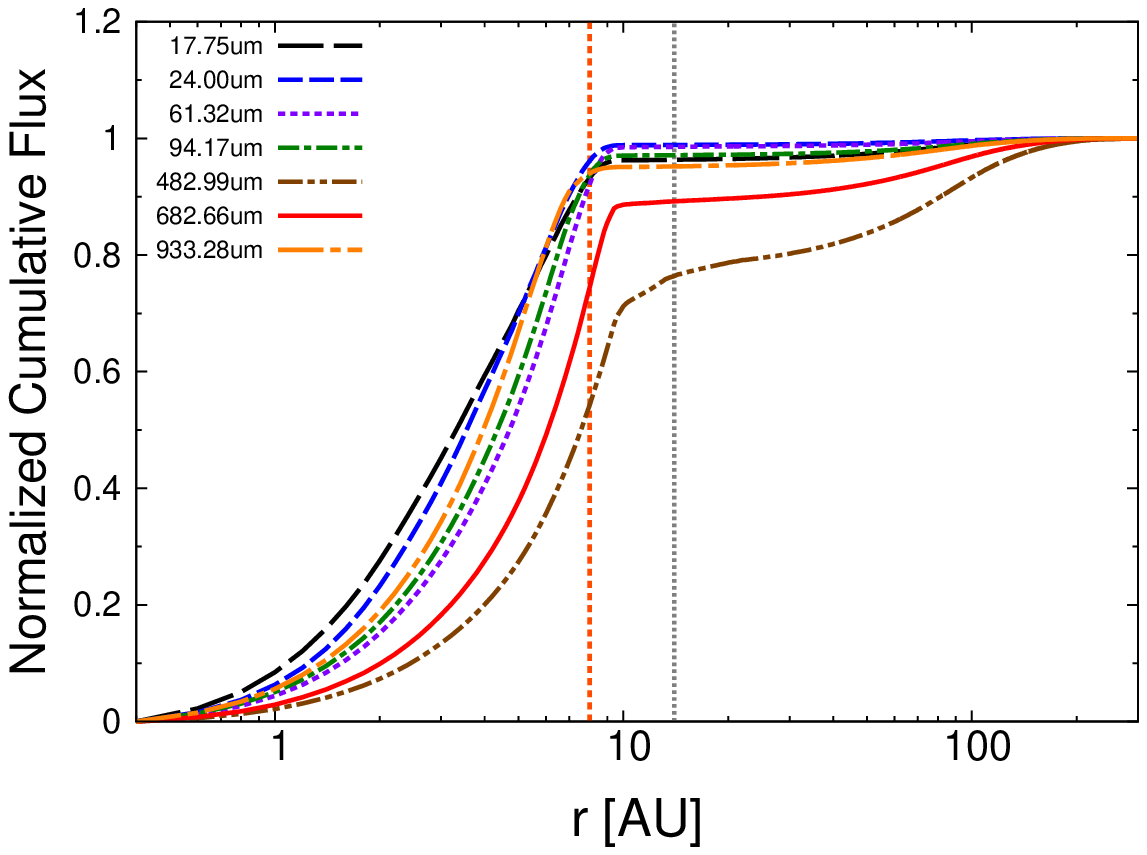}
\end{center}
\vspace{0.4cm}
\caption{\noindent The radial distributions of the normalized cumulative flux for seven pure rotational ortho-$\mathrm{H_2O}$ lines at $\lambda$=17.75$\mu$m ({\it black long dashed line}), 24.00$\mu$m ({\it blue dashed line}),  61.32$\mu$m ({\it violet dotted line}), 94.17$\mu$m ({\it green dashed dotted line}), 482.99$\mu$m ({\it brown dashed two dotted line}), 682.66$\mu$m ({\it red solid line}), and 933.28$\mu$m ({\it orange long dashed dotted line}). 
These are candidate ortho-$\mathrm{H_2O}$ lines to trace the emission from the hot water vapor within the $\mathrm{H_2O}$ snowline.
The vertical straight lines display the positions of $r$=8 au ({\it orange dotted line}) and 14 au ({\it grey dotted line}), respectively.
We normalized the cumulative flux of each line using the values at $r=30$ au (top panel) and at $r=300$ au (bottom panel).
We assume that the inclination angle of the disk $i$ is 0 degree in making these figures.}\label{Figure12_HerbigAe}
\end{figure}
\\ \\
Figure \ref{Figure12_HerbigAe} shows the normalized radial cumulative fluxes for seven rotational ortho-$\mathrm{H_2O}$ lines that trace the hot water vapor within the $\mathrm{H_2O}$ snowline: $\lambda$=17.75$\mu$m, 24.00$\mu$m,  61.32$\mu$m, 94.17$\mu$m, 482.99$\mu$m, 682.66$\mu$m, and 933.28$\mu$m.
We discussed the properties of these seven lines in Sections 3.2.1 and 3.2.3.
In the cases of the $\mathrm{H_2O}$ 482.99$\mu$m and 682.66$\mu$m lines, most of the flux is emitted from the region with a high $\mathrm{H_2O}$ gas abundance ($\sim$$10^{-4}$, $r<$8 au), with some emission flux emitted from the region with a relatively high $\mathrm{H_2O}$ gas abundance ($\sim$$10^{-8}$, $r=$8-14 au). 
On the other hand, for the other $\mathrm{H_2O}$ lines, almost all of the emission flux comes from the region with a high $\mathrm{H_2O}$ gas abundance ($\sim$$10^{-4}$, $r<$8 au).
Such differences in line properties come from the differences in $E_{up}$ and the dust opacity at different wavelengths. The $\mathrm{H_2O}$ 482.99$\mu$m and 682.66$\mu$m lines have relatively smaller values of $E_{up}$ ($<$1100K) and longer wavelengths ($>$400$\mu$m), as discussed in Section 3.2.1.
In Figure \ref{Figure12_HerbigAe}, lines which have larger $E_{up}$ and shorter wavelengths tend to be emitted from the innermost region within the $\mathrm{H_2O}$ snowline.
 \\
\section{Discussion}
\subsection{Influence of model assumptions on the properties of $\mathrm{H_2O}$ emission lines}
\noindent In Section 3.2.4 of paper I \citep{Notsu2016}, we discussed the uncertainties in our model predictions in detail, and we also discussed the behavior of the $\mathrm{H_2O}$ lines for some cases in which we artificially changed the distribution of $\mathrm{H_2O}$ vapor, the position of the $\mathrm{H_2O}$ snowline, and the fractional abundance of $\mathrm{H_2O}$ gas in the outer disk surface layer (see Figures 8 and 9 of paper I, \citealt{Notsu2016}).
We explored different values of the $\mathrm{H_2O}$ snowline radius to simulate the effects of (i) increased and decreased viscous heating (e.g., \citealt{Oka2011, Harsono2015}), (ii) using different dust opacities due to dust-grain growth (e.g., \citealt{Aikawa2006, Oka2011}).
We varied the abundance of water vapor in the disk atmosphere to simulate the effects of (iii) increasing/decreasing the strength of UV photoreactions.
We found that in the cases for which the $\mathrm{H_2O}$ snowline is more distant from the central star, and the fractional abundance of $\mathrm{H_2O}$ gas in the disk atmosphere is lower than that in the originally adopted T Tauri disk, the line fluxes from the hot inner disk midplane inside the $\mathrm{H_2O}$ snowline are more dominant from those from the outer disk (e.g., \citealt{Walsh2012}).
\\ \\
An analysis of spectral energy distributions (SEDs) classified Herbig Ae/Be stars into two groups (group I/II, \citealt{Meeus2001}). Group I sources show both power-law and blackbody components up to far-infrared wavelengths in their SEDs. SEDs of group II sources can be well modeled with only a single power law from mid- to far-infrared wavelengths. 
\citet{Meeus2001} suggested that group I sources have a flaring disk while the group II disks are geometrically flat.
\citet{Meeus2001} proposed a possible evolutionary scenario where a group I flaring disk evolves into a group II flat disk through grain growth and settling of grains onto the disk midplane (see also \citealt{Dominik2003, Dullemond2004}). 
\\ \\
However, recent high-spatial resolution observations at various wavelengths have revealed a more complex structure in disks, with inner holes and/or gaps toward many group I sources such as HD100546 (e.g., \citealt{Benisty2010, Panic2014, Walsh2014b}), HD142527 (e.g., \citealt{Fujiwara2006, Fukagawa2006, Fukagawa2013}), and HD169142 (e.g., \citealt{Benisty2010, Honda2012}).
On the other hand, there is little evidence for inner holes and/or gaps reported toward group II disks, and they seem to have a radially continuous structure (e.g., \citealt{Honda2015}).
\citet{Honda2012} and \citet{Maaskant2013} suggested that most group I sources can be classified as (pre-)transitional disks.
Transitional and pre-transitional disks
are protoplanetary disks with an inner hole and/or gaps indicated by a lack of near-infrared/mid-infrared
excess in their SEDs (e.g., \citealt{Strom1989, Espaillat2007}).
They pointed out that there is no significant difference in age between groups I and II sources \citep{Meeus2001, Honda2015}.
Therefore, recent studies (e.g., \citealt{Maaskant2013, Honda2015}) suggested that both group I and II sources had experienced different evolutionary paths from some common primordial and continuous flaring disks.
Here we note that \citet{Menu2015}, \citet{Isella2016}, and \citet{Zhang2016} suggested that some geometrically flat disks (group II disk) have gaps.
\citet{Menu2015} suggested flat disks with gaps are most likely descendants of flat disks without gaps.
\citet{Banzatti2016} discovered a correlation between water line fluxes at $\lambda =$2.9$\mu$m and $10-33\mu$m and the size of inner disk gaps/holes as probed by 4.7$\mu$m CO rovibrational emission lines. They described that the lower detection frequency of near- and mid-infrared water vapor lines in disks around intermediate mass stars ($M_{\mathrm{*}}$=1.5-4$M_{\bigodot}$) is linked to inner gaps/holes with depleted molecular gas content of the disks out to close to or beyond the $\mathrm{H_2O}$ snowline.
\\ \\
Our Herbig Ae disk model adopted in this paper has a radially continuous structure with no inner hole and/or gap.
If we adopt a disk model with an inner hole and/or gap, the emission fluxes of $\mathrm{H_2O}$ lines, especially from the hot water vapor inside the $\mathrm{H_2O}$ snowline is expected to decrease.
Recently ALMA observations with high spatial resolution have been conducted towards protoplanetary disks with various central star masses/ages (e.g., \citealt{ALMA2015, Andrews2016, Isella2016, Tsukagoshi2016}), and will help understand these evolutionary scenarios.
 \\
\subsection{Critical density and the assumption of LTE}
\noindent
In this subsection, we discuss the validity of the assumption of LTE.
As we discussed in paper I \citep{Notsu2016}, the assumption of LTE is valid in calculating the emission flux of the 682.66$\mu$m line. 
This is because this line mainly comes from the hot region around $z/r \sim 0.1$ within the $\mathrm{H_2O}$ snowline where
the total gas density ($\sim 10^{11}-10^{14}$ cm$^{-3}$) is much larger than the critical density for this line (n$_{\mathrm{cr}}$$=1.0\times 10^{6}$ cm$^{-3}$, see also Table \ref{tab:T1}).
On the other hand, in our LTE calculations it remains possible that we have overestimated the emission flux of strong $\mathrm{H_2O}$ lines with large $A_{ul}$ ($\sim$ $10^{-1}-10^{0}$ s$^{-1}$) which trace the hot surface layer of the inner/outer disk (e.g., the 12.40 $\mu$m and the 63.32 $\mu$m lines) and lines which trace the cold water vapor in the photodesorbed layer (e.g., the 538.29 $\mu$m line). The values of n$_{\mathrm{cr}}$ of these lines (e.g., n$_{\mathrm{cr}}$$=1.1\times 10^{11}$, $1.5\times 10^{10}$, and $2.9\times 10^{7}$ cm$^{-3}$ for the 12.40 $\mu$m, the 63.32 $\mu$m, and the 538.29 $\mu$m lines, respectively) are larger than or similar to the total gas density in the hot surface layer of the inner/outer disk ($\sim 10^{7}-10^{8}$ cm$^{-3}$) and the photodesorbed layer ($\sim 10^{8}-10^{10}$ cm$^{-3}$).
Previous studies which model such $\mathrm{H_2O}$ lines (e.g., \citealt{Meijerink2009, Woitke2009b, Banzatti2012, Antonellini2015, Antonellini2016}) showed that non-LTE calculations are important for the latter lines. 
They suggested that non-LTE effects may, however, alter line fluxes by factors of only a few for moderate excitation lines ($E_{up} <$ a few thousand K). 
Moreover, current non-LTE calculations are likely to remain inaccurate, due to the incompleteness and uncertainty of collisional rates (e.g., \citealt{Meijerink2009, Banzatti2012, Kamp2013, Zhang2013, Antonellini2015}). 
\\ \\
We calculated the critical density $n_{\mathrm{cr}}$ of the six other lines discussed in Section 3.2.1 and Figures \ref{Figure4_HerbigAe}, \ref{Figure5_HerbigAe}, and \ref{Figure6_HerbigAe} as probes interior to the $\mathrm{H_2O}$ snowline (the 17.75, 24.00, 61.32, 94.17, 482.99, and 933.28 $\mu$m lines, see Table \ref{tab:T1}).
$n_{\mathrm{cr}}$ for the sub-millimeter lines at 482.99, 682.66, and 933.28 $\mu$m ($3.3\times 10^{6}$, $1.0\times 10^{6}$, and $4.7\times 10^{6}$ cm$^{-3}$, respectively) are lower than the values of the total gas density in the hot surface layer of the outer disk ($\sim10^{7}-10^{8}$ $\mathrm{cm}^{-3}$) and in the photodesorbed water layer ($\sim10^{8}-10^{10}$ $\mathrm{cm}^{-3}$).
In contrast, $n_{\mathrm{cr}}$ for the mid- to far-infrared lines ($8.3\times 10^{10}$, $1.9\times 10^{9}$, $4.1\times 10^{8}$, and $3.1\times 10^{8}$ cm$^{-3}$ for the 17.75, 24.00, 61.32, and 94.17 $\mu$m lines, respectively) are similar to and larger than the values of the total gas density in those regions.
This is because the latter infrared lines have larger Einstein $A$ coefficients and shorter wavelengths compared with the aforementioned sub-millimeter lines.
If the wavelength of a line is shorter, the energy difference between the upper and lower state is larger and the value of collisional rates $<\sigma v>$ tends to be smaller \citep{Faure2008}.
However, emission from these lines mainly comes from the hot water vapor within the $\mathrm{H_2O}$ snowline where total gas density is much larger ($\sim 10^{11}-10^{14}$ cm$^{-3}$) than the values of $n_{\mathrm{cr}}$, and thus it is valid to use LTE to calculate their fluxes.
 \\
\subsection{Previous $\mathrm{H_2O}$ line observations in Herbig Ae disks}
\noindent Since $\mathrm{H_2O}$ line emission from the disk midplane is likely obscured by dust grains at near- to mid-infrared wavelengths \citep{Walsh2015}, the ``visible" $\mathrm{H_2O}$ gas column density at these wavelengths is smaller than the total $\mathrm{H_2O}$ column density integrated over the disk in the vertical direction (see e.g., Figure \ref{Figure3_HerbigAe}). For example, in \citet{Walsh2015}, the visible column density at 14$\mu$m in the Herbig Ae disk case is $\sim 10^{19}-10^{20}$ $\mathrm{cm}^{-2}$ within the $\mathrm{H_2O}$ snowline. 
In the bottom panel of Figure \ref{Figure3_HerbigAe}, the visible $\mathrm{H_2O}$ gas column densities at 17.75, 61.32, and 682.66$\mu$m in the Herbig Ae disk are $\sim 10^{18}-10^{19}$ $\mathrm{cm}^{-2}$ within the $\mathrm{H_2O}$ snowline, which are lower than those of \citet{Walsh2015}.
This is because the total integrated column density of water vapor in our model ($\sim 10^{20}-10^{22}$ $\mathrm{cm}^{-2}$) is lower than that of \citet{Walsh2015} ($\sim 10^{21}-10^{23}$ $\mathrm{cm}^{-2}$), and because absorption by dust grains is dominant in the disk midplane and the disk surface ($\tau_{ul} \lesssim 1$) compared to that by excited gas molecules, especially remarkable in the cases of infrared lines (see also Section 3.2.1).
Previous near- and mid-infrared spectroscopic observations using instruments on ground and space telescopes (e.g., VLT/CRIRES and $Spitzer$/IRS) for Herbig Ae disks \citep{Pontoppidan2010a, Fedele2011, Salyk2011} have not detected the $\mathrm{H_2O}$ lines, and they derive upper limits for $\mathrm{H_2O}$ gas column densities ($\lesssim 10^{18}$ $\mathrm{cm}^{-2}$). 
However, such near- and mid-infrared $\mathrm{H_2O}$ lines are observed in many T Tauri disks (e.g., \citealt{Pontoppidan2010a}, see also Section 1 of paper I, \citealt{Notsu2016}).
$\mathrm{H_2O}$ lines at far-infrared wavelengths have been detected with $Herschel$/PACS only in the disk around HD163296, although this emission originates in the hot surface layer of the outer disk ($r>$ 15 au), and it is farther out than that expected for emission at shorter wavelengths \citep{Fedele2012, Fedele2013, Meeus2012}.
From these observational results, there is an important question as to why the detection rate for near- and mid-infrared $\mathrm{H_2O}$ lines for T Tauri disks is higher than that for Herbig Ae disks.
\\ \\
Previous studies (e.g., \citealt{Woitke2009b, Pontoppidan2010a, Fedele2011, Meeus2012, Walsh2015, Antonellini2015, Antonellini2016}) discussed some answers to the above question. Here we highlight some important ideas (see also Section 3.2.2).
First, there may be additional destruction routes for gas-phase water in the inner disk atmosphere, not yet included in the chemical networks we have adopted, for example the reaction to produce OH via photodissociation of $\mathrm{H_2O}$ by Lyman-$\alpha$ photons \citep{Walsh2015}.
Second, dust-grain settling and dust-grain growth can reduce the total dust-grain surface area and possibly increase the UV irradiation rates in the upper disk (e.g., \citealt{Vasyunin2011, Akimkin2013}), which can push the molecular layer deeper into the disk atmosphere. Hence, a higher fraction of the gas-phase water may be hidden from view (e.g., \citealt{Walsh2015, Krijt2016}).
We note that HD100546, for which far-infrared $\mathrm{H_2O}$ lines have not been detected, has very high UV flux from the central star and even at 30 au the UV field is
expected to be too strong for gas-phase water to survive \citep{Meeus2012}. 
\citet{Tilling2012} modeled the disk of HD163296 and pointed out that the dust material is settled.
Meanwhile, dust-grain growth and the dust-grain shape also affect the UV field through scattering efficiency in the disk atmosphere. If the dust-grain radius is large enough compared with the wavelength of radiation from the central star, forward scattering by dust grains becomes efficient
and the UV field decreases in the disk atmosphere (e.g., \citealt{Bethell2011}). 
The gas temperature in the disk atmosphere, which affects the $\mathrm{H_2O}$ line fluxes, is also controlled by the UV radiation field. If the UV radiation field increases/decreases, the gas temperature in the disk atmosphere will become higher/lower. 
We have assumed isotropic dust scattering and the grain size distribution of the dark cloud model with compact and spherical dust grains (For more details, see \citealt{NomuraMillar2005} and paper I, \citealt{Notsu2016}), and these assumptions will affect the resulting $\mathrm{H_2O}$ line fluxes.
Third, if the disk is transitional and has a significant gap/hole in the inner disk (e.g., HD100546), the line fluxes from the inner disk atmosphere will be decreased (e.g., \citealt{Banzatti2016}, see also Section 4.1).
\\ \\
Fourth, in Herbig Ae disks the strong infrared excess of dust emission might veil the faint emission of molecular lines at infrared wavelengths (e.g., \citealt{Lahuis2007, Pontoppidan2010a, Fedele2011, Antonellini2015, Antonellini2016}).
Previous line modeling calculations such as \citet{Du2014} and \citet{Antonellini2015, Antonellini2016} concluded the infrared and sub-millimeter water line fluxes are affected by many parameters related to disk physical structure, such as dust-size distribution, dust-to-gas mass ratio, disk gas mass, maximum dust size, and luminosity of the central star. \citet{Antonellini2015, Antonellini2016} showed that the sensitivity and spectral resolution of previous mid-infrared observations (e.g., $Spitzer$/IRS) were not sufficient to detect the detailed profiles of even strong $\mathrm{H_2O}$ lines (with large $A_{ul}$) in many disks, especially disks around high-mass Herbig Ae/Be stars. This was because of the presence of noise in the spectra which can mask the line emission, combined with the high dust continuum flux (the noise level is proportional to the dust continuum flux).
 \\
\subsection{Requirement for the observations of candidate $\mathrm{H_2O}$ lines to trace the $\mathrm{H_2O}$ snowline}
\noindent Since the velocity width between the emission peaks is $\sim$20 km s$^{-1}$, high-dispersion spectroscopic observations (R=$\lambda$/$\delta \lambda$$>$ tens of thousands) of the $\mathrm{H_2O}$ lines in Table \ref{tab:bT1}.1 are needed to trace emission from the hot water vapor within the $\mathrm{H_2O}$ snowline.
Their profiles contain information which can be used to locate the position of the $\mathrm{H_2O}$ snowline.
The area of the emitting regions are small ($r$$<$ 2 au for a T Tauri disk and $r<$ 14 au for a Herbig Ae disk) compared with the total disk size.
The spectral resolution (of many instruments) and sensitivity used for previous mid-infrared, far-infrared, and sub-millimeter observations (e.g., $Spitzer$/IRS, $Herschel$/PACS, $Herschel$/HIFI) were not sufficient to detect and resolve the candidate lines we identified in Table \ref{tab:bT1}.1 which trace emission from the hot water vapor within $\mathrm{H_2O}$ snowline.  
\\ \\
Among the various $\mathrm{H_2O}$ lines in ALMA band 8, the $\mathrm{H_2O}$ 682.66$\mu$m line is the most suitable to trace emission from the hot water vapor within the $\mathrm{H_2O}$ snowline. Other candidate sub-millimeter $\mathrm{H_2O}$ lines to trace the $\mathrm{H_2O}$ snowline, having the same order-of-magnitude fluxes, exist in ALMA bands 7, 9 and 10 ($\sim$ $300-1000$ $\mu$m).
The $\mathrm{H_2O}$ 933.28$\mu$m and 482.99$\mu$m lines are the most suitable lines in ALMA band 7 and 9, respectively.
Here we note that although there is no candidate ortho-$\mathrm{H_2O}$ line in ALMA band 10, some candidate para-$\mathrm{H_2O}$ lines do fall in this band.
With ALMA, we can now conduct high sensitivity $\bigl($$\sim 10^{-21}-10^{-20}$ W $\mathrm{m}^{-2}$ (5$\sigma$, 1 hour)$\bigr)$, high-dispersion (R$>$ 100,000), and even high spatial resolution ($<$ 100 mas) spectroscopic observations.
Since the total fluxes of the candidate sub-millimeter lines which trace the emission from hot water vapor within the $\mathrm{H_2O}$ snowline are low in T Tauri disks ($\sim 10^{-22}-10^{-21}$ W $\mathrm{m}^{-2}$), they remain challenging to detect with current ALMA sensitivity unless we have an unrealistically long integration time (see paper I, \citealt{Notsu2016}).
However, in hotter Herbig Ae disks, in younger T Tauri disks (e.g., HL Tau, \citealt{ALMA2015, Banzatti2015, Harsono2015, Zhang2015, Okuzumi2016}), and in disks around FU Orionis type stars (e.g., V883 Ori, \citealt{Cieza2016}), the $\mathrm{H_2O}$ snowline exists at a larger radius and the fluxes of these lines will be stronger compared with those in our fiducial T Tauri disk.
\\ \\
Our calculations for a Herbig Ae disk predict the fluxes of the 482.99$\mu$m (band 9), 682.66$\mu$m (band 8), and 933.28$\mu$m (band 7) lines to be around $10^{-21}-10^{-20}$ W $\mathrm{m}^{-2}$ if we assume that the distance to the object $d$ is 140pc ($\sim$ the distance of Taurus molecular cloud), and the inclination angle of the disk $i$ is 30 deg. 
Thus the possibility of a successful detection is expected to increase in such Herbig Ae disks and could be achieved with current ALMA capabilities.
Here we mention that the 933.28 $\mu$m line has been detected with high spectral resolution in the disk and outflow around the massive protostar candidate, Source I in Orion KL \citep{Hirota2014ApJL}, using ALMA, and around the embedded low mass protostar (Class I), HL Tau, using SMA \citep{Kristensen2016}.
\citet{Kristensen2016} suggested that future observations at higher sensitivity and angular resolution with ALMA will clarify the origin of this water emission from HL Tau and resolve the disk structures.
\\ \\
Candidate $\mathrm{H_2O}$ lines to trace the $\mathrm{H_2O}$ snowline exist over a wide wavelength range, from mid-infrared to sub-millimeter.
As we discuss in Section 3.2.3, the values of the total fluxes tend to increase as the wavelengths of the candidate $\mathrm{H_2O}$ lines which trace emission from the hot water vapor within $\mathrm{H_2O}$ snowline become shorter (see Table \ref{tab:T1}, \ref{tab:bT1}.1, and Figure \ref{Figure10_HerbigAe}).
There are future mid-infrared instruments covering the part of Q band which will enable high sensitivity and high-dispersion spectroscopic observations: the HRS on SPICA\footnote[6]{\url{http://www.ir.isas.jaxa.jp/SPICA/SPICA_HP/research-en.html}} Mid-infrared Instrument (SPICA/SMI) and Mid-Infrared Camera High-disperser \& IFU spectrograph on the Thirty Meter Telescope (TMT/MICHI, e.g., \citealt{Packham2012}).
HRS on SPICA/SMI will have a relatively high spectral resolution (R$\sim$ 28,000), and especially high sensitivity $\bigl($$\sim 10^{-20}$ W $\mathrm{m}^{-2}$ (5$\sigma$, 1 hour)$\bigr)$ compared with previous mid-infrared instruments at the same wavelengths.
TMT/MICHI will have a high spectral resolution (R$\sim$ 60,000$-$120,000), high sensitivity $\bigl($$\sim 10^{-19}$ W $\mathrm{m}^{-2}$ (5$\sigma$, 1 hour)$\bigr)$.
The $\mathrm{H_2O}$ 17.75$\mu$m and 24.00 $\mu$m lines are in the Q band at mid-infrared wavelengths, and the former line falls in the wavelength coverage of SPICA/SMI-HRS and TMT/MICHI.
Here we note that since TMT is a ground-based telescope, the effect of atmospheric absorption has to be considered carefully in selecting lines from the candidate line list.
Figure \ref{Figure14_HerbigAe} in Appendix C shows the profiles of mid-infrared candidate lines ($\lambda \sim 11-25 $$\mu$m) which trace emission from the hot water vapor within $\mathrm{H_2O}$ snowline. All of the lines in this Figure are also listed in Table \ref{tab:bT1}.1.
Our calculations for a Herbig Ae disk suggest that
the fluxes of the stronger candidate $\mathrm{H_2O}$ lines in Q band (including 17.75$\mu$m and 24.00$\mu$m lines) are $\sim 10^{-17}-10^{-18}$ W $\mathrm{m}^{-2}$.
Since HRS on SPICA/SMI has a high sensitivity in Q band, we predict not only successful detections for some Herbig Ae disks, but also suggest the possibility of a survey 
of the locations of $\mathrm{H_2O}$ snowlines in Herbig Ae disks in nearby ($\lesssim$150 pc) star-forming regions for the first time.
Moreover, since HRS on SPICA/SMI has an especially high sensitivity, successful detections are expected even for T Tauri disks in nearby ($\lesssim$150pc) star-forming regions, as well as Herbig Ae disks in the Orion star-forming region ($\sim$410 pc), with several hours of observations.
Our calculations for a T Tauri disk (see Table \ref{tab:T1} and \ref{tab:bT1}.1) show that the fluxes of the strongest candidate $\mathrm{H_2O}$ lines at Q band (including 17.75$\mu$m and 24.00$\mu$m lines) are $\sim 10^{-20}-10^{-21}$ W $\mathrm{m}^{-2}$.
\\ \\
We also expect to detect emission from those candidate lines (some of which are not accessible from the ground) which trace emission from the hot water vapor within the $\mathrm{H_2O}$ snowline using other high sensitivity mid-infrared and far-infrared instruments on future space telescopes, such as The Mid-Infrared Instrument on board the James Webb Space Telescope (JWST/MIRI\footnote[7]{\url{http://ircamera.as.arizona.edu/MIRI/index.htm}}), SPICA FAR-infrared Instrument (SPICA/SAFARI), and MRS of SPICA/SMI.
Since the spectral resolution of these instruments is not so high ($\sim$ a few thousands), we cannot resolve the velocity profiles of these candidate lines at sufficient resolution to locate the position of the $\mathrm{H_2O}$ snowline.
However, it will be possible to detect the total fluxes of these candidate lines with these high sensitivity instruments. Our results suggest that these lines mainly trace emission from the hot water vapor within the $\mathrm{H_2O}$ snowline. Moreover, since the sensitivity of these instruments is very high, meaning that the time necessary for a detection is not very long, we expect to detect emission from the hot $\mathrm{H_2O}$ gas inside the $\mathrm{H_2O}$ snowline for various protoplanetary disks, which are suitable candidates for high-dispersion spectroscopic observations with future instruments (e.g., ALMA, SPICA/SMI-HRS).
\\
\section{Conclusion} 
\noindent In this paper, we calculated the disk water vapor distribution and corresponding $\mathrm{H_2O}$ line profiles for a Herbig Ae disk, and identified candidate water lines which can locate the position of the $\mathrm{H_2O}$ snowline across a wide wavelength range from mid-infrared to sub-millimeter.
\\ \\
First we calculated the chemical composition using a self-consistent physical model of a Herbig Ae disk, and investigated the abundance distributions of $\mathrm{H_2O}$ gas and ice, and the position of the $\mathrm{H_2O}$ snowline. 
We found that the abundance of $\mathrm{H_2O}$ is high (up to $10^{-4}$) in the inner region with higher temperature ($\gtrsim$ 170K) within $\sim 7-8$ au, relatively high ($\sim 10^{-8}$) between $7-8$ au and 14 au ($=$ the position of the $\mathrm{H_2O}$ snowline, $\sim$120K) near the equatorial plane.
In addition, it is relatively high ($\sim 10^{-8}-10^{-7}$) in the hot surface layer and the photodesorbed region of the outer disk, compared to its value ($\sim 10^{-12}$) in the regions outside the $\mathrm{H_2O}$ snowline near the equatorial plane.
The position of the $\mathrm{H_2O}$ snowline in the Herbig Ae disk is further from the central star compared to that in the T Tauri disk, in agreement with previous studies (e.g., \citealt{Woitke2009b}).
\\ \\
Second, we calculated the $\mathrm{H_2O}$ line profiles, and showed that $\mathrm{H_2O}$ emission lines with small Einstein $A$ coefficients ($A_{ul}\sim10^{-3}-10^{-6}$ s$^{-1}$) and relatively high upper state energies (E$_{\mathrm{up}}\sim1000$K) are dominated by emission from the region inside the $\mathrm{H_2O}$ snowline, and therefore their profiles potentially contain information which can be used to locate the position of the snowline.
Since the fluxes of these lines from Herbig Ae disks are larger than those from T Tauri disks, the possibility of a successful detection is expected to increase for a Herbig Ae disk. 
The wavelengths of those lines which are the best candidates to locate the position of the $\mathrm{H_2O}$ snowline range from mid-infrared to sub-millimeter.
The values of total fluxes tend to be larger as the wavelengths of the $\mathrm{H_2O}$ lines become shorter. This is because the peak wavelength of the Planck function at the gas temperature around the $\mathrm{H_2O}$ snowline ($T_{g}$$\sim 100-200$ K) is in the mid-infrared region.
\\ \\
In addition, we investigated the properties of water lines which have been detected by previous spectroscopic observations (e.g., 63.32$\mu$m, 538.29$\mu$m). These lines are less suited to locate the $\mathrm{H_2O}$ snowline, because they are not dominated by emission from the region within the $\mathrm{H_2O}$ snowline. The properties of near-, and mid-infrared $\mathrm{H_2O}$ emission lines which do not trace emission from the hot water vapor within the $\mathrm{H_2O}$ snowline are also discussed.
\\ \\
The wavelengths of such candidate lines which trace emission from the hot water vapor within $\mathrm{H_2O}$ snowline overlap with the capabilities of ALMA and future mid-infrared high-dispersion spectrographs (e.g., SPICA/SMI-HRS).
The successful detection in a Herbig Ae disk could be achieved with current ALMA capabilities using several lines.
Mid-infrared instruments such as HRS on SPICA/SMI would have a high sensitivity in the Q band (e.g., $\sim$ 16-18$\mu$m), and we predict not only successful detections for some Herbig Ae disks, but also suggest the possibility of a survey of $\mathrm{H_2O}$ snowline locations in many Herbig Ae disks in nearby ($\lesssim$150pc) star-forming regions for the first time.
\\ \\
\acknowledgments
\noindent We are grateful to Dr. Itsuki Sakon, Dr. Chris Packham, Dr. Hiroshi Shibai, Dr. Takao Nakagawa, Dr. Takashi Onaka, Dr. Satoshi Okuzumi, Dr. Inga Kamp, and Dr. Eiji Akiyama for their useful comments.
We would like to thank the referee for comments that improved our paper.
The numerical calculations in this study were carried out on SR16000 at Yukawa Institute for Theoretical Physics (YITP) and computer systems at Kwasan and Hida Observatory (KIPS) in 
Kyoto University, and PC cluster at Center for Computational Astrophysics, National Astronomical Observatory of Japan.
This work is supported by JSPS (Japan Society for the Promotion of Science) Grants-in-Aid for Scientific Research (Grant Number; 23103005, 25108004, 25108005, 25400229, 15H03646),
by Grants-in-Aid for JSPS fellows (Grant Number; 16J06887), and by the Astrobiology Center Program of National Institutes of Natural Sciences (NINS) (Grant Number; AB281013).
S. N. is grateful for the support from the educational program organized by Unit of Synergetic Studies for Space, Kyoto University.
C. W. acknowledges support from the Netherlands Organization for Scientific Research (NWO, program number 639.041.335).
Astrophysics at Queen's University Belfast is supported by a grant from the STFC.
\appendix
 \section{The vertical distributions of normalized cumulative line emissivity}
\begin{figure}[htbp]
\begin{center}
\includegraphics[scale=0.5]{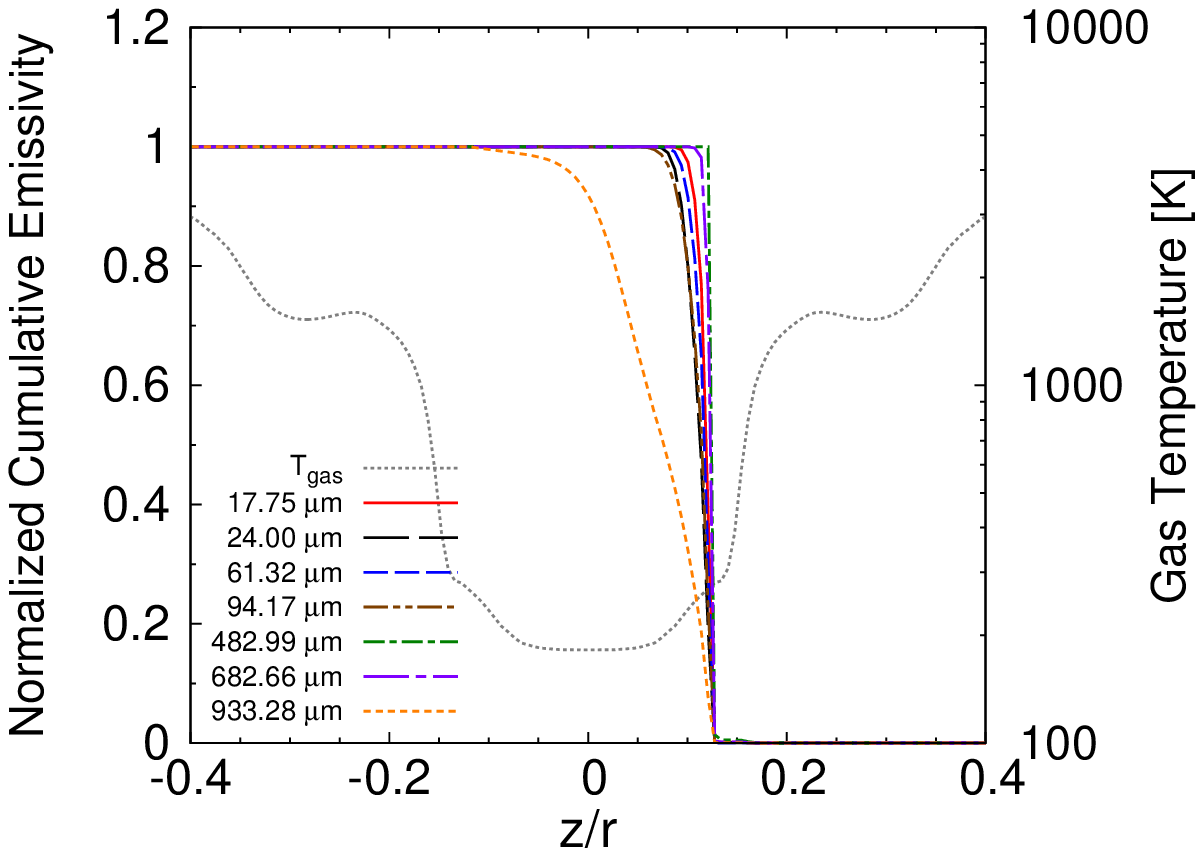}
\includegraphics[scale=0.5]{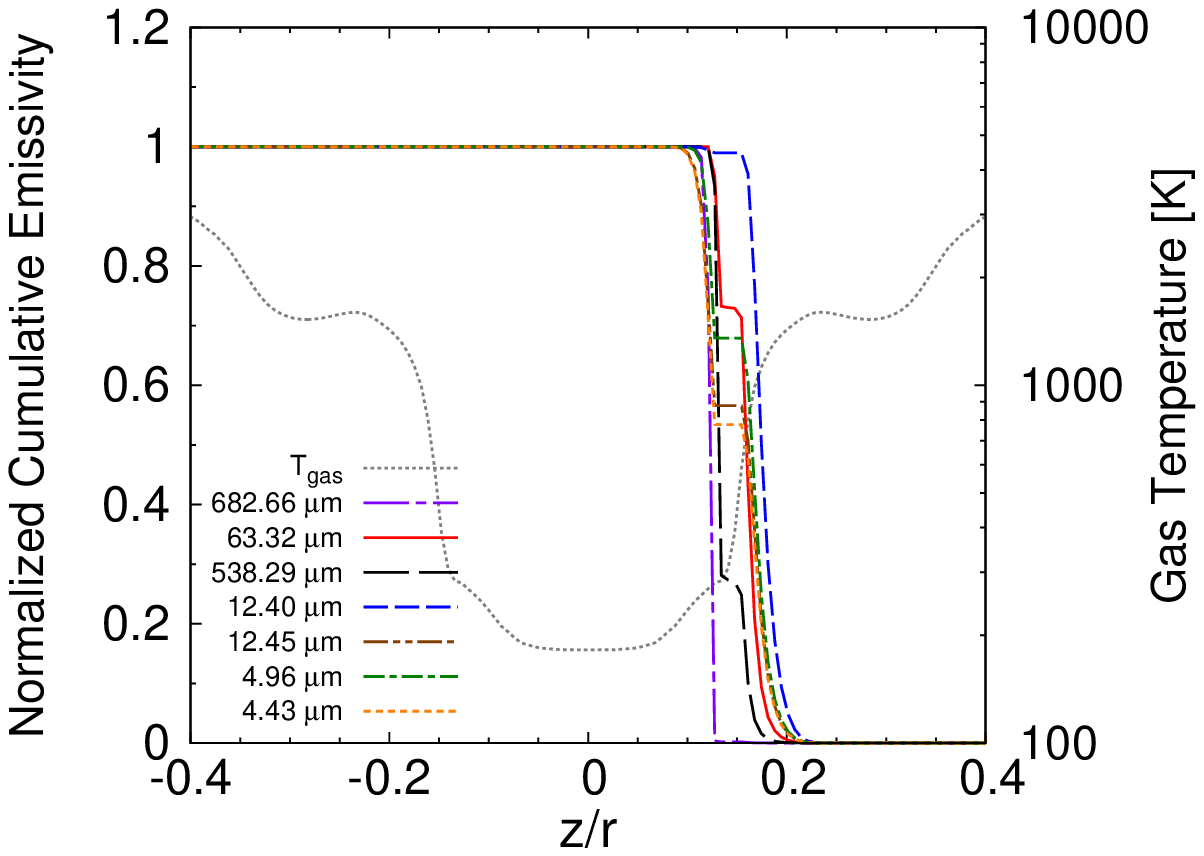}
\includegraphics[scale=0.5]{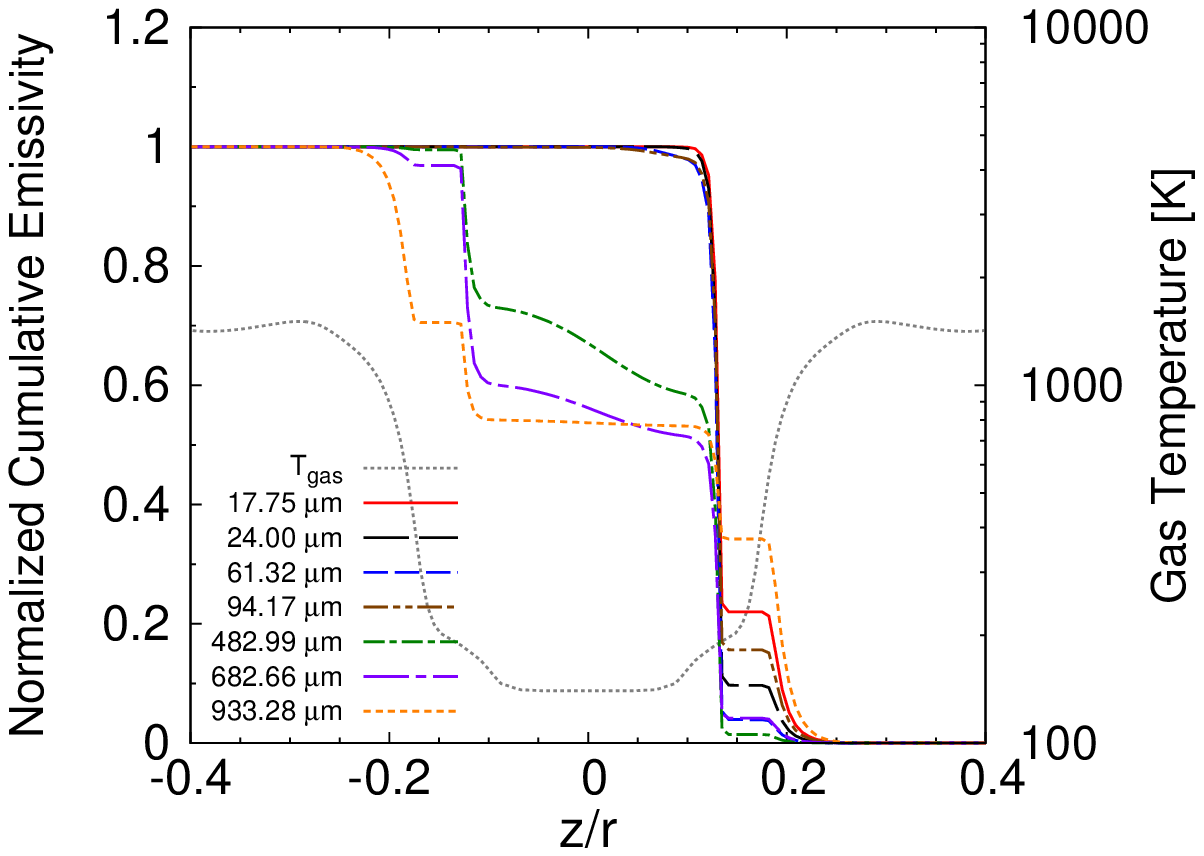}
\includegraphics[scale=0.5]{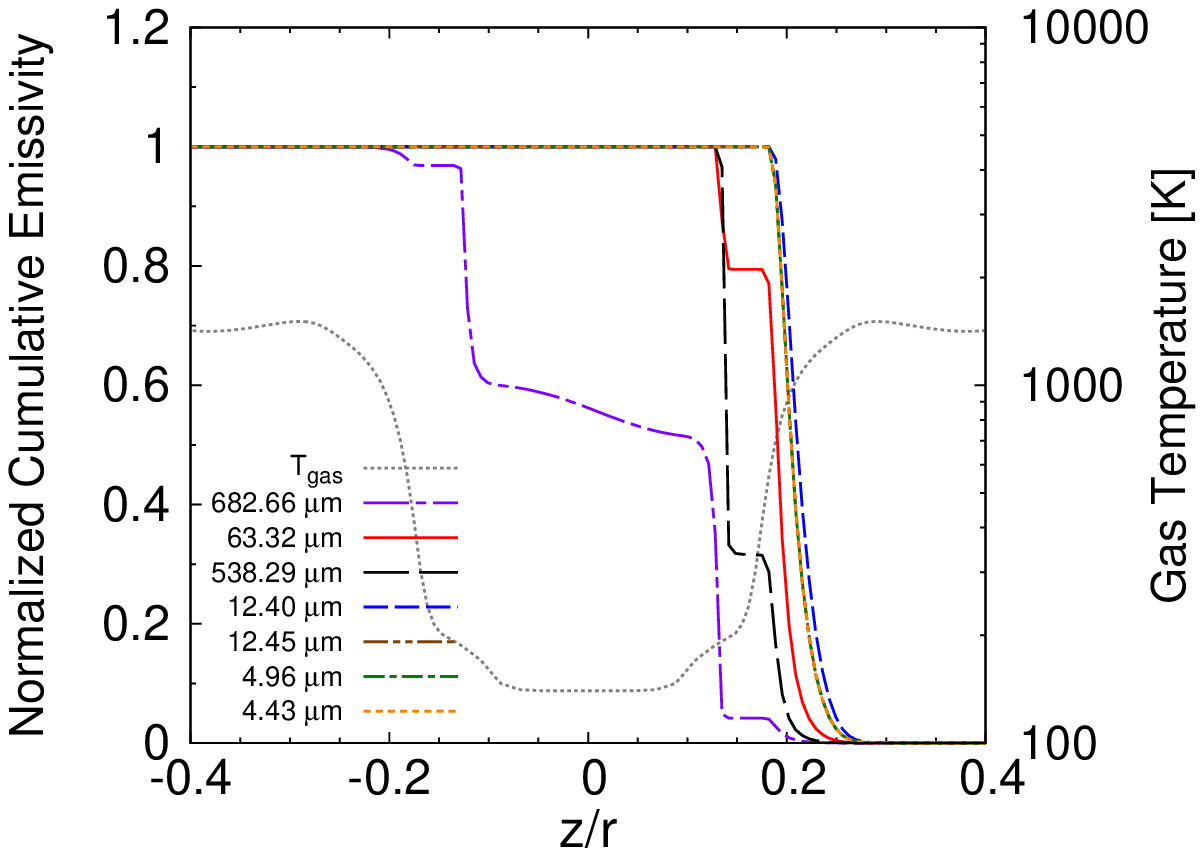}
\includegraphics[scale=0.5]{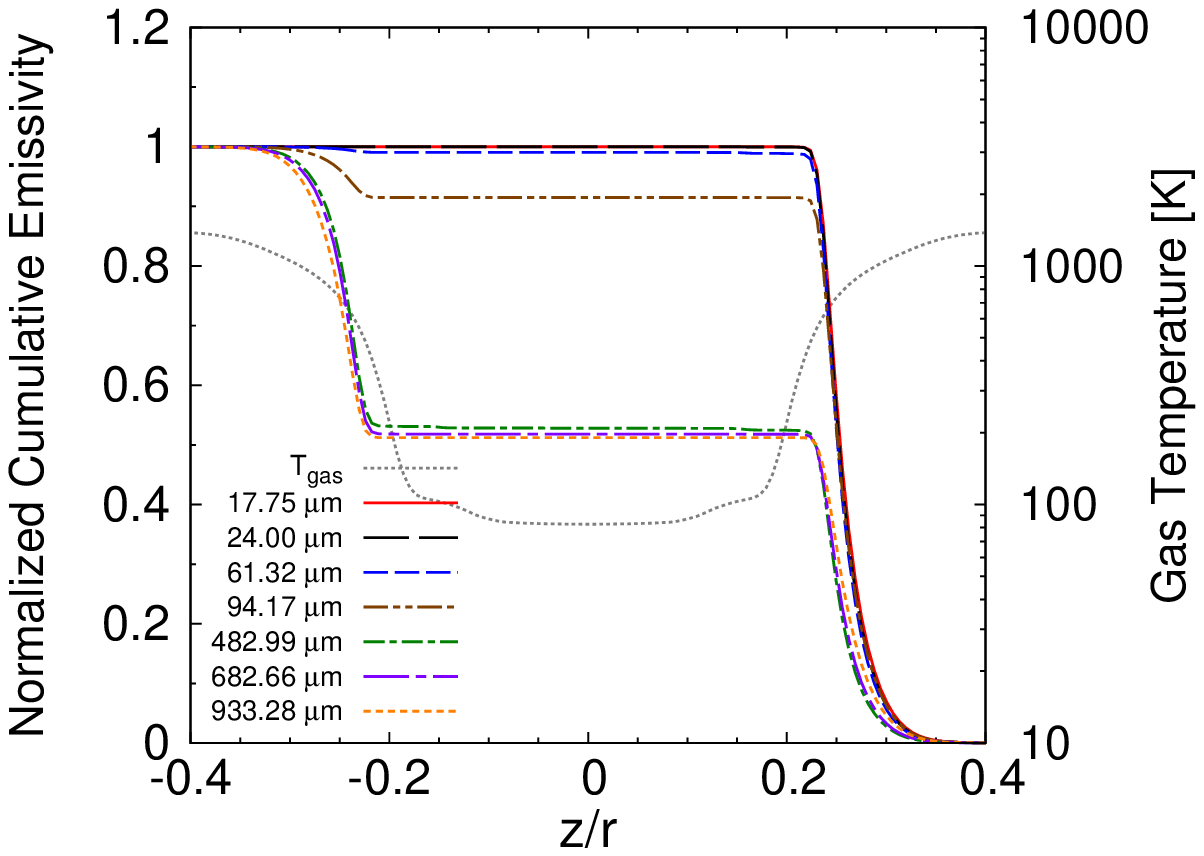}
\includegraphics[scale=0.5]{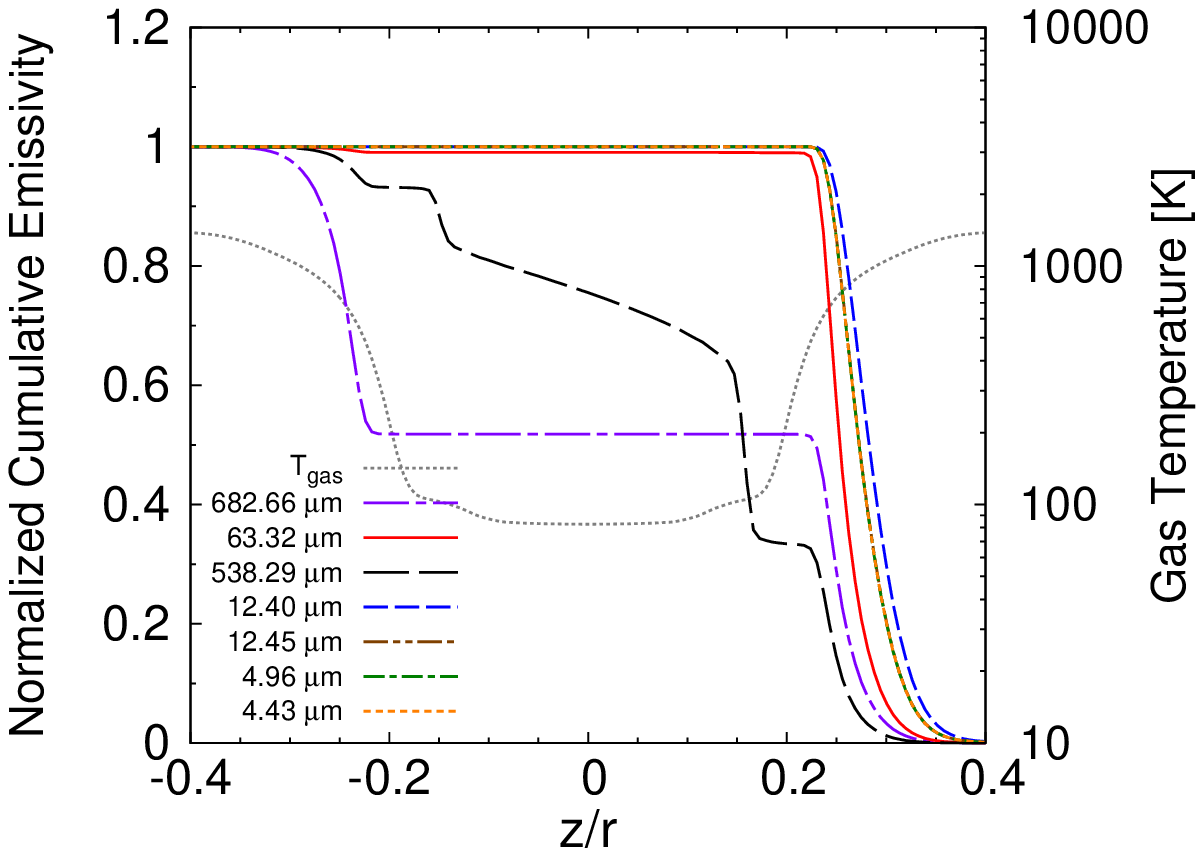}
\end{center}
\vspace{1cm}
\caption{\noindent The vertical distributions of the normalized cumulative line emissivity at $r=$5 au (top two panels), $r=$10 au (middle two panels), and $r=$30 au (bottom two panels), and of the gas temperature $T_{g}$ at Kelvin ({\it gray dotted line}).
The left three panels show the distributions for seven pure rotational ortho-$\mathrm{H_2O}$ lines at $\lambda$=17.75$\mu$m ({\it red solid line}), 24.00$\mu$m ({\it black long dashed line}),  61.32$\mu$m ({\it blue dashed line}), 94.17$\mu$m ({\it brown dashed two dotted line}), 482.99$\mu$m ({\it green dashed dotted line}), 682.66$\mu$m ({\it violet long dashed dotted line}), and 933.28$\mu$m ({\it orange dotted line}.
The right three panels show the distributions for seven pure rotational ortho-$\mathrm{H_2O}$ lines at $\lambda$=682.66$\mu$m ({\it violet long dashed dotted line}), 63.32$\mu$m ({\it red solid line}), 538.29$\mu$m ({\it black long dashed line}), 12.40$\mu$m ({\it blue dashed line}), 12.45$\mu$m ({\it brown dashed two dotted line}), 4.96$\mu$m ({\it green dashed dotted line}), and 4.43$\mu$m ({\it orange dotted line}).
We normalized the cumulative emissivity of each line using the values at $z=-\infty$.
We assume that the inclination angle of the disk $i$ is 0 degree in making these figures.
}\label{Figure13_HerbigAe}
\end{figure}

\noindent Figure \ref{Figure13_HerbigAe} shows the vertical distributions of the normalized cumulative line emissivity at $r=$5 au (top two panels), $r=$10 au (middle two panels), and $r=$30 au (bottom two panels), and of the gas temperature $T_{g}$.
The left three panels show the distributions for seven $\mathrm{H_2O}$ lines at $\lambda$=17.75$\mu$m, 24.00$\mu$m,  61.32$\mu$m, 94.17$\mu$m, 482.99$\mu$m, 682.66$\mu$m, and 933.28$\mu$m, for the Herbig Ae disk.
The right three panels show the distributions for seven $\mathrm{H_2O}$ lines at $\lambda$=682.66$\mu$m, 63.32$\mu$m, 538.29$\mu$m, 12.40$\mu$m, 12.45$\mu$m, 4.96$\mu$m, and 4.43$\mu$m. for the Herbig Ae disk.
We normalize the cumulative emissivity of each line using the values at $z=-\infty$. 
\\ \\
The differences in the properties of the line profiles (see also Figures \ref{Figure4_HerbigAe}, \ref{Figure5_HerbigAe}, \ref{Figure6_HerbigAe}, \ref{Figure7_HerbigAe}, \ref{Figure8_HerbigAe}, and \ref{Figure9_HerbigAe}) come from the differences in $A_{ul}$, $E_{up}$, and wavelengths among lines.
For the lines with similar wavelengths, the emitting regions tend to be in the upper region of the disk as values of $A_{ul}$ of the lines become larger, since the absorption by excited $\mathrm{H_2O}$ molecules increases.
In addition, in the sub-millimeter lines, the values of normalized cumulative line emissivity are smaller than unity at $z/r\sim0$.
It is because the values of dust opacity at sub-millimeter wavelengths are smaller than those at infrared wavelengths (see also Figures \ref{Figure5_HerbigAe}, \ref{Figure6_HerbigAe}, \ref{Figure8_HerbigAe}, and \ref{Figure9_HerbigAe}).
\section{$\mathrm{H_2O}$ line parameters and fluxes}
\renewcommand{\thetable}{\Alph{section}.\arabic{table}}
\setcounter{table}{0}
\noindent The detailed parameters, such as transitions ($J_{K_{a}K_{c}}$), wavelength, frequency, $A_{ul}$, $E_{up}$, and total line fluxes of the candidate ortho-$\mathrm{H_2O}$ lines which trace emission from the hot water vapor within the $\mathrm{H_2O}$ snowline are listed in Table \ref{tab:bT1}.1. 
In Table \ref{tab:bT1}.1, we show the values of the total fluxes from both disks around the Herbig Ae star and the T Tauri star.
Some of these lines are also listed in Table \ref{tab:T1}. The line selection method is described in Section 3.2.3.   
\begin{deluxetable}{rrrrrrr}
\tablewidth{0pt}
\tablecaption{{Candidate ortho-$\mathrm{H_2O}$ line parameters and total line fluxes (11$-$1000$\mu\textrm{m}$)}}\label{tab:bT1}
\tablehead{
\colhead{$J_{K_{a}K_{c}}$}&\colhead{\ $\lambda$\tablenotemark{1}}&
\colhead{\ Freq.}& \colhead{\ $A_{ul}$}&
\colhead{$E_{up}$} & \colhead{HAe flux\tablenotemark{2,3}} & \colhead{TT flux\tablenotemark{3,4}}\\
\colhead{}&\colhead{[$\mu$m]}&
\colhead{[GHz]}& \colhead{[s$^{-1}$]}&
\colhead{[K]}& \colhead{[W $\mathrm{m}^{-2}$]}& \colhead{[W $\mathrm{m}^{-2}$]}}
\startdata   
      7$_{70}$-6$_{25}$ & 11.878 & 25239.615 & 6.653$\times10^{-3}$ & 2006.8 & $3.4\times10^{-19}$ & $1.8\times10^{-22}$\\
      7$_{61}$-6$_{16}$ & 13.005 & 23052.310 & 2.585$\times10^{-4}$ & 1749.8 & $2.9\times10^{-18}$ & $1.1\times10^{-21}$\\      
      6$_{61}$-5$_{14}$ & 15.489 & 19354.618 & 4.138$\times10^{-4}$ & 1503.6 & $6.9\times10^{-18}$ & $2.6\times10^{-21}$\\     
      7$_{70}$-7$_{25}$ & 16.329 & 18359.435 & 7.756$\times10^{-5}$ & 2006.8 & $3.8\times10^{-19}$ & $1.6\times10^{-22}$\\    
      6$_{61}$-6$_{16}$ & 16.728 & 17921.765 & 2.153$\times10^{-5}$ &1503.6 & $3.2\times10^{-19}$  & $1.1\times10^{-22}$\\         
      6$_{52}$-5$_{05}$ & 17.754 & 16885.840 & 2.909$\times10^{-3}$ & 1278.5 & $4.1\times10^{-17}$ & $2.3\times10^{-20}$\\     
      7$_{61}$-7$_{16}$ & 19.532 & 15348.791 & 1.269$\times10^{-3}$ &1749.8& $8.3\times10^{-18}$  & $3.5\times10^{-21}$\\         
      7$_{70}$-8$_{27}$ & 19.638 & 15265.853 & 1.900$\times10^{-6}$ &2006.8& $7.6\times10^{-21}$  & $3.1\times10^{-24}$\\
      7$_{52}$-7$_{07}$ & 21.115 & 14197.932 & 2.078$\times10^{-3}$ & 1524.8 & $2.0\times10^{-17}$  & $9.9\times10^{-21}$\\       
      7$_{61}$-8$_{18}$ & 21.185 & 14151.161 & 6.154$\times10^{-6}$ & 1749.8 & $4.8\times10^{-20}$ & $1.7\times10^{-23}$\\   
      7$_{70}$-7$_{43}$ & 21.571 & 13897.693 & 3.474$\times10^{-3}$ & 2006.8 & $1.1\times10^{-17}$ & $5.5\times10^{-21}$\\       
      5$_{50}$-5$_{05}$ & 23.996 & 12493.205 & 2.696$\times10^{-4}$ & 1067.7 & $9.4\times10^{-18}$ & $6.4\times10^{-21}$\\       
       6$_{61}$-6$_{34}$ & 25.247 & 11874.160 & 3.803$\times10^{-3}$ & 1503.6 & $2.6\times10^{-17}$ & $1.9\times10^{-20}$\\        
       6$_{61}$-7$_{16}$ & 29.339 & 10218.246 & 1.859$\times10^{-5}$ & 1503.6 & $2.5\times10^{-19}$ & $1.3\times10^{-22}$\\     
       8$_{54}$-9$_{09}$ & 29.851 & 10043.001 & 1.605$\times10^{-4}$ & 1805.9 & $1.2\times10^{-18}$ & $4.7\times10^{-22}$\\  
       6$_{52}$-7$_{07}$ & 33.074 & 9064.381 & 5.942$\times10^{-5}$ & 1278.5 & $1.4\times10^{-18}$ & $1.3\times10^{-21}$\\ 
       5$_{50}$-5$_{23}$ & 33.833 & 8860.835 & 4.552$\times10^{-3}$ & 1067.7 & $3.2\times10^{-17}$ & $7.1\times10^{-20}$\\        
       7$_{70}$-8$_{45}$ & 36.750 & 8157.523 & 1.123$\times10^{-4}$ & 2006.8 & $4.3\times10^{-19}$ & $1.6\times10^{-22}$\\     
       4$_{41}$-4$_{14}$ & 37.984 & 7892.616 & 2.629$\times10^{-3}$ & 702.3 & $3.1\times10^{-17}$ & $1.0\times10^{-19}$\\  
       9$_{45}$-10$_{1 10}$ & 40.702 & 7365.464 & 4.815$\times10^{-4}$ & 1957.1 & $2.1\times10^{-18}$ & $1.1\times10^{-21}$\\ 
       7$_{61}$-8$_{36}$ & 47.601 & 6297.997 & 4.268$\times10^{-4}$ & 1749.8 & $2.3\times10^{-18}$ & $2.2\times10^{-21}$\\           
       6$_{61}$-7$_{34}$ & 49.334 & 6076.834 & 1.195$\times10^{-4}$ & 1503.6 & $1.3\times10^{-18}$ & $1.8\times10^{-21}$\\ 
       5$_{50}$-6$_{25}$ & 52.864 & 5671.021 & 1.442$\times10^{-4}$ & 1067.7 & $3.7\times10^{-18}$ & $1.4\times10^{-20}$\\ 
       7$_{43}$-8$_{18}$ & 53.455 & 5608.351 & 5.030$\times10^{-4}$ & 1339.8 & $5.2\times10^{-18}$ & $1.6\times10^{-20}$\\ 
       7$_{52}$-8$_{27}$ & 57.394 & 5223.441 &7.203$\times10^{-4}$ & 1524.8 & $4.1\times10^{-18}$  & $1.1\times10^{-20}$\\           
       5$_{41}$-6$_{16}$ & 61.316 & 4889.280 & 2.686$\times10^{-4}$ & 878.1 & $5.9\times10^{-18}$ & $3.5\times10^{-20}$\\
       6$_{52}$-7$_{25}$ & 94.172 & 3183.464 & 3.387$\times10^{-4}$ & 1278.5 & $1.8\times10^{-18}$ & $1.6\times10^{-20}$\\       
       4$_{41}$-5$_{14}$ & 112.803 & 2657.666 & 1.670$\times10^{-4}$ & 702.3 & $1.8\times10^{-18}$ & $2.3\times10^{-20}$\\ 
       9$_{27}$-10$_{1 10}$ & 114.454 & 2619.334 & 6.768$\times10^{-4}$ & 1729.3 & $1.0\times10^{-18}$ & $8.1\times10^{-21}$\\ 
       8$_{36}$-9$_{09}$ & 116.350 & 2576.644 & 7.255$\times10^{-4}$ & 1447.5 & $1.4\times10^{-18}$ & $1.5\times10^{-20}$\\ 
       8$_{45}$-7$_{52}$ & 159.051& 1884.888 & 7.403$\times10^{-3}$ & 1615.3 & $1.1\times10^{-18}$ & $1.7\times10^{-20}$\\ 
       6$_{34}$-7$_{07}$ & 159.400 & 1880.753 & 3.911$\times10^{-4}$ & 933.7 & $9.0\times10^{-19}$ & $1.5\times10^{-20}$\\ 
       8$_{54}$-9$_{27}$ & 187.810 & 1595.252 & 2.947$\times10^{-4}$ & 1805.9 & $2.5\times10^{-19}$ & $3.8\times10^{-21}$\\ 
       6$_{43}$-7$_{16}$ & 190.437 & 1574.232 & 2.927$\times10^{-4}$ & 1088.7 & $4.9\times10^{-19}$ & $8.9\times10^{-21}$\\              
       6$_{25}$-5$_{32}$ & 226.761 & 1322.065 & 2.334$\times10^{-3}$ & 795.5 & $6.0\times10^{-19}$ & $1.2\times10^{-20}$\\ 
       8$_{45}$-9$_{18}$ & 229.206 & 1307.963 & 2.599$\times10^{-4}$ & 1615.3 & $2.0\times10^{-19}$ & $3.9\times10^{-21}$\\  
       8$_{27}$-7$_{34}$ & 231.248 & 1296.411 & 1.074$\times10^{-3}$ & 1274.1 & $3.7\times10^{-19}$ & $6.8\times10^{-21}$\\     
       7$_{43}$-6$_{52}$ & 234.531 & 1278.266 & 1.552$\times10^{-3}$ & 1399.8 & $3.6\times10^{-19}$  & $6.7\times10^{-21}$\\      
       8$_{54}$-7$_{61}$ & 256.593 & 1168.358 & 9.459$\times10^{-4}$ & 1805.9 & $1.8\times10^{-19}$ & $3.5\times10^{-21}$\\     
       6$_{34}$-5$_{41}$ & 258.816 & 1158.324 & 1.417$\times10^{-3}$ & 933.7 & $3.5\times10^{-19}$  & $7.0\times10^{-21}$\\
       7$_{25}$-8$_{18}$ & 261.457 & 1147.621& 1.087$\times10^{-4}$ & 1125.7 & $1.9\times10^{-19}$  & $3.9\times10^{-21}$\\
       5$_{32}$-4$_{41}$ & 482.990 & 620.701 & 1.106$\times10^{-4}$ & 732.1 & $5.3\times10^{-20}$ & $1.1\times10^{-21}$\\
       7$_{52}$-6$_{61}$ & 676.704 & 443.018 & 2.231$\times10^{-5}$ & 1524.8 & $9.4\times10^{-21}$  & $2.6\times10^{-22}$\\   
       6$_{43}$-5$_{50}$ & 682.664 & 439.151 & 2.816$\times10^{-5}$ & 1088.7 & $1.4\times10^{-20}$  & $3.1\times10^{-22}$\\
      10$_{29}$-9$_{36}$ & 933.277 & 321.226 & 6.165$\times10^{-6}$ & 1861.2 & $2.3\times10^{-21}$ & $7.8\times10^{-23}$\\     
\enddata
\tablenotetext{1}{In calculating the value of line wavelength from the value of line frequency, we use the value of speed of light $c$$=$$2.99792458\times 10^{8}$ m s$^{-1}$.}
\tablenotetext{2}{The total flux of each emission line from the Herbig Ae disk.}
\tablenotetext{3}{In calculating the total fluxes of these $\mathrm{H_2O}$ lines, we use a distance $d=140$pc and an inclination angle of $i=$30 degree.}
\tablenotetext{4}{The total flux of each emission line from the T Tauri disk (see also paper I, \citealt{Notsu2016}).}
\end{deluxetable}
\clearpage
\section{The profiles of candidate mid-infrared $\mathrm{H_2O}$ lines to trace the $\mathrm{H_2O}$ snowline}
\noindent Figure \ref{Figure14_HerbigAe} shows the profiles of mid-infrared candidate $\mathrm{H_2O}$ lines which trace the hot water vapor within the $\mathrm{H_2O}$ snowline ($\lambda \sim 11-25 $$\mu$m). 
 All of the lines in this Figure are also listed in Table \ref{tab:bT1}.1. We discuss the properties of these lines in Sections 3.2.3 and 4.4.
\begin{figure}[htbp]
\begin{center}
\includegraphics[scale=0.5]{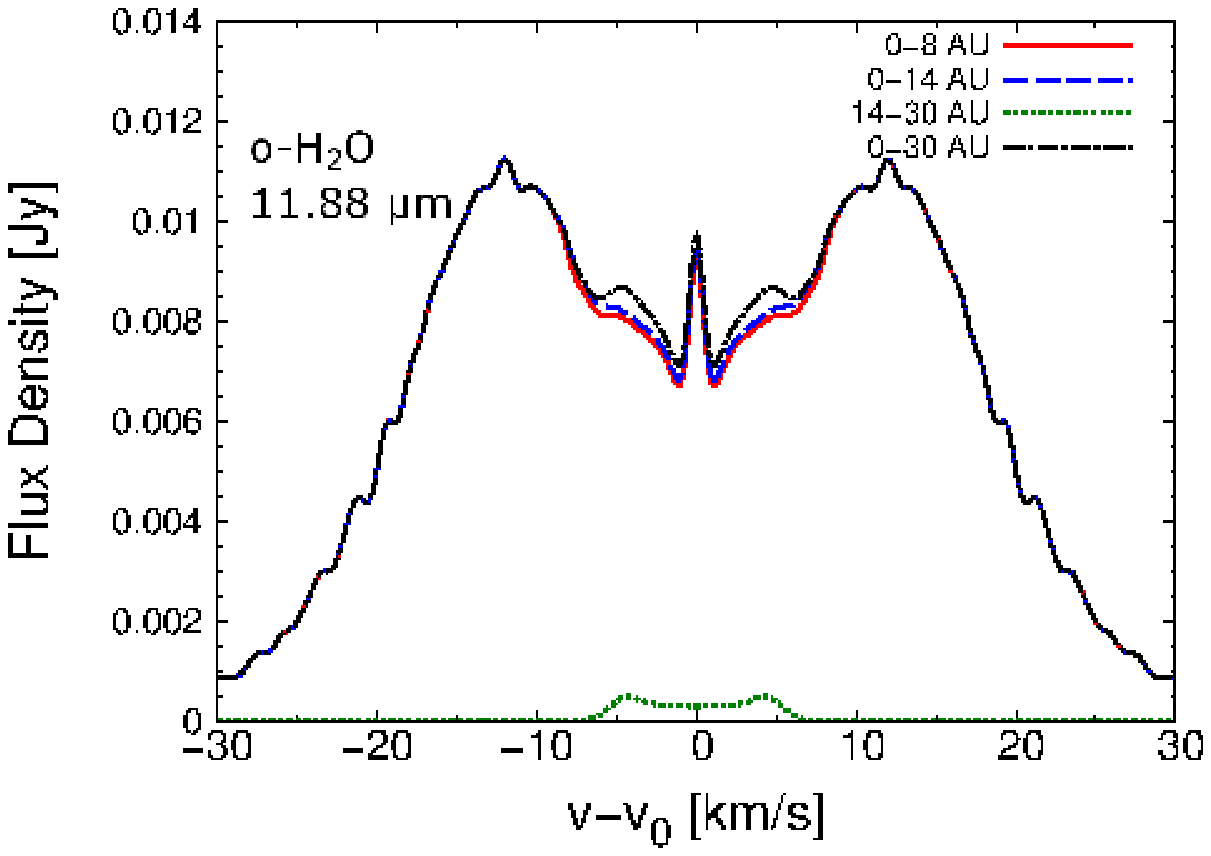}
\includegraphics[scale=0.5]{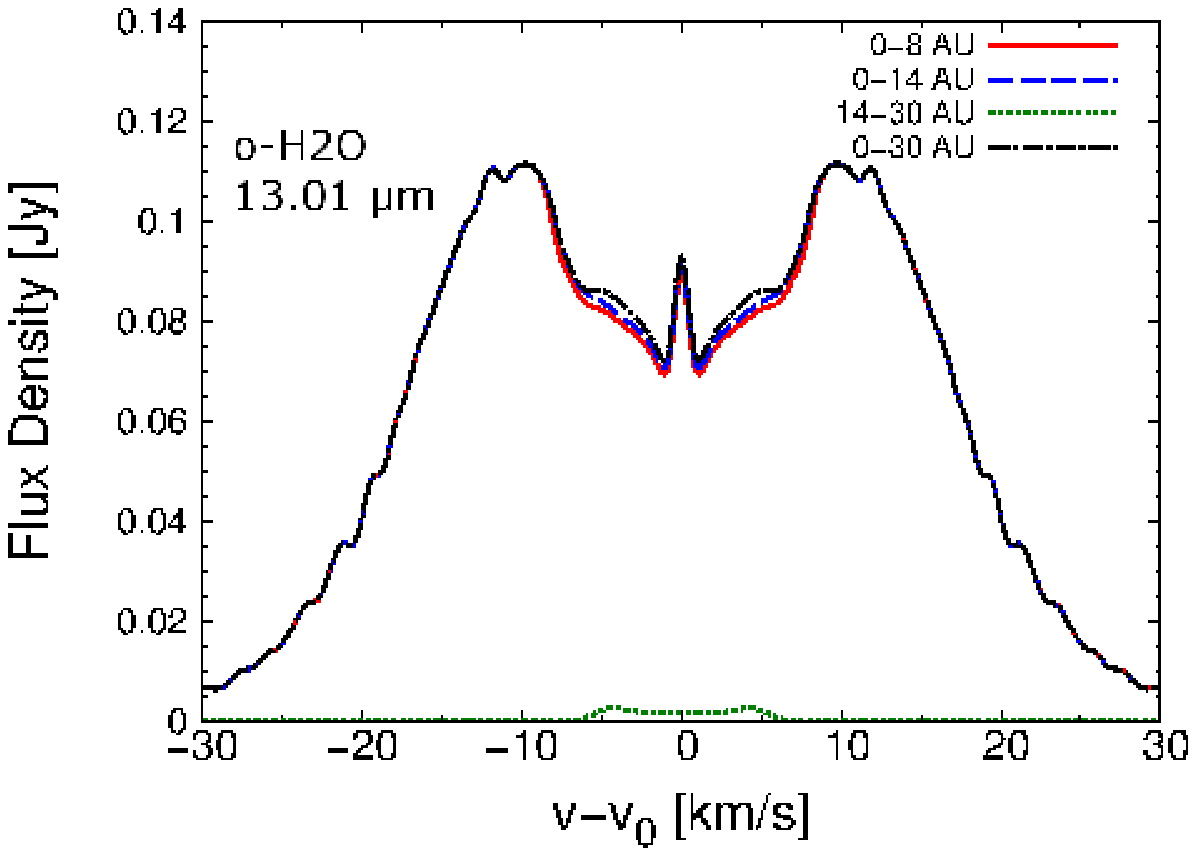}
\includegraphics[scale=0.5]{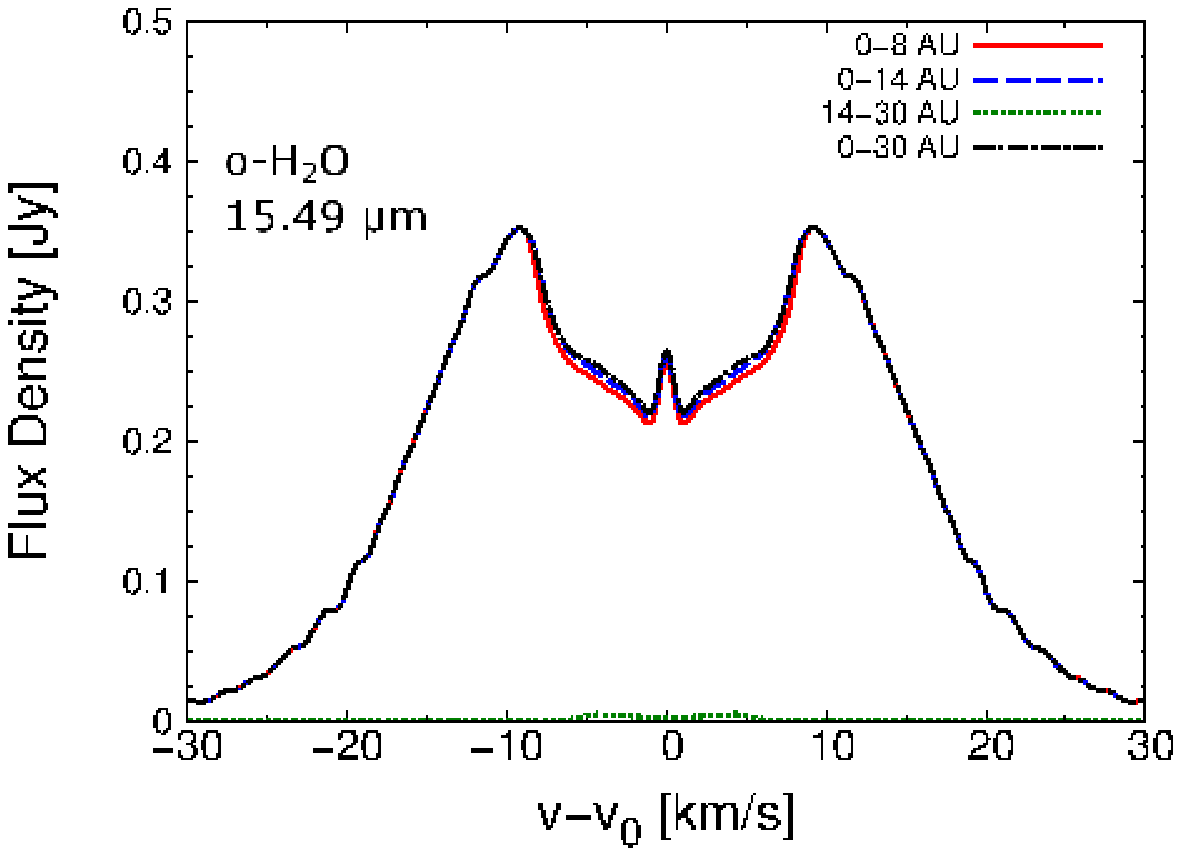}
\includegraphics[scale=0.5]{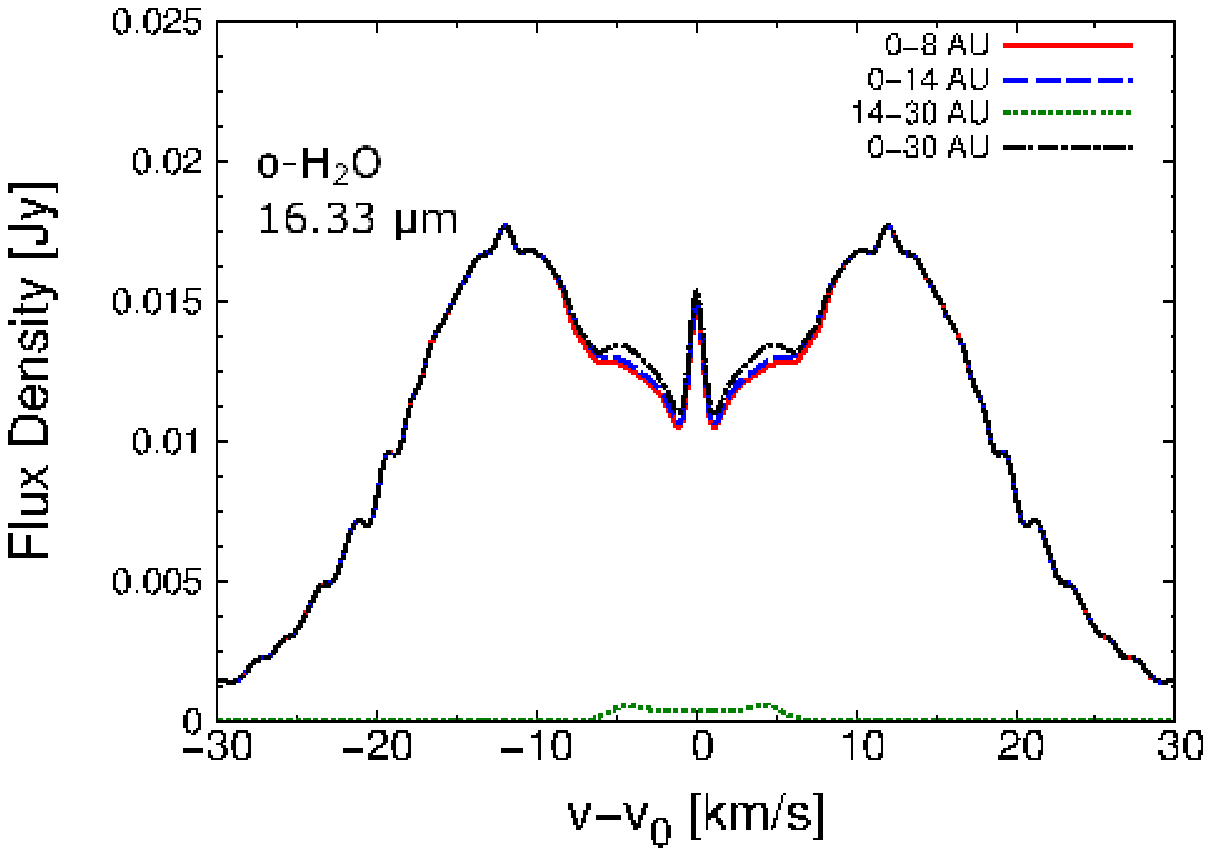}
\includegraphics[scale=0.5]{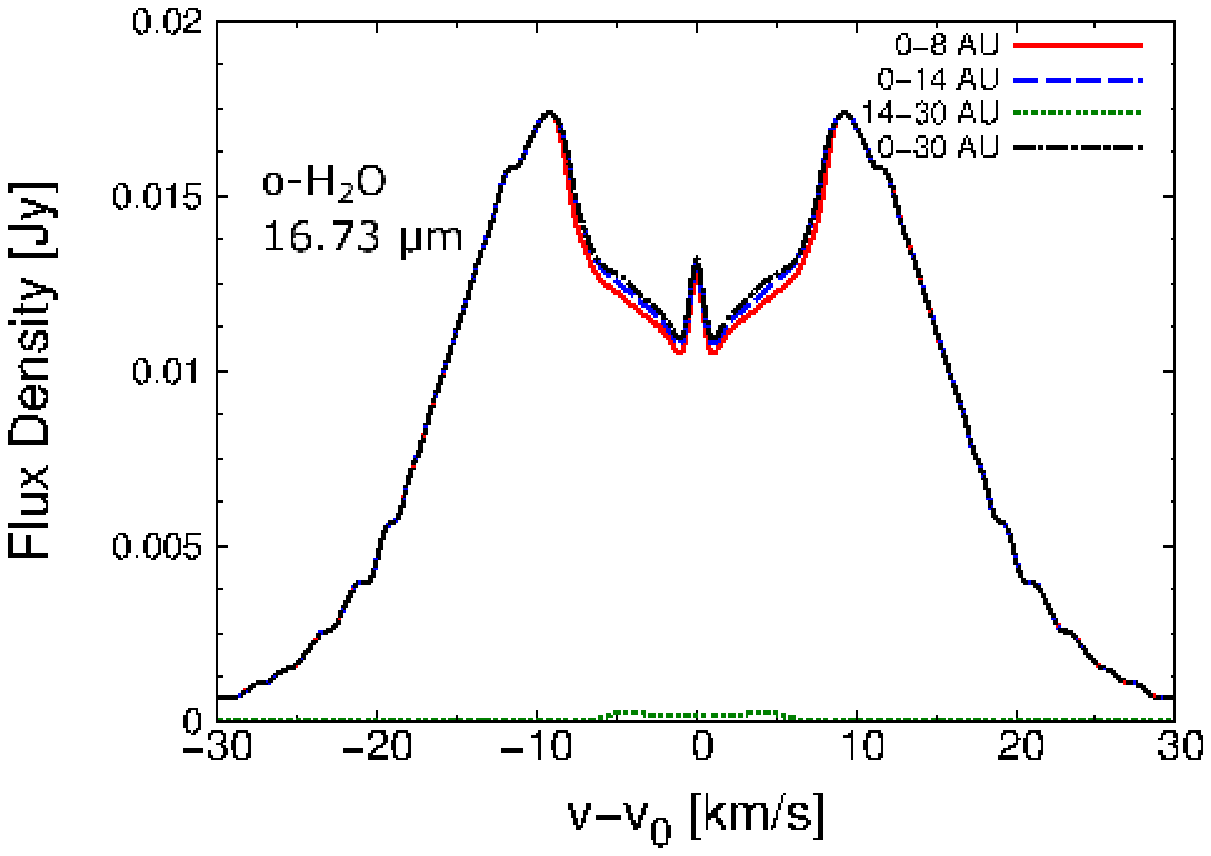}
\includegraphics[scale=0.5]{Figure4a_paper2_HAe_17.75um_trans79_20160911ver3r.eps}
\end{center}
\caption{\noindent The velocity profiles of mid-infrared ortho-$\mathrm{H_2O}$ lines at $\lambda$=11.88$\mu$m (top left), 13.01$\mu$m (top right), 15.49$\mu$m (middle left), 16.33$\mu$m (middle right), 16.73$\mu$m (bottom left), and 17.75$\mu$m (bottom right), from the Herbig Ae disk. These exist between 11-25$\mu$m and are also the best mid-infrared candidate ortho-$\mathrm{H_2O}$ lines to trace the hot water vapor within the $\mathrm{H_2O}$ snowline.
The parameters and total fluxes of these ortho-$\mathrm{H_2O}$ lines are reported in Table \ref{tab:bT1}.1.
{\it Red solid lines} are the emission line profiles from inside 8 au ($=$the inner high temperature region), {\it blue dashed lines} are those from inside 14 au ($\sim$inside the $\mathrm{H_2O}$ snowline), {\it green dotted lines} are those from 14-30 au ($\sim$outside the $\mathrm{H_2O}$ snowline), and {\it black dashed dotted lines} are those from the total area inside 30au. 
}\label{Figure14_HerbigAe}
\end{figure} 
\setcounter{figure}{13}
\begin{figure}[htbp]
\begin{center}
\includegraphics[scale=0.5]{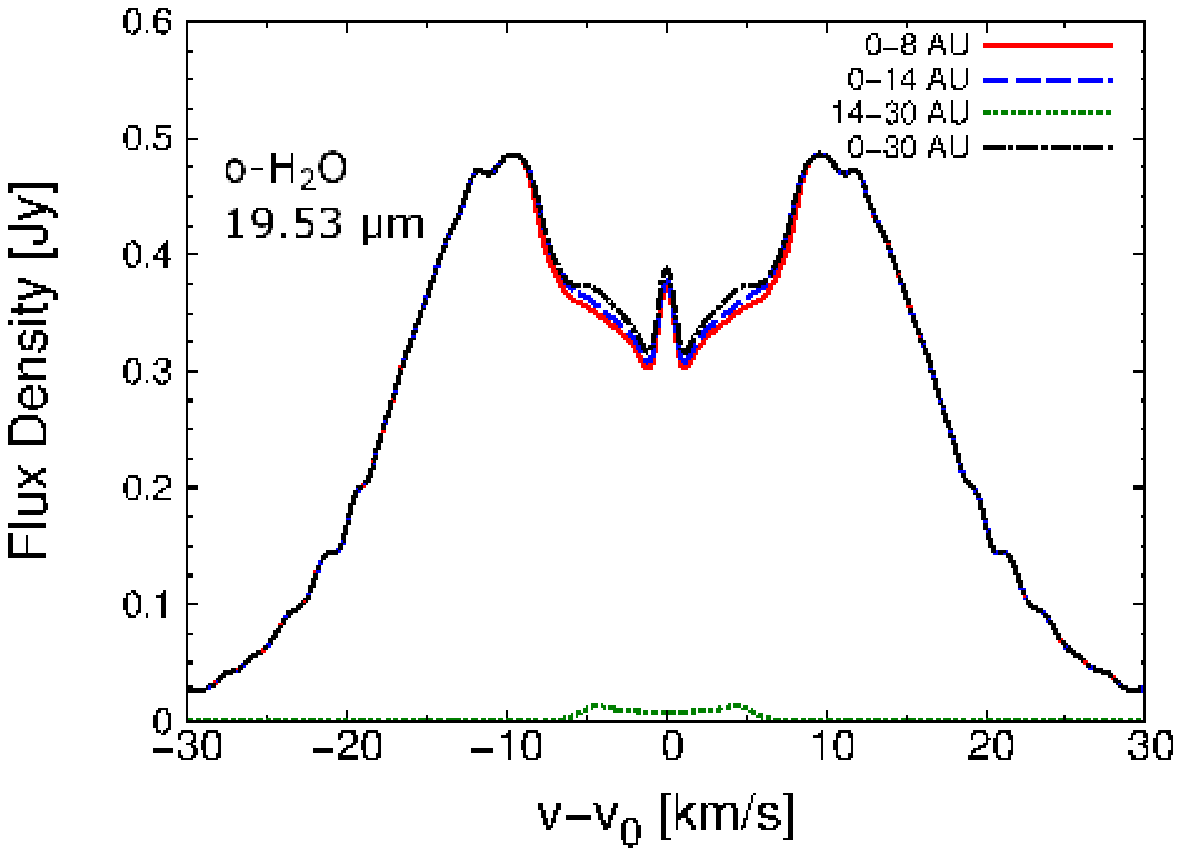}
\includegraphics[scale=0.5]{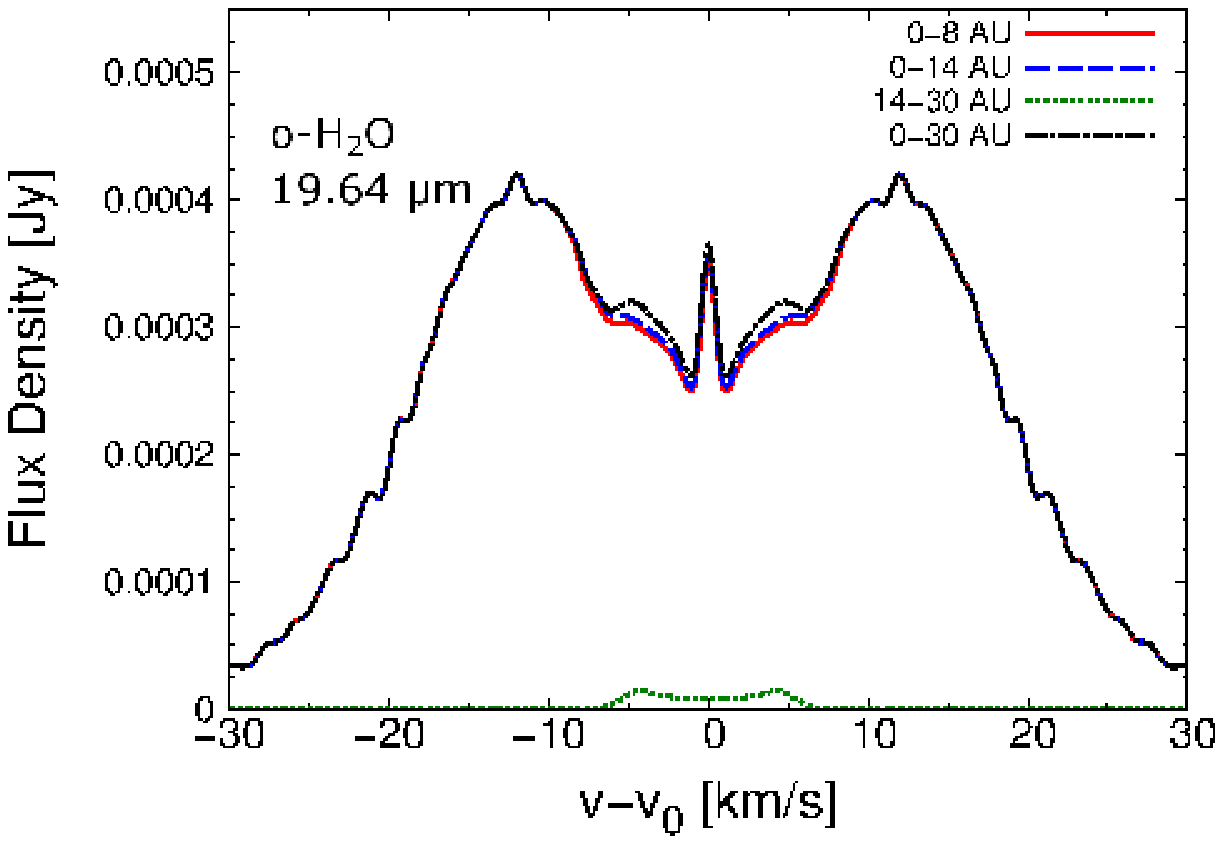}
\includegraphics[scale=0.5]{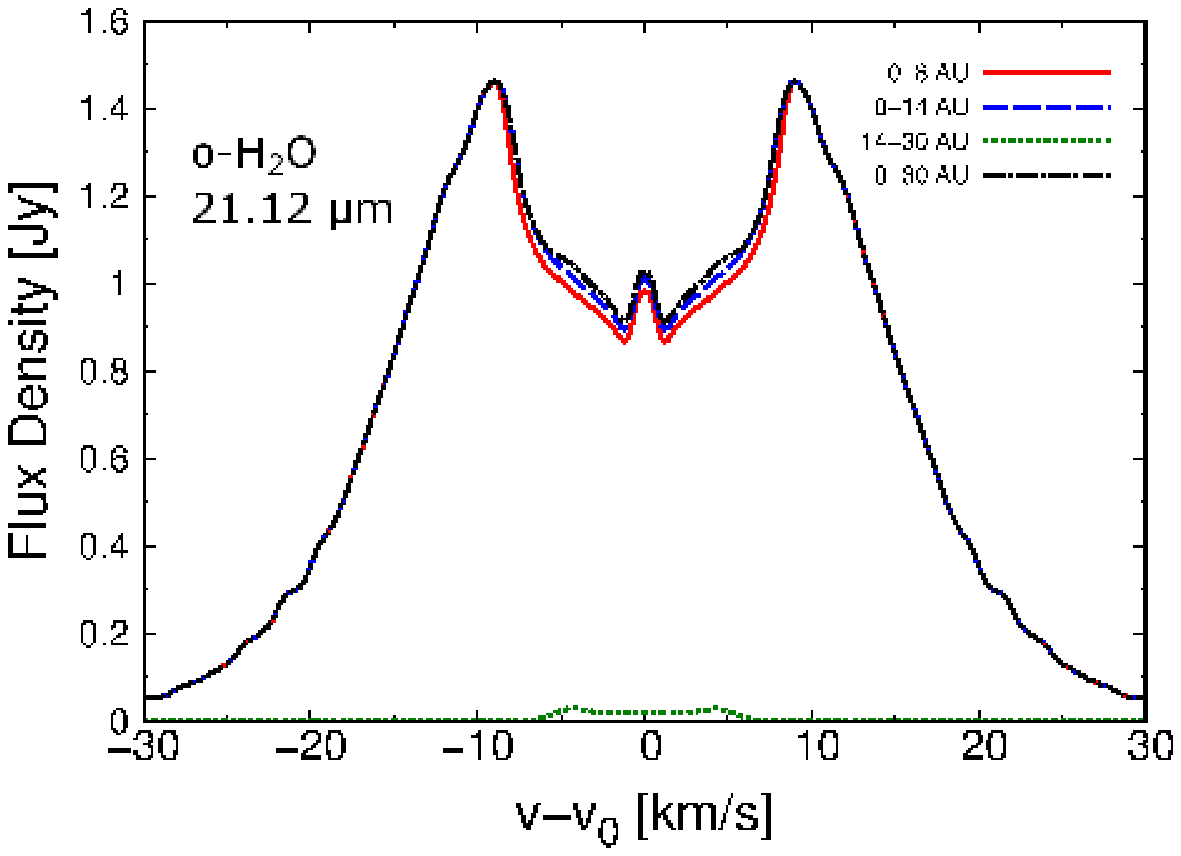}
\includegraphics[scale=0.5]{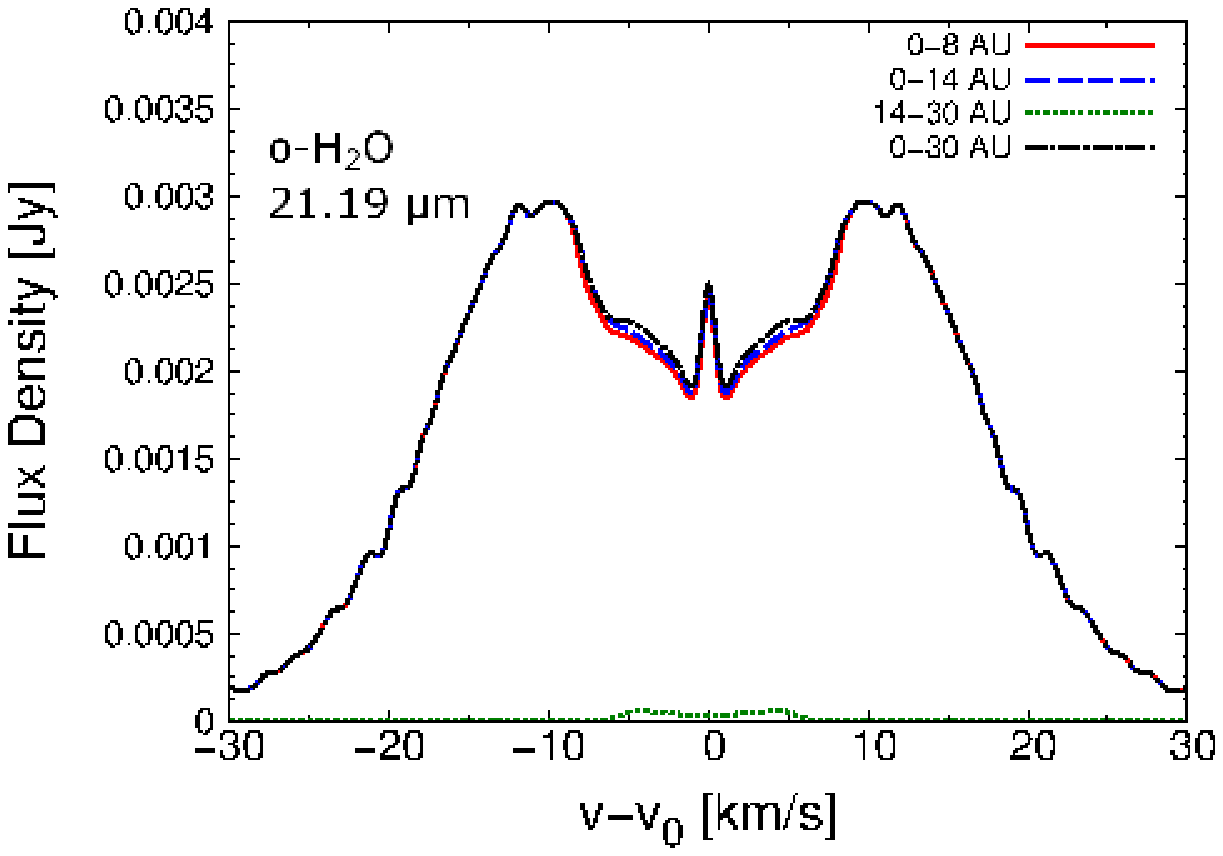}
\includegraphics[scale=0.5]{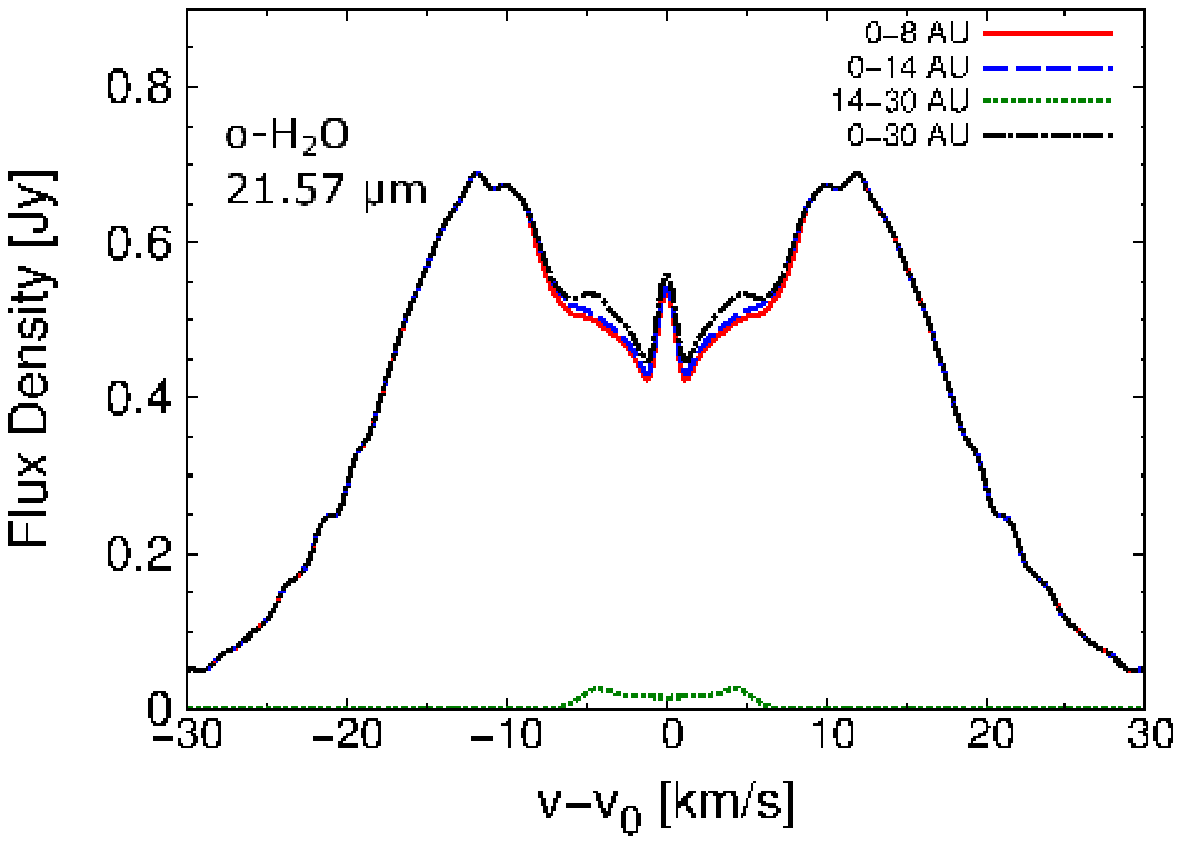}
\includegraphics[scale=0.5]{Figure4b_paper2_HAe_24.00um_trans56_20160911ver3r.eps}
\end{center}
\caption{\noindent (Continued.) The velocity profiles of mid-infrared ortho-$\mathrm{H_2O}$ lines at $\lambda$=19.53$\mu$m (top left), 19.64$\mu$m (top right),  21.12$\mu$m (middle left), 21.19$\mu$m (middle right), 21.57$\mu$m (bottom left), and 24.00$\mu$m (bottom right), from the Herbig Ae disk. 
}\label{Figure14_HerbigAe}
\end{figure} 


\begin{thebibliography}{}
\bibitem[Aikawa \& Nomura(2006)]{Aikawa2006} Aikawa, Y., \& Nomura, H.\ 2006, \apj, 642, 1152
\bibitem[Akimkin et al.(2013)]{Akimkin2013} Akimkin, V., Zhukovska, S., Wiebe, D., et al.\ 2013, \apj, 766, 8
\bibitem[ALMA Partnership et al.(2015)]{ALMA2015} ALMA Partnership, Brogan, C.~L., P{\'e}rez, L.~M., et al.\ 2015, \apjl, 808, L3 
\bibitem[Andrews et al.(2016)]{Andrews2016} Andrews, S.~M., Wilner, D.~J., Zhu, Z., et al.\ 2016, \apjl, 820, L40
\bibitem[Antonellini et al.(2016)]{Antonellini2016} Antonellini, S., Kamp, I., Lahuis, F., et al.\ 2016, \aap, 585, A61
\bibitem[Antonellini et al.(2015)]{Antonellini2015} Antonellini, S., Kamp, I., Riviere-Marichalar, P., et al.\ 2015, \aap, 582, A105
\bibitem[Banzatti et al.(2012)]{Banzatti2012} Banzatti, A., Meyer, M.~R., Bruderer, S., et al.\ 2012, \apj, 745, 90
\bibitem[Banzatti et al.(2015)]{Banzatti2015} Banzatti, A., Pinilla, P., Ricci, L., et al.\ 2015, \apjl, 815, L15
\bibitem[Banzatti et al.(2016)]{Banzatti2016} Banzatti, A., Pontoppidan, K.~M., Salyk, C., et al.\ 2016, \apj \ in press, arXiv:1611.06230
\bibitem[Benisty et al.(2010)]{Benisty2010} Benisty, M., Tatulli, E., M{\'e}nard, F., \& Swain, M.~R.\ 2010, \aap, 511, A75
\bibitem[Bethell \& Bergin(2011)]{Bethell2011} Bethell, T.~J., \& Bergin, E.~A.\ 2011, \apj, 739, 78
\bibitem[Blevins et al.(2016)]{Blevins2016} Blevins, S.~M., Pontoppidan, K.~M., Banzatti, A., et al.\ 2016, \apj, 818, 22
\bibitem[Cieza et al.(2016)]{Cieza2016} Cieza, L.~A., Casassus, S., Tobin, J., et al.\ 2016, \nat, 535, 258
\bibitem[Davis(2005)]{Davis2005} Davis, S.~S.\ 2005, \apj, 620, 994 
\bibitem[Dent et al.(2013)]{Dent2013} Dent, W.~R.~F., Thi, W.~F., Kamp, I., et al.\ 2013, \pasp, 125, 477 
\bibitem[Dominik et al.(2003)]{Dominik2003} Dominik, C., Dullemond, C.~P., Waters, L.~B.~F.~M., \& Walch, S.\ 2003, \aap, 398, 607
\bibitem[Du \& Bergin(2014)]{Du2014} Du, F., \& Bergin, E.~A.\ 2014, \apj, 792, 2
\bibitem[Dullemond \& Dominik(2004)]{Dullemond2004} Dullemond, C.~P., \& Dominik, C.\ 2004, \aap, 417, 159
\bibitem[Dutrey et al.(2014)]{Dutrey2014} Dutrey, A., Semenov, D., Chapillon, E., et al.\ 2014, Protostars and Planets VI, University of Arizona Press, 317
\bibitem[Espaillat et al.(2007)]{Espaillat2007} Espaillat, C., Calvet, N., D'Alessio, P., et al.\ 2007, \apjl, 670, L135
\bibitem[Eistrup et al.(2016)]{Eistrup2016} Eistrup, C., Walsh, C., \& van Dishoeck, E.~F.\ 2016, \aap, 595, A83 
\bibitem[Faure \& Josselin(2008)]{Faure2008} Faure, A., \& Josselin, E.\ 2008, \aap, 492, 257 
\bibitem[Fedele et al.(2012)]{Fedele2012} Fedele, D., Bruderer, S., van Dishoeck, E.~F., et al.\ 2012, \aap, 544, LL9 
\bibitem[Fedele et al.(2013)]{Fedele2013} Fedele, D., Bruderer, S., van Dishoeck, E.~F., et al.\ 2013, \aap, 559, AA77 
\bibitem[Fedele et al.(2011)]{Fedele2011} Fedele, D., Pascucci, I., Brittain, S., et al.\ 2011, \apj, 732, 106
\bibitem[Fujiwara et al.(2006)]{Fujiwara2006} Fujiwara, H., Honda, M., Kataza, H., et al.\ 2006, \apjl, 644, L133
\bibitem[Fukagawa et al.(2006)]{Fukagawa2006} Fukagawa, M., Tamura, M., Itoh, Y., et al.\ 2006, \apjl, 636, L153
\bibitem[Fukagawa et al.(2013)]{Fukagawa2013} Fukagawa, M., Tsukagoshi, T., Momose, M., et al.\ 2013, \pasj, 65, L14 
\bibitem[Furuya et al.(2013)]{Furuya2013} Furuya, K., Aikawa, Y., Nomura, H., Hersant, F., \& Wakelam, V.\ 2013, \apj, 779, 11 
\bibitem[Garaud \& Lin(2007)]{Garaud2007} Garaud, P., \& Lin, D.~N.~C.\ 2007, \apj, 654, 606
\bibitem[Graedel et al.(1982)]{Graedel1982} Graedel, T.~E., Langer, W.~D., \& Frerking, M.~A.\ 1982, \apjs, 48, 321
\bibitem[Harsono et al.(2015)]{Harsono2015} Harsono, D., Bruderer, S., \& van Dishoeck, E.~F.\ 2015, \aap, 582, A41
\bibitem[Hasegawa et al.(1992)]{Hasegawa1992} Hasegawa, T.~I., Herbst, E., \& Leung, C.~M.\ 1992, \apjs, 82, 167
\bibitem[Haworth et al.(2016)]{Haworth2016} Haworth, T.~J., Ilee, J.~D., Forgan, D.~H., et al.\ 2016, \pasa, 33, e053
\bibitem[Hayashi(1981)]{Hayashi1981} Hayashi, C.\ 1981, Progress of Theoretical Physics Supplement, 70, 35 
\bibitem[Hayashi et al.(1985)]{Hayashi1985} Hayashi, C., Nakazawa, K., \& Nakagawa, Y.\ 1985, Protostars and Planets II, University of Arizona Press, 1100
\bibitem[Heinzeller et al.(2011)]{Heinzeller2011} Heinzeller, D., Nomura, H., Walsh, C., \& Millar, T.~J.\ 2011, \apj, 731, 115
\bibitem[Henning \& Semenov(2013)]{Henning2013} Henning, T., \& Semenov, D.\ 2013, Chemical Reviews, 113, 9016
\bibitem[Hirota et al.(2014)]{Hirota2014ApJL} Hirota, T., Kim, M.~K., Kurono, Y., \& Honma, M.\ 2014, \apjl, 782, L28
\bibitem[Hogerheijde et al.(2011)]{Hogerheijde2011} Hogerheijde, M.~R., Bergin, E.~A., Brinch, C., et al.\ 2011, Science, 334, 338
\bibitem[Hogerheijde \& van der Tak(2000)]{Hogerheijde2000} Hogerheijde, M.~R., \& van der Tak, F.~F.~S.\ 2000, \aap, 362, 697
\bibitem[Honda et al.(2012)]{Honda2012} Honda, M., Maaskant, K., Okamoto, Y.~K., et al.\ 2012, \apj, 752, 143
\bibitem[Honda et al.(2015)]{Honda2015} Honda, M., Maaskant, K., Okamoto, Y.~K., et al.\ 2015, \apj, 804, 143
\bibitem[Isella et al.(2016)]{Isella2016} Isella, A., Guidi, G., Testi, L., et al.\ 2016, Phys. Rev. Lett. 117, 251101 
\bibitem[Kamp et al.(2013)]{Kamp2013} Kamp, I., Thi, W.-F., Meeus, G., et al.\ 2013, \aap, 559, A24 
\bibitem[Kenyon \& Hartmann(1995)]{KenyonHartmann1995} Kenyon, S.~J., \& Hartmann, L.\ 1995, \apjs, 101, 117
\bibitem[Krijt et al.(2016)]{Krijt2016} Krijt, S., Ciesla, F.~J., \& Bergin, E.~A.\ 2016, \apj \ in press, arXiv:1610.06463
\bibitem[Kristensen et al.(2016)]{Kristensen2016} Kristensen, L.~E., Brown, J.~M., Wilner, D., \& Salyk, C.\ 2016, \apjl, 822, L20 
\bibitem[Lahuis et al.(2007)]{Lahuis2007} Lahuis, F., van Dishoeck, E.~F., Blake, G.~A., et al.\ 2007, \apj, 665, 492
\bibitem[Maaskant et al.(2013)]{Maaskant2013} Maaskant, K.~M., Honda, M., Waters, L.~B.~F.~M., et al.\ 2013, \aap, 555, A64
\bibitem[Mandell et al.(2012)]{Mandell2012} Mandell, A.~M., Bast, J., van Dishoeck, E.~F., et al.\ 2012, \apj, 747, 92
\bibitem[Mathis et al.(1977)]{Mathis1977} Mathis, J.~S., Rumpl, W., \& Nordsieck, K.~H.\ 1977, \apj, 217, 425
\bibitem[Meeus et al.(2012)]{Meeus2012} Meeus, G., Montesinos, B., Mendigut{\'{\i}}a, I., et al.\ 2012, \aap, 544, AA78 
\bibitem[Meeus et al.(2001)]{Meeus2001} Meeus, G., Waters, L.~B.~F.~M., Bouwman, J., et al.\ 2001, \aap, 365, 476
\bibitem[Meijerink et al.(2008)]{Meijerink2008} Meijerink, R., Poelman, D.~R., Spaans, M., Tielens, A.~G.~G.~M., \& Glassgold, A.~E.\ 2008, \apjl, 689, L57
\bibitem[Meijerink et al.(2009)]{Meijerink2009} Meijerink, R., Pontoppidan, K.~M., Blake, G.~A., Poelman, D.~R., \& Dullemond, C.~P.\ 2009, \apj, 704, 1471
\bibitem[Menu et al.(2015)]{Menu2015} Menu, J., van Boekel, R., Henning, T., et al.\ 2015, \aap, 581, A107
\bibitem[Min et al.(2016)]{Min2016} Min, M., Bouwman, J., Dominik, C., et al.\ 2016, \aap, 593, A11
\bibitem[Min et al.(2011)]{Min2011} Min, M., Dullemond, C.~P., Kama, M., \& Dominik, C.\ 2011, \icarus, 212, 416 
\bibitem[Morbidelli et al.(2016)]{Morbidelli2016} Morbidelli, A., Bitsch, B., Crida, A., et al.\ 2016, \icarus, 267, 368
\bibitem[Morbidelli et al.(2000)]{Morbidelli2000} Morbidelli, A., Chambers, J., Lunine, J.~I., et al.\ 2000, Meteoritics and Planetary Science, 35, 1309
\bibitem[Morbidelli et al.(2012)]{Morbidelli2012} Morbidelli, A., Lunine, J.~I., O'Brien, D.~P., Raymond, S.~N., \& Walsh, K.~J.\ 2012, Annual Review of Earth and Planetary Sciences, 40, 251
\bibitem[Mulders et al.(2015)]{Mulders2015} Mulders, G.~D., Ciesla, F.~J., Min, M., \& Pascucci, I.\ 2015, \apj, 807, 9
\bibitem[Nomura \& Millar(2005)]{NomuraMillar2005} Nomura, H., \& Millar, T.~J.\ 2005, \aap, 438, 923
\bibitem[Nomura et al.(2007)]{Nomura2007} Nomura, H., Aikawa, Y., Tsujimoto, M., Nakagawa, Y., \& Millar, T.~J.\ 2007, \apj, 661, 334 
\bibitem[Notsu et al.(2016)]{Notsu2016}Notsu, S., Nomura, H., Ishimoto, D., Walsh, C., Honda, M., Hirota, T., \& Millar, T.~J. \ 2016, \apj, 827, 113
\bibitem[Notsu et al.(2015)]{Notsu2015} Notsu, S., Nomura, H., Ishimoto, D., et al.\ 2015, Revolution in Astronomy with ALMA: The Third Year, ASP Conference Series, 499, 289
\bibitem[{\"O}berg et al.(2011)]{Oberg2011} {\"O}berg, K.~I., Murray-Clay, R., \& Bergin, E.~A.\ 2011, \apjl, 743, L16 
\bibitem[Oka et al.(2011)]{Oka2011} Oka, A., Nakamoto, T., \& Ida, S.\ 2011, \apj, 738, 141
\bibitem[Okuzumi et al.(2016)]{Okuzumi2016} Okuzumi, S., Momose, M., Sirono, S.-i., Kobayashi, H., \& Tanaka, H.\ 2016, \apj, 821, 82 
\bibitem[Okuzumi et al.(2012)]{Okuzumi2012} Okuzumi, S., Tanaka, H., Kobayashi, H., \& Wada, K.\ 2012, \apj, 752, 106
\bibitem[Packham et al.(2012)]{Packham2012} Packham, C., Honda, M., Richter, M., et al.\ 2012, \procspie, 8446, 84467G
\bibitem[Pani{\'c} et al.(2014)]{Panic2014} Pani{\'c}, O., Ratzka, T., Mulders, G.~D., et al.\ 2014, \aap, 562, A101
\bibitem[Piso et al.(2015)]{Piso2015} Piso, A.-M.~A., {\"O}berg, K.~I., Birnstiel, T., \& Murray-Clay, R.~A.\ 2015, \apj, 815, 109
\bibitem[Podio et al.(2013)]{Podio2013} Podio, L., Kamp, I., Codella, C., et al.\ 2013, \apjl, 766, L5 
\bibitem[Pontoppidan et al.(2010a)]{Pontoppidan2010a} Pontoppidan, K.~M., Salyk, C., Blake, G.~A., et al.\ 2010a, \apj, 720, 887
\bibitem[Pontoppidan et al.(2010b)]{Pontoppidan2010b} Pontoppidan, K.~M., Salyk, C., Blake, G.~A., {\ K\&auml}ufl, H.~U.\ 2010b, \apjl, 722, L173  
\bibitem[Ros \& Johansen(2013)]{Ros2013} Ros, K., \& Johansen, A.\ 2013, \aap, 552, A137 
\bibitem[Rybicki \& Lightman(1986)]{Rybicki1986} Rybicki, G.~B., \& Lightman, A.~P.\ 1986, Radiative Processes in Astrophysics, by George B.~Rybicki, Alan P.~Lightman, pp.~400.~ISBN 0-471-82759-2.~Wiley-VCH, June 1986
\bibitem[Salinas et al.(2016)]{Salinas2016} Salinas, V.~N., Hogerheijde, M.~R., Bergin, E.~A., et al.\ 2016, \aap, 591, A122
\bibitem[Salyk et al.(2008)]{Salyk2008} Salyk, C., Pontoppidan, K.~M., Blake, G.~A., et al.\ 2008, \apjl, 676, L49
\bibitem[Salyk et al.(2011)]{Salyk2011} Salyk, C., Pontoppidan, K.~M., Blake, G.~A., Najita, J.~R., \& Carr, J.~S.\ 2011, \apj, 731, 130
\bibitem[Sato et al.(2016)]{Sato2016} Sato, T., Okuzumi, S., \& Ida, S.\ 2016, \aap, 589, A15
\bibitem[Sch{\"o}ier et al.(2005)]{Schoier2005} Sch{\"o}ier, F.~L., van der Tak, F.~F.~S., van Dishoeck, E.~F., \& Black, J.~H.\ 2005, \aap, 432, 369
\bibitem[Strom et al.(1989)]{Strom1989} Strom, K.~M., Strom, S.~E., Edwards, S., Cabrit, S., \& Skrutskie, M.~F.\ 1989, \aj, 97, 1451
\bibitem[Tilling et al.(2012)]{Tilling2012} Tilling, I., Woitke, P., Meeus, G., et al.\ 2012, \aap, 538, A20
\bibitem[Tsukagoshi et al.(2016)]{Tsukagoshi2016} Tsukagoshi, T., Nomura, H., Muto, T., et al.\ 2016, \apjl, 829, L35
\bibitem[van Dishoeck et al.(2014)]{vanDishoeck2014} van Dishoeck, E.~F., Bergin, E.~A., Lis, D.~C., \& Lunine, J.~I.\ 2014, Protostars and Planets VI, University of Arizona Press, 835 
\bibitem[Vasyunin et al.(2011)]{Vasyunin2011} Vasyunin, A.~I., Wiebe, D.~S., Birnstiel, T., et al.\ 2011, \apj, 727, 76 
\bibitem[Wada et al.(2013)]{Wada2013} Wada, K., Tanaka, H., Okuzumi, S., et al.\ 2013, \aap, 559, AA62 
\bibitem[Walsh et al.(2014b)]{Walsh2014b} Walsh, C., Juh{\'a}sz, A., Pinilla, P., et al.\ 2014b, \apjl, 791, L6
\bibitem[Walsh et al.(2010)]{Walsh2010} Walsh, C., Millar, T.~J., \& Nomura, H.\ 2010, \apj, 722, 1607
\bibitem[Walsh et al.(2014a)]{Walsh2014a} Walsh, C., Millar, T.~J., Nomura, H., et al.\ 2014a, \aap, 563, AA33
\bibitem[Walsh et al.(2012)]{Walsh2012} Walsh, C., Nomura, H., Millar, T.~J., \& Aikawa, Y.\ 2012, \apj, 747, 114
\bibitem[Walsh et al.(2015)]{Walsh2015} Walsh, C., Nomura, H., \& van Dishoeck, E.\ 2015, \aap, 582, A88 
\bibitem[Weingartner \& Draine(2001)]{WeingartnerDraine2001} Weingartner, J.~C., \& Draine, B.~T.\ 2001, \apj, 548, 296
\bibitem[Woitke et al.(2009b)]{Woitke2009b} Woitke, P., Thi, W.-F., Kamp, I., \& Hogerheijde, M.~R.\ 2009b, \aap, 501, L5
\bibitem[Woodall et al.(2007)]{Woodall2007} Woodall, J., Ag{\'u}ndez, M., Markwick-Kemper, A.~J., \& Millar, T.~J.\ 2007, \aap, 466, 1197
\bibitem[Zhang et al.(2016)]{Zhang2016} Zhang, K., Bergin, E.~A., Blake, G.~A., et al.\ 2016, \apjl, 818, L16
\bibitem[Zhang et al.(2015)]{Zhang2015} Zhang, K., Blake, G.~A., \& Bergin, E.~A.\ 2015, \apjl, 806, L7
\bibitem[Zhang et al.(2013)]{Zhang2013} Zhang, K., Pontoppidan, K.~M., Salyk, C., \& Blake, G.~A.\ 2013, \apj, 766, 82
\end{thebibliography}
\end{document}